\documentclass[journal, onecolumn]{IEEEtran}
\usepackage{cite}
\usepackage{tikz}
\usetikzlibrary{arrows.meta,positioning,calc,shapes.geometric}
\usepackage[caption=false,font=footnotesize]{subfig}
\usepackage{graphicx}
\usepackage{amssymb}
\usepackage{array}
\usepackage[cmex10]{amsmath, nccmath}
\usepackage{mathtools}
\usepackage{upgreek}
\usepackage{enumitem}
\usepackage{bm}
\usepackage{orcidlink}
\newtheorem{theorem}{Theorem}

\newtheorem{lemma}{Lemma}
\newtheorem{definition}{Definition}

\newcommand{\E}[1]{\mathbb E\left[#1\right]}
\newcommand{\Prob}[1]{\mathbb P\left[#1\right]}
\newcommand{\1}[1]{1\left\{#1\right\}}
\newcommand{\mc}[1]{\mathcal{#1}}
\newcommand{\mb}[1]{\mathbf{#1}}
\newcommand{\mr}[1]{\mathrm{#1}}
\newcommand{\ms}[1]{\mathsf{#1}}
\newcommand{\nn}[1]{\widehat{#1}}
\newcommand{\CG}{C}
\newcommand{\dG}{d}
\newcommand{\thmref}[1]{Theorem~\ref{#1}}
\newcommand{\lemref}[1]{Lemma~\ref{#1}}

\newcommand{\secref}[1]{Section~\ref{#1}}
\newcommand{\appref}[1]{Appendix~\ref{#1}}
\newcommand{\defnref}[1]{Definition~\ref{#1}}

\newcommand{\SE}{\zeta}
\newcommand{\OP}{\Xi}
\newcommand{\ntr}{\overline{n}}

\DeclareMathOperator*{\argmax}{arg\,max}
\DeclareMathOperator*{\argmin}{arg\,min}
\usepackage{hyperref}
\usepackage{url}
\hypersetup{colorlinks=true, linkcolor=blue, citecolor=blue, urlcolor = blue}
\renewcommand{\IEEEproofindentspace}{1\parindent}
\allowdisplaybreaks

\begin{document}
\title{Variable-Length Feedback Coding With\\Universal Decoding and Bandit Channel Selection}
\author{Recep Can Yavas~\orcidlink{0000-0002-5640-515X},~\IEEEmembership{Member,~IEEE}
\thanks{R. C. Yavas is with the Department of Electrical and Electronics Engineering at Bilkent University, 06800, Ankara, Turkey (e-mail: ryavas@bilkent.edu.tr). This work was supported by T{\"U}B{\.I}TAK under the 2232-B International Fellowship for Early Stage Researchers Programme, Project No. 125C979.}}
\IEEEoverridecommandlockouts
\maketitle

\begin{abstract}
We study variable-length feedback (VLF) coding over $K$ additive white Gaussian noise (AWGN) channels whose noise variances are unknown to both terminals but lie in a known interval. The receiver selects one channel at each time, conveys the selection over noiseless feedback, and must learn the channel qualities while decoding, to avoid overusing suboptimal channels. This setting, which we call the VLF-Bandit problem, combines VLF coding, universal decoding, and bandit channel selection. We develop two joint coding and channel-selection schemes. The VLSF-Bandit scheme is a single-phase variable-length stop-feedback (VLSF) scheme based on a Gaussian random codebook, a score formed from a ratio of likelihood mixtures, and the asymptotically optimal $\text{kl-UCB}^{++}$ bandit algorithm applied to the received power. The VLF-Bandit scheme sends the message in packets by Schalkwijk--Kailath transmission, uses its normalized outputs as noise samples to learn the variances, and lets its exploration packets carry message bits. We prove second-order achievability bounds for both schemes, which are first-order optimal. In the logarithm of the achievable codebook size, the magnitude of the $\log N$ coefficient, with $N$ the average decoding time, is a baseline of one plus a channel-exploration term for VLSF-Bandit. For VLF-Bandit, this magnitude is smaller, with a universal-decoding term replacing the baseline and a smaller channel-exploration term. The VLF-Bandit bound improves on the universal VLF bound of Yavas and Tan (2025) for $K=1$ and on the bound of an explore-then-commit strategy whenever $K\geq2$.

\end{abstract}


\section{Introduction} \label{sec:intro}

Feedback does not increase the capacity of memoryless channels~\cite{shannon1956zero}. Yet it can simplify capacity-achieving schemes~\cite{horstein, schalkwijk}. The benefits of feedback are more pronounced for variable-length feedback (VLF) codes, in which transmission stops at a random time determined by the noise realization. Burnashev~\cite{burnashev1976data} establishes the optimal error exponent of VLF coding over a discrete memoryless channel (DMC), given by
\begin{align}
    \lim_{\epsilon \to 0}
    \frac{\log \frac{1}{\epsilon}}{\E{\tau}}
    = C_1 \left(1 - \frac{R}{C}\right),
    \label{eq:burnashev}
\end{align}
where $C$ is the capacity of the DMC, $R \in (0,C)$ is the rate, $\epsilon$ is the average error probability, $\E{\tau}$ is the average decoding time, and
\begin{align}
    C_1
    &\triangleq \max_{x,x' \in \mc{X}}
    D(P_{Y|X=x}\|P_{Y|X=x'})
    \label{eq:C1intro}
\end{align}
is the Kullback--Leibler (KL) divergence between the conditional output distributions induced by the two most distinguishable input symbols. For every rate below capacity, the exponent in~\eqref{eq:burnashev} exceeds the random-coding error exponent for fixed-length codes without feedback~\cite{Gallager1965}.

\subsection{VLF Literature}

Yamamoto and Itoh~\cite{YamamotoItoh} propose a simple alternative to Burnashev's scheme that still achieves the optimal exponent in~\eqref{eq:burnashev}. It alternates fixed-length communication and confirmation phases, and in each confirmation phase, the transmitter sends one of two control sequences, built from the two input symbols that attain~\eqref{eq:C1intro}, to tell the receiver whether to accept its estimate. Chen \emph{et al.}~\cite{chen2023} derive non-asymptotic achievability bounds for a variant with a finite number of feedback instances, in which the phases may have different lengths and the outputs of a rejected phase are kept.

Error-exponent analysis says little about the maximum achievable rate in the finite-blocklength regime, where the target rate is close to capacity. To address this, Polyanskiy \emph{et al.}~\cite{polyanskiy2011feedback} extend Burnashev's work to non-vanishing error probabilities and derive achievability and converse bounds on the logarithm of the maximum achievable codebook size $\log M_{\mr{VLF}}^*(N,\epsilon)$ under an average decoding-time constraint $N$ and an average error probability constraint $\epsilon$. They show that
\begin{align}
    \frac{NC}{1-\epsilon} - \log N + O(1)
    \leq \log M_{\mr{VLF}}^*(N,\epsilon)
    \leq \frac{NC}{1-\epsilon}
    + \frac{h(\epsilon)}{1-\epsilon},
    \label{eq:pol_upper}
\end{align}
where $h(\epsilon) \triangleq -\epsilon \log \epsilon - (1-\epsilon)\log(1-\epsilon)$ is the binary entropy function. The bounds in~\eqref{eq:pol_upper} imply that the $\epsilon$-capacity of VLF codes is $\frac{C}{1-\epsilon}$ and that the achievable second-order term in the rate scales as $O(\frac{\log N}{N})$. The lower bound in~\eqref{eq:pol_upper} is achieved by variable-length stop-feedback (VLSF) codes, in which a single feedback bit indicates whether the transmitter should stop or continue. VLSF codes, a subset of VLF codes, use codewords that are fixed before communication starts. In \cite[Th.~1]{truong2016gaussian}, Truong and Tan establish the corresponding bounds for VLSF codes over the additive white Gaussian noise (AWGN) channel under an average power constraint. Variants of VLSF codes with a finite number of feedback instances appear in~\cite{kim2015VLF, williamson2015VLConvolution, Vakilinia2016, yavas2021VLF, yavas2023VLF, Yang2022}, and multi-transmitter extensions in~\cite{yavas2023VLF, truong2018Journal}.  For VLF codes, Yavas and Tan~\cite{yavasTan2025} improve the coefficient of the $\log N$ term in the lower bound in~\eqref{eq:pol_upper} from $-1$ to $-\frac{C}{C_1}$ by modifying the Yamamoto--Itoh construction to let the communication and confirmation phases have variable lengths. Their scheme does not belong to the VLSF subset.

A complementary line of work studies deterministic schemes built on posterior-matching partitions. For symmetric binary-input channels, Naghshvar \emph{et al.}~\cite{naghshvar} achieve Burnashev's exponent with a single-phase small-enough-difference (SED) encoder, which splits the message set at each time into two subsets of nearly equal posterior probability, and they extend the idea to arbitrary DMCs in~\cite{naghshvar2015EJS}. Yang \emph{et al.}~\cite{yang2022SED} extend the SED encoder to binary asymmetric channels, Antonini \emph{et al.}~\cite{antonini2024} propose a posterior-matching encoder of polynomial complexity for the binary symmetric channel (BSC), and Yavas~\cite{yavas2026PM} extends both the non-asymptotic analysis and the low-complexity encoder to binary-input memoryless symmetric channels, including the binary-input AWGN channel after output quantization. With the refined analyses in~\cite{yang2022SED, antonini2024, yavas2026PM}, the SED encoder attains the same second-order term $-\frac{C}{C_1}\log N$ as in~\cite{yavasTan2025}, with sharper non-asymptotic bounds. Yavas~\cite{yavas2026converse} has recently shown that no VLF code can do better over a DMC with $C>0$ and $C_1<\infty$, so that
\begin{align}
    \log M_{\mr{VLF}}^*(N,\epsilon)=\frac{NC}{1-\epsilon}-\frac{C}{C_1}\log N+o(\log N).
    \label{eq:vlf_converse_dmc}
\end{align}
For the AWGN channel with an average power constraint, the best VLF converse known to us still has the form of the upper bound in~\eqref{eq:pol_upper}, with no $\log N$ term~\cite{truong2016gaussian}.

For the AWGN channel with noiseless feedback, the Schalkwijk--Kailath (SK) scheme~\cite{schalkwijk, schalkwijk1966bandlimited} achieves every rate below capacity, with an error probability that decays doubly exponentially in the blocklength. The transmitter first sends the message as a point of a pulse-amplitude constellation. At each later time, the transmitter learns the receiver's current estimation error through the feedback and sends this error multiplied by a gain that depends on the noise variance. Gallager and Nakibo\u{g}lu~\cite{gallagerNakiboglu2010} derive the scheme from Elias's transmission of a Gaussian random variable over a Gaussian channel with feedback. Shayevitz and Feder~\cite{shayevitz} identify the SK scheme as the posterior-matching scheme of the AWGN channel. Truong \emph{et al.}~\cite{truongfong2017logL} combine the scheme with power control to determine the $\epsilon$-capacity of fixed-length feedback codes under an expected power constraint.

The preceding achievability results assume that the channel law is known to the code designer. Universal codes require only that the channel belong to a prescribed family. For DMCs, Goppa's maximum mutual information decoder~\cite{goppa} attains capacity universally, and in~\cite[Th.~10.2]{csiszarbook}, Csisz\'ar and K\"orner show that the same decoder also achieves the random-coding error exponent of constant-composition codes.

There are also several works on universal coding with feedback. Tchamkerten and Telatar~\cite{tchamkerten2006variable} achieve Burnashev's exponent universally over families of BSCs and Z channels through a universal Yamamoto--Itoh scheme. Lomnitz and Feder adapt the rate to the empirical mutual information over a fixed block without a channel model~\cite{lomnitzfeder}, and vary the transmit power of the SK scheme with the empirical power of an unknown individual noise sequence so that the rate and the signal-to-noise ratio asymptotically satisfy the Gaussian capacity formula~\cite{lomnitzFederPower2009}. Yavas and Tan~\cite{yavasTan2025} analyze universal VLF codes for DMCs and for AWGN channels with unknown noise variance. For both the BSC and the AWGN channel, replacing the channel-dependent information density of a known-channel scheme by a universal score incurs an additional $\frac{1}{2}\log N$ cost in the second-order term of their achievability bounds.
All of these universal results operate over a single unknown channel; we now consider settings in which several channels are available simultaneously.

\subsection{Motivation and the VLF-Bandit Problem} \label{sec:motivation}

Modern wireless systems often operate over multiple orthogonal resources of varying quality. A user may communicate through several frequency bands, spatial streams, or subcarriers, each modeled as a memoryless channel, while hardware and scheduling constraints allow only one of these resources per channel use. The quality of each resource depends on physical parameters such as noise power, fading, path loss, and interference, which may be unknown to both terminals before transmission starts. The receiver, however, observes the output statistics and knows which resources are available, so it is natural to let the receiver act as the channel scheduler: after observing the past outputs, it selects the next channel and sends this selection to the transmitter through the noiseless feedback link.

There is a second reason to let the receiver select the channels. If the transmitter chose the channel index as an additional message-dependent input, the index sequence would itself carry information, since the receiver observes which channel is in use. With $K$ distinguishable resources, it could carry up to $\log K$ nats per channel use, even when the individual channels have zero capacity. The problem would then become the universal version of Problem~7.28 in Cover and Thomas~\cite{coverthomas}. We therefore require the channel-selection rule to reside at the receiver, depending only on the common randomness, the past selections, and the past outputs, but not directly on the message.

Under this convention, the receiver has two coupled tasks: it schedules the channels so that high-capacity channels are used more often, and it decodes the message from observations collected under an adaptively chosen channel sequence. We call this adaptive multi-channel setting the \emph{VLF-Bandit} problem and consider two code classes, VLSF-Bandit and VLF-Bandit, which differ in how the transmitter uses the feedback. In the VLSF-Bandit setting, the transmitted codeword is fixed before communication starts, and the feedback link carries only the channel selections and the stop signal. In the VLF-Bandit setting, the transmitter also observes the past outputs through noiseless feedback and may adapt the transmitted symbols to the feedback history.

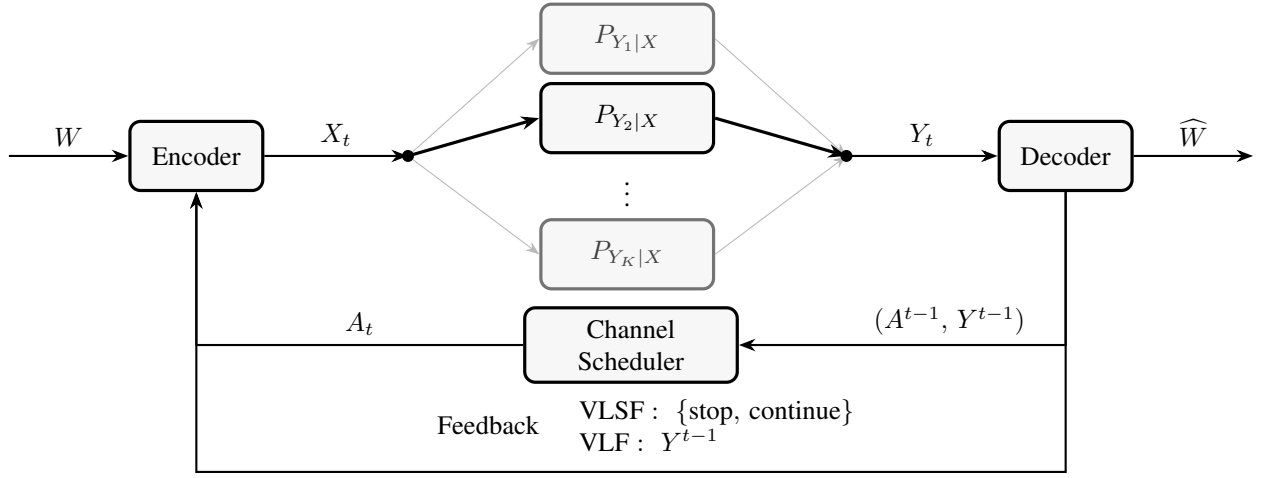
\begin{figure*}[t]
\centering
\begin{tikzpicture}[
    >={Stealth[length=2.2mm]},
    thick,
    execute at begin node={\linespread{1}\selectfont},
    box/.style      = {rectangle, rounded corners, draw=black, very thick,
                       minimum width=5em, minimum height=2.6em, align=center, fill=black!3},
    chbox/.style    = {rectangle, rounded corners, draw=black, very thick,
                       minimum width=6.5em, minimum height=2.6em, align=center, fill=black!3},
    chboxoff/.style = {rectangle, rounded corners, draw=black!55, very thick,
                       minimum width=6.5em, minimum height=2.6em, align=center,
                       fill=black!3, text=black!75},
    pivot/.style    = {circle, fill=black, inner sep=0pt, minimum size=4.5pt},
    line/.style     = {draw, thick, ->},
    candA/.style    = {draw, thin, gray!55, -{Stealth[length=1.8mm]}},
    arm/.style      = {draw, very thick, black}
]

\node[box] (ENC) at (0,    0) {Encoder};
\node[box] (DEC) at (11.5, 0) {Decoder};
\node[box, minimum width=8.0em, minimum height=2.8em] (SCH) at (5.75, -2.5) {Channel\\Scheduler};

\node[chboxoff] (CH1) at (5.7,  1.55) {$P_{Y_1\mid X}$};
\node[chbox]    (CH2) at (5.7,  0.5) {$P_{Y_2\mid X}$};
\node           (DTS) at (5.7, -0.4) {$\vdots$};
\node[chboxoff] (CHK) at (5.7, -1.3) {$P_{Y_K\mid X}$};

\coordinate (PL) at (2.8, 0);
\coordinate (PR) at (8.6, 0);
\coordinate (DSH) at ($(DEC.south |- SCH.east)$);
\coordinate (SCHIN) at ($(SCH.east)+(1.25,0)$);
\coordinate (FBR) at ($(DEC.south)+(0,-3.7)$);
\coordinate (FBL) at ($(ENC.south |- FBR)$);

\draw[candA] (PL) -- (CH1.west);
\draw[candA] (PL) -- (CHK.west);
\draw[candA] (CH1.east) -- (PR);
\draw[candA] (CHK.east) -- (PR);

\draw[arm, -{Stealth[length=2.4mm]}] (PL) -- (CH2.west);
\draw[arm, -{Stealth[length=2.4mm]}] (CH2.east) -- (PR);

\node[pivot] at (PL) {};
\node[pivot] at (PR) {};

\draw[line] ($(ENC.west) + (-1.58, 0)$) -- node[above] {$W$} (ENC.west);
\draw[line] (ENC.east) -- node[above] {$X_t$} (PL);
\draw[line] (PR) -- node[above] {$Y_t$} (DEC.west);
\draw[line] (DEC.east) -- node[above] {$\nn{W}$} ($(DEC.east) + (1.58, 0)$);

\draw[line] (DEC.south) -- (DSH) -- node[midway, above] {$(A^{t-1},\, Y^{t-1})$} (SCHIN) -- (SCH.east);

\coordinate (ALEFT) at ($(ENC.south |- SCH.west)$);
\draw[line] (SCH.west) -- node[midway, above] {$A_t$} (ALEFT) -- (ENC.south);

\draw[line] (DEC.south) -- (FBR) -- (FBL) -- (ENC.south);

\node[anchor=east, inner sep=0pt] at (4.55, -3.56) {Feedback};
\node[anchor=west, align=left, inner sep=0pt] at (5.05, -3.56) {$\text{VLSF}:\ \{\text{stop},\,\text{continue}\}$\\$\text{VLF}:\ Y^{t-1}$};

\end{tikzpicture}
\caption{VLF-Bandit system model, with the selected channel $A_t = 2$ in solid black.}
\label{fig:vlf_bandit_system}
\end{figure*}

A natural modular baseline is an explore-then-commit (ETC) strategy, which first sends pilot symbols to estimate the channel qualities and then runs a single-channel VLF code over the declared best channel. This separation has a cost, since the pilot observations probe the channels but carry no information about the message. A joint scheme instead lets exploration and communication share the channel uses, so that a use of a channel refines the receiver's knowledge of that channel and also carries information about the message. See Figure~\ref{fig:vlf_bandit_system} for our system model.

\subsection{Multi-Armed Bandit Literature}
The channel-selection component of the VLF-Bandit problem connects to the multi-armed bandit (MAB) literature; see~\cite{lattimore2020} for a textbook treatment. In the stochastic MAB model, a learner repeatedly pulls one of $K$ arms and observes a sample from that arm's unknown distribution. In \emph{regret minimization}~\cite{lai1985,auer2002}, the learner maximizes its expected cumulative reward over a fixed time horizon; in \emph{best-arm identification}~\cite{garivier2016}, it seeks to identify the arm with the largest mean as quickly as possible while satisfying an error probability criterion.

The scheduling problem considered here is closer in spirit to regret minimization than to best-arm identification. The receiver does not identify the best channel in a separate sensing phase; it keeps selecting channels while it decodes the message. Selecting a suboptimal channel therefore creates an opportunity cost analogous to regret, even though the performance criterion is the average decoding time rather than cumulative reward.

A common approach to regret minimization is optimism in the face of uncertainty: each arm receives an index equal to the largest mean that remains statistically plausible given its observations, and the arm with the largest index is pulled. The regret of such upper confidence bound (UCB) algorithms~\cite{auer2002} grows logarithmically in the horizon $T$ for bounded rewards, and Lai and Robbins~\cite{lai1985} lower-bound the coefficient of $\log T$ through the KL divergences between the arms' reward distributions. For one-parameter exponential families, the KL-UCB rule of Garivier and Capp\'e~\cite{garivier2011kl,cappe2013kl} attains this lower bound asymptotically, and the $\text{kl-UCB}^{++}$ variant of M\'enard and Garivier~\cite{menardGarivier2017} attains it and is also minimax optimal.

Bandit methods have been applied to channel access and selection in wireless networks~\cite{zhao2019bandits}. Closer to the multi-channel coding setting, Egger \emph{et al.}~\cite{egger2025} bound the number of samples needed to identify the maximal-capacity DMC, and Weinberger and Yemini~\cite{weinbergerYemini2023} study MABs with self-information rewards, in which the expected reward of an arm is the entropy of its source. Active sequential hypothesis testing~\cite{naghshvar2013active}, in which a decision maker adaptively selects sensing actions to discriminate among finitely many hypotheses, is another related framework; VLF coding over a known DMC is an instance of it.

The setting studied here falls outside the standard VLF, universal-coding, channel-identification, and bandit literature. Universal VLF coding treats a single unknown channel, bandit formulations do not include a variable-length code, and channel-identification methods learn the best channel from samples that carry no coded information. Extending a universal VLF code with a bandit channel-selection rule meets three obstacles. The decoder must aggregate evidence collected under different unknown parameters into one message score. Adaptive scheduling makes each score increment depend on the history, so the analyses of previous VLF codes that rely on independent and identically distributed (i.i.d.) observations do not apply directly. Regret bounds do not immediately translate into bounds on the average decoding time, the expectation of a stopping time. A full-feedback design based on the SK scheme needs no message score, but its gains depend on the noise variance, which must be learned from the same outputs that refine the message estimate, and the estimation error enters the second-order term. We address these obstacles through the contributions listed below.

\subsection{Contributions}
We study the VLF-Bandit problem for $K$ AWGN channels whose noise variances are unknown but lie in a known interval. The output of channel $k$ is $Y^{(k)} = X + Z^{(k)}$ with $Z^{(k)} \sim \mc{N}(0, \sigma_k^2)$, where $\sigma_k^2 \in [\sigma_{\min}^2,\sigma_{\max}^2]$ is unknown, and the inputs $X_t$ satisfy the average power constraint $\E{\sum_{t=1}^{\tau}X_t^2}\leq\E{\tau}$. Dividing the output by $\sigma_k$ gives the form of a scalar quasi-static fading AWGN channel~\cite{yang2014quasistatic} with gain $\frac{1}{\sigma_k}$. Feedback lets the decoding time adapt to this fixed gain, so the first-order term of our results is governed by the capacity of the best channel rather than by an outage capacity.

We propose two schemes, one for each code class. Our main results are second-order achievability bounds for them, with explicit coefficients of $\log N$.
\begin{enumerate}[leftmargin=*]
    \item We formulate the VLF-Bandit problem, a model that couples VLF coding, universal decoding over unknown noise variances, and adaptive channel selection. The formulation includes both VLSF-Bandit and VLF-Bandit codes.

    \item For the VLSF-Bandit setting, we develop a single-phase scheme based on a Gaussian random codebook, a score formed from a ratio of likelihood mixtures for the candidate residuals and the outputs, and an output-power $\text{kl-UCB}^{++}$ channel-selection rule. In \thmref{thm:vlsf_main}, the magnitude of the coefficient of $\log N$ consists of a baseline of one for the single-phase architecture, as in the lower bound in~\eqref{eq:pol_upper}, and a channel-exploration term that weights the capacity gap of each suboptimal channel by the reciprocal of the KL divergence between its output distribution and that of the best channel.

    \item For the VLF-Bandit setting, we develop a scheme that divides the message into packets and sends each by SK transmission over one channel selected by the receiver. After the first symbol of a packet, each output carries the receiver's estimation error, which depends only on past noise, so the normalized outputs are distributed as i.i.d.\ noise samples of the channel. Both terminals use them to estimate the noise variance, which sets the next gain, and to eliminate channels across packets. Exploration packets carry message bits at reduced power, and the final packet, sent over the selected channel, spends the energy they save. In \thmref{thm:vlf_main}, the magnitude of the coefficient of $\log N$ has no baseline, and its universal-decoding term, due to the variance estimates, is $\frac{1}{2(1+\sigma^2)^2}$ for the noise variance $\sigma^2$ of the best channel, smaller than the baseline of one in item~2. The channel-exploration term is also smaller than in item~2, because of the energy saved in exploration and because noise samples separate the channels better than output powers do. For $K=1$ and $\sigma^2=1$, this magnitude drops from about $0.673$ in~\cite[Th.~5]{yavasTan2025} to $0.125$.

    \item We further compare the joint design with an ETC strategy that identifies a channel from zero-input pilot observations by an asymptotically optimal fixed-confidence identification rule~\cite{kaufmann2021mixture} and then sends the whole message over the declared channel by the SK transmission of item~3. In every problem instance with $K\geq2$, the magnitude of the coefficient of $\log N$ in \thmref{thm:vlf_main} is strictly smaller than in the resulting ETC bound.
\end{enumerate}

\subsection{Paper Organization}

\secref{sec:notation} introduces the notation, and \secref{sec:problem} formulates the VLF-Bandit problem and defines the two code classes. \secref{sec:scheme} describes the coding schemes, a single-phase VLSF-Bandit scheme and a packetized SK VLF-Bandit scheme. \secref{sec:main} states the main achievability results. Sections~\ref{sec:vlsf_proof} and~\ref{sec:vlf_proof} prove the results for the two schemes. \secref{sec:etc_comparison} compares the VLF-Bandit scheme with the ETC strategy. \secref{sec:conclusion} concludes the paper. Appendices~\ref{app:Gaussian} and~\ref{app:vlf_lemmas} contain deferred proofs.

\section{Notation and Preliminaries} \label{sec:notation}

For $n \in \mathbb{N}$, let $[n] \triangleq \{1, \dots, n\}$ and $x^n \triangleq (x_1, \dots, x_n)$. For real numbers $a$ and $b$, we write $a \wedge b \triangleq \min\{a, b\}$ and $a \vee b \triangleq \max\{a,b\}$, and we let $x^+ \triangleq \max\{0, x\}$. For $a\leq b$, $\Pi_{[a,b]}(x)\triangleq b\wedge(a\vee x)$ denotes the projection of $x\in\mathbb{R}$ onto $[a,b]$. Ties in $\argmin$ and $\argmax$ are broken in favor of the smallest index.

The distribution of a random variable $X$ is denoted by $P_X$. The density of the Gaussian distribution $\mc{N}(0, v)$, $v > 0$, is denoted by $\phi_v(\cdot)$. We use the standard asymptotic notation $o(\cdot)$ and $O(\cdot)$.

All logarithms have base $e$. We use $\log_+(x) \triangleq \max\{\log x,0\}$ for $x > 0$. The binary entropy function $h(\cdot)$ is defined as
\begin{align}
    h(\epsilon) \triangleq - \epsilon \log (\epsilon) - (1-\epsilon) \log (1-\epsilon). \label{eq:hfunc}
\end{align}
For distributions $P$ and $Q$ with $P$ absolutely continuous with respect to $Q$, the KL divergence is denoted by
\begin{align}
     D(P \| Q) \triangleq \int\log \frac{dP}{dQ} dP. \label{eq:KLdef}
\end{align}

An AWGN channel with noise variance $\sigma^2 > 0$, denoted by $\mr{AWGN}(\sigma^2)$, has input and output in $\mathbb{R}$, and its output is
\begin{align}
    Y = X + Z, \label{eq:AWGNchannel}
\end{align}
where $Z \sim \mc{N}(0, \sigma^2)$ is independent of the input $X$.
Throughout the paper, the available average power is $P=1$; an average power constraint $P>0$ reduces to this case after dividing inputs and outputs by $\sqrt{P}$ and all noise variances by $P$. For a signal-to-noise ratio $x$, define
\begin{align}
    \CG(x) &\triangleq \frac{1}{2}\log\left(1+x\right),\qquad x\geq0, \label{eq:CGdef}\\
    C_1(x) &\triangleq D\left(\mc{N}\left(1,\frac{1}{x}\right)\,\middle\|\,\mc{N}\left(-1,\frac{1}{x}\right)\right)=2x, \label{eq:C1Gdef}
\end{align}
where $x>0$. The capacity of $\mr{AWGN}(\sigma^2)$ under this constraint is $\CG\left(\frac{1}{\sigma^2}\right)$, achieved by the Gaussian input distribution $X\sim\mc{N}(0,1)$. For the antipodal inputs $\pm1$, the KL divergence between the corresponding conditional output distributions is $C_1\left(\frac{1}{\sigma^2}\right)$.

The KL divergence between $\mc{N}(0, u)$ and $\mc{N}(0, v)$, $u, v > 0$, is denoted by
\begin{align}
    \dG(u \| v) &\triangleq D\bigl(\mc{N}(0, u) \,\|\, \mc{N}(0, v)\bigr)
    = \frac{1}{2}\Bigl(\frac{u}{v} - 1 - \log\frac{u}{v}\Bigr). \label{eq:dGdef}
\end{align}

\section{Problem Formulation} \label{sec:problem}

\subsection{Channel Models and Code Definitions} \label{sec:models_codes}

Consider $K$ memoryless AWGN channels $(\mr{AWGN}(\sigma_1^2), \dots, \mr{AWGN}(\sigma_K^2))$. Channel~$k \in [K]$ has noise variance $\sigma_k^2$, unknown to both terminals but known to lie in the interval $[\sigma_{\min}^2, \sigma_{\max}^2]$, where $0 < \sigma_{\min}^2 < \sigma_{\max}^2 < \infty$ are known to the code designer. Under the precision parameterization $\eta=\frac{1}{\sigma^2}$, this interval corresponds to
\begin{align}
    \eta_{\min}\triangleq\frac{1}{\sigma_{\max}^2}, \qquad \eta_{\max}\triangleq\frac{1}{\sigma_{\min}^2}.
    \label{eq:eta_range}
\end{align}

At each time $t\in \mathbb{N}$, the receiver selects a channel $A_t \in [K]$ as a function of the common randomness $U$, the past selections $A^{t-1}$, and the past channel outputs $Y^{t-1}$, and communicates this selection to the transmitter over a noiseless feedback link. The transmitter then sends
a symbol $X_t \in \mathbb{R}$ over channel $A_t$. In the VLSF setting, $X_t$ is a preselected function of the message and the common randomness alone. In the VLF setting, the transmitter additionally observes the past outputs $Y^{t-1}$ through full noiseless feedback and may adapt $X_t$ to $Y^{t-1}$ and $A^t$. Given $X_t = x$ and $A_t = k$, the receiver observes $Y_t \sim \mc{N}(x, \sigma_{k}^2)$.

The following definitions adapt the VLF and VLSF codes in~\cite{polyanskiy2011feedback, yavasTan2025, truong2018Journal} to bandit channel selection over $K$ memoryless channels.

\begin{definition}[VLF-Bandit code with an average power constraint] \label{def:VLFcode}
Fix $\epsilon \in (0, 1)$, $N > 0$, and positive integers $K$ and $M$, and let the message $W$ be uniformly distributed on $[M]$. An $(N, M, \epsilon, K)$-VLF-Bandit code for $K$ unknown AWGN channels consists of the following components.
    \begin{enumerate}
        \item \emph{Common randomness:} A random variable $U$, independent of $W$, taking values on an alphabet $\mc{U}$, with distribution $P_U$, whose realization is revealed to both the transmitter and the receiver before the start of transmission.
        \item \emph{Channel selection functions:} $\ms{a}_t \colon \mc{U}  \times [K]^{t-1} \times \mathbb{R}^{t-1} \to [K]$ such that
\begin{align}
            A_t = \ms{a}_t(U, A^{t-1}, Y^{t-1}), \quad \forall\, t \geq 1. \label{eq:channelselect}
        \end{align}
\item \emph{Encoding functions:} $\ms{f}_t \colon \mc{U} \times [M] \times [K]^{t} \times \mathbb{R}^{t-1} \to \mathbb{R}$ such that
\begin{align}
            X_t = \ms{f}_t(U, W, A^{t}, Y^{t-1}), \quad \forall\, t \geq 1. \label{eq:encode}
        \end{align}
\item \emph{Stopping time:} A random time $\tau \in \mathbb{Z}_{\geq 0}$ that is a stopping time with respect to the filtration $\{\sigma(U, A^t, Y^t)\}_{t \geq 0}$, satisfying $\E{\tau} \leq N$.
        \item \emph{Decoding function:} A function $\ms{g}_\tau \colon \mc{U} \times [K]^\tau \times \mathbb{R}^\tau \to [M]$ producing the estimate $\nn{W} = \ms{g}_\tau(U, A^\tau, Y^\tau)$, which satisfies $\Prob{\nn{W} \neq W} \leq \epsilon$.
        \item \emph{Average power constraint:} The transmitted symbols satisfy
\begin{align}
            \E{\sum_{t=1}^{\tau} X_t^2} \leq \E{\tau}. \label{eq:powerconstraint}
        \end{align}
\end{enumerate}

\end{definition}

\begin{definition}[VLSF-Bandit code with an average power constraint] \label{def:VLSFcode}
An $(N, M, \epsilon, K)$-VLSF-Bandit code, where $\epsilon \in (0, 1)$, $N > 0$, and $K$ and $M$ are positive integers, is defined as in \defnref{def:VLFcode}, except that the encoding functions in item 3) are replaced by
\begin{itemize}
    \item[3')] \emph{Encoding functions:} $\ms{f}_t \colon \mc{U} \times [M] \to \mathbb{R}$ such that
\begin{align}
            X_t = \ms{f}_t(U, W), \quad \forall\, t \geq 1. \label{eq:VLSFencode}
        \end{align}
\end{itemize}

\end{definition}

The VLSF restriction~\eqref{eq:VLSFencode} fixes the input sequence before transmission. Every $(N,M,\epsilon,K)$-VLSF-Bandit code is therefore an $(N,M,\epsilon,K)$-VLF-Bandit code.

In both code classes, the common-randomness distribution $P_U$ and the channel-selection, encoding, stopping, and decoding rules are specified without knowledge of $(\sigma_1^2,\dots,\sigma_K^2)$. They may depend on $N$, $M$, $\epsilon$, $K$, and the known variance bounds $\sigma_{\min}^2$ and $\sigma_{\max}^2$.

\subsection{Performance Metrics} \label{sec:perf}

The problem instance is the noise-variance vector $\bm{\sigma}^2\triangleq(\sigma_1^2,\dots,\sigma_K^2)$. We require the error-probability constraint in item~5) of \defnref{def:VLFcode} and the average power constraint~\eqref{eq:powerconstraint} at every noise-variance vector in $[\sigma_{\min}^2,\sigma_{\max}^2]^K$ with a unique smallest entry, and we evaluate the average decoding time at each such instance. Thus, one code must perform well at every instance, as one bandit algorithm must in regret minimization~\cite{lai1985}.

\subsection{Instance-Dependent Quantities} \label{sec:AWGN}

We let $v_k \triangleq 1 + \sigma_k^2$ denote the output variance on channel $k$ under a Gaussian $\mc{N}(0,1)$ input and define
\begin{align}
    C^{(k)} &\triangleq \CG\left(\frac{1}{\sigma_k^2}\right), \label{eq:CkGdef} \\
    k^\star &\triangleq \argmin_{k \in [K]} \sigma_k^2 = \argmax_{k \in [K]} C^{(k)}, \label{eq:kstarGdef} \\
    C_\star &\triangleq C^{(k^\star)}, \qquad v_\star \triangleq v_{k^\star} = 1 + \sigma_{k^\star}^2, \label{eq:CstarGdef} \\
    \Delta_k &\triangleq C_\star - C^{(k)}, \quad k \neq k^\star, \label{eq:DeltaGdef} \\
    D_{k,k^\star}
    &\triangleq \dG(v_k \| v_\star), \quad k \neq k^\star,
    \label{eq:DkkstarGdef} \\
    C_1^{(k)} &\triangleq C_1\left(\frac{1}{\sigma_k^2}\right), \label{eq:C1kGdef} \\
    C_{1,\star} &\triangleq C_1^{(k^\star)}. \label{eq:C1starGdef}
\end{align}
We assume throughout the paper that the best channel $k^\star$ is unique, which implies $\Delta_k > 0$ and $D_{k,k^\star} > 0$ for $k \neq k^\star$.

\section{Coding Schemes} \label{sec:scheme}

We describe a VLSF-Bandit scheme and a VLF-Bandit scheme, which instantiate the code definitions in \secref{sec:problem}.

\subsection{\texorpdfstring{VLSF-Bandit Components}{VLSF-Bandit Components}} \label{sec:scheme_common}

The VLSF-Bandit scheme uses a random Gaussian codebook, a score formed from the ratio of two likelihood mixtures, and an output-power-based channel-selection rule.

\subsubsection{Random Gaussian codebook} \label{sec:scheme_common_codebook}

The common randomness $U$ includes $M$ independent infinite-length codewords $\mb{c}(m) \triangleq \bigl(\mb{c}_1(m),\mb{c}_2(m),\ldots\bigr)$, $m\in[M]$, whose entries $\mb{c}_t(m)$, $t\geq1$, are independent $\mc{N}(0,1)$ random variables. At time $t$, the transmitter sends $X_t=\mb{c}_t(W)$ over channel $A_t$. Since the channel-selection rule uses only past selections and outputs, under the random-coding ensemble the output satisfies $Y_t\sim\mc{N}(0,v_k)$ conditioned on $(A^{t-1},Y^{t-1})$ and $A_t=k$.

\subsubsection{Per-channel statistics and exploration budget} \label{sec:scheme_common_stats}

For each channel $k$, define its set and number of uses up to time $n$ by
\begin{align}
    \mc{T}_k(n) &\triangleq \{i\in[n]: A_i=k\}, \qquad N_k(n)\triangleq|\mc{T}_k(n)|.
    \label{eq:Nkdef}
\end{align}
For a candidate message $m\in[M]$, a channel $k$, and a time $n$, define
\begin{align}
    \SE_{k,m}(n) &\triangleq \sum_{t\in\mc{T}_k(n)} \bigl(Y_t-\mb{c}_t(m)\bigr)^2,
    \qquad
    \OP_k(n) \triangleq \sum_{t\in\mc{T}_k(n)} Y_t^2.
    \label{eq:SEOPdef}
\end{align}
The channel-selection rule uses the exploration budget function in~\cite[eq.~(3)]{menardGarivier2017}, which is defined for a
design horizon $T\in\mathbb{N}$ with $T\geq K$ and an integer $s\geq1$ by
\begin{align}
    g_T(s) &\triangleq \log_+\!\left(\frac{T}{Ks}\left(1 + \log_+^2\!\left(\frac{T}{Ks}\right)\right)\right),
    \qquad G_K(T) \triangleq g_T(1).
    \label{eq:budget_def}
\end{align}

\subsubsection{Ratio-of-mixtures score} \label{sec:scheme_common_score}

For a candidate message $m$ and a channel $k$, the receiver forms two Gaussian likelihoods from the uses $t\in\mc{T}_k(n)$, one of the residuals $Y_t-\mb{c}_t(m)$ with variance $s$ and one of the outputs $Y_t$ with variance $1+s$, and mixes each over the unknown noise variance $s$. Both mixtures use the log-uniform prior on the enlarged interval $[\frac{\sigma_{\min}^2}{2},2\sigma_{\max}^2]$, which is the Jeffreys prior of the AWGN channel family restricted to this interval. The per-channel and total scores are
\begin{align}
    S_{k,n}(m)
    &\triangleq \log\frac{
    \displaystyle\int_{\frac{\sigma_{\min}^2}{2}}^{2\sigma_{\max}^2}
    s^{-\frac{N_k(n)}{2}}\exp\!\left\{-\frac{\SE_{k,m}(n)}{2s}\right\}\frac{ds}{s}
    }{
    \displaystyle\int_{\frac{\sigma_{\min}^2}{2}}^{2\sigma_{\max}^2}
    (1+s)^{-\frac{N_k(n)}{2}}\exp\!\left\{-\frac{\OP_k(n)}{2(1+s)}\right\}\frac{ds}{s}
    },
    \label{eq:Skn_def} \\
    S_n(m) &\triangleq \sum_{k=1}^K S_{k,n}(m).
    \label{eq:Sn_def}
\end{align}
The normalizing factors of the prior and of the Gaussian densities cancel in~\eqref{eq:Skn_def}. If the prior were a point mass at the true noise variance $\sigma_k^2$, $S_{k,n}(m)$ would be the information density between the candidate codeword and the outputs, accumulated over the uses of channel~$k$.

\subsubsection{Output-power $\text{kl-UCB}^{++}$ channel selection} \label{sec:scheme_common_select}

Because $\CG\left(\frac{1}{\sigma^2}\right)$ is strictly decreasing in $\sigma^2$, the largest-capacity channel is the one with
the smallest output variance $v_k = 1 + \sigma_k^2$ under the random-coding ensemble. For $N_k(t)\geq1$, define the constrained empirical output variance
\begin{align}
    \nn{v}_k(t) &\triangleq
    \Pi_{[1+\sigma_{\min}^2,\,1+\sigma_{\max}^2]}\!\left(\frac{\OP_k(t)}{N_k(t)}\right).
    \label{eq:vhat_def}
\end{align}
Set $b_k(t)=\sigma_{\min}^2$ for $N_k(t)=0$. For $N_k(t)\geq1$ and the design horizon $T$ in~\eqref{eq:budget_def}, let $b_k(t)$ be the smallest noise variance whose induced output variance remains plausible
under the Gaussian KL confidence constraint, given by
\begin{align}
    b_k(t) &\triangleq \inf\left\{ \widetilde\sigma^2\in[\sigma_{\min}^2,\sigma_{\max}^2] \colon
    N_k(t)\,\dG\!\left(\nn{v}_k(t)\,\big\|\,1+\widetilde\sigma^2\right)\leq g_T\left(N_k(t)\right)\right\}.
    \label{eq:lcb_def}
\end{align}
After the round-robin initialization $A_t=t$ for $t\in[K]$, the receiver selects, for $t\geq K+1$ and until the decoder stops,
\begin{align}
    A_t &\triangleq \argmax_{k\in[K]} \CG\left(\frac{1}{b_k(t-1)}\right).
    \label{eq:klucb_rule}
\end{align}
The rule~\eqref{eq:klucb_rule} selects the channel with the largest optimistic capacity estimate, which is the $\text{kl-UCB}^{++}$ index in~\cite{menardGarivier2017} applied to the output power.

\subsection{VLSF-Bandit Scheme} \label{sec:scheme_vlsf}

The VLSF-Bandit scheme realizes \defnref{def:VLSFcode}. The receiver forms the scores $S_n(m)$ in~\eqref{eq:Sn_def} for each candidate message $m\in[M]$ and selects channels according to~\eqref{eq:klucb_rule}. Fix $M\geq2$, $\epsilon\in(0,1)$, and $\epsilon_0\in[0,\epsilon)$. Set
\begin{align}
 \epsilon_{\mr{dec}}&\triangleq\frac{\epsilon-\epsilon_0}{1-\epsilon_0},
 \label{eq:mixture_eps_dec}\\
 \gamma&\triangleq\log\frac{M-1}{\epsilon_{\mr{dec}}}
 +\sqrt{\log\left(1+\log\frac{M-1}{\epsilon_{\mr{dec}}}\right)},
 \label{eq:mixture_threshold}\\
 T&\triangleq K\vee\lceil\gamma\log^2\gamma\rceil.
 \label{eq:mixture_horizon}
\end{align}
The decoder stops at
\begin{align}
 \tau&\triangleq\inf\Bigl\{n\geq1\colon\max_{m\in[M]}S_n(m)\geq\gamma\Bigr\}
 \label{eq:vlsf_tau_def}
\end{align}
and decodes to
\begin{align}
 \nn{W}&\triangleq\argmax_{m\in[M]}S_\tau(m).
 \label{eq:vlsf_decoder_def}
\end{align}

Following the stop-at-time-zero strategy in~\cite[Proof of Th.~2]{polyanskiy2011feedback}, the decoder uses a component of the common randomness, independent of the codebook, the message, and the channel noises, to stop at time zero and output message~1 with probability $\epsilon_0$. With probability $1-\epsilon_0$, the terminals run the procedure above.

\subsection{VLF-Bandit Scheme} \label{sec:scheme_vlf}

\begin{figure*}[t]
\centering
\definecolor{vlfBest}{HTML}{0072B2}
\definecolor{vlfSub}{HTML}{D55E00}
\definecolor{vlfSave}{HTML}{009E73}
\begin{tikzpicture}[
  >={Stealth[length=2mm]},
  font=\footnotesize,
  execute at begin node={\linespread{1}\selectfont},
  lab/.style = {font=\footnotesize, align=center},
  sml/.style = {font=\scriptsize, align=center}
]
\node[circle, draw=black, thick, fill=black!3, inner sep=1.5pt, font=\small] (U) at (0.3,0) {$U$};
\node[draw=black, thick, rounded corners=1pt, fill=black!3, lab, inner sep=3pt] (STOP) at (0.12,1.55) {stop at time zero,\\ output message $1$};
\draw[thick, ->] (U.north) -- node[left, font=\footnotesize] {$\epsilon_0$} (STOP.south);
\draw[thick, ->] (U.east) -- node[below, font=\footnotesize] {$1-\epsilon_0$} (1.62,0);
\draw[thick, ->] (1.62,0) -- (1.62,1.72);
\node[rotate=90, sml, anchor=south] at (1.62,0.42) {power};
\draw (1.55,1) -- (1.69,1);
\node[sml, anchor=east] at (1.55,1) {$1$};
\fill[black!50] (1.75,0) rectangle (2.6,0.07);
\fill[vlfBest!75] (2.75,0) rectangle (5.2,1);
\fill[vlfSub!70] (5.35,0) rectangle (6.2,0.5);
\fill[vlfSub!70] (6.3,0) rectangle (7.15,0.78);
\node at (7.55,0.3) {$\cdots$};
\fill[vlfSub!70] (7.95,0) rectangle (8.8,0.5);
\fill[vlfBest!35] (8.95,0) rectangle (14.8,1);
\fill[vlfSave!75] (9.07,1) rectangle (10.4,1.34);
\node[sml] at (5.775,0.25) {$J_1$};
\node[sml] at (6.725,0.39) {$J_2$};
\node[sml] at (8.375,0.25) {$J_B$};
\node[sml, anchor=south] at (9.735,1.34) {$1+\frac{\Omega}{n_{\mr{b}}}$};
\fill[white] (11.82,-0.02) -- (12.02,-0.02) -- (12.18,1.04) -- (11.98,1.04) -- cycle;
\draw[black!70] (11.82,-0.02) -- (11.98,1.04);
\draw[black!70] (12.02,-0.02) -- (12.18,1.04);
\draw[dashed, black!60] (1.69,1) -- (14.8,1);
\draw[thick, ->] (1.62,0) -- (15.15,0) node[below left, sml, inner sep=1pt] {time};
\node[lab, anchor=east] at (2.67,2.17) {bits of $W-1$};
\fill[vlfBest!75] (2.75,2.0) rectangle (5.2,2.34);
\fill[vlfSub!70] (5.35,2.0) rectangle (6.2,2.34);
\fill[vlfSub!70] (6.3,2.0) rectangle (7.15,2.34);
\node at (7.55,2.17) {$\cdots$};
\fill[vlfSub!70] (7.95,2.0) rectangle (8.8,2.34);
\fill[vlfBest!35] (8.95,2.0) rectangle (14.8,2.34);
\fill[white] (11.82,1.98) -- (12.02,1.98) -- (12.08,2.36) -- (11.88,2.36) -- cycle;
\node[sml, text=white] at (3.975,2.17) {$\lceil\psi^2\rceil$ bits};
\node[sml] at (5.775,2.17) {$\ell_0$};
\node[sml] at (6.725,2.17) {$\ell_0$};
\node[sml] at (8.375,2.17) {$\ell_0$};
\node[sml] at (10.4,2.17) {all remaining bits};
\node[lab] at (2.175,-0.3) {$K\ell_0$};
\node[lab] at (3.975,-0.3) {$O(\psi^2)$};
\node[lab] at (7.075,-0.3) {$O(\psi^{\frac{2}{3}})$ each};
\node[lab] at (12.0,-0.33) {$\approx\frac{\log M}{\CG_\star}$};
\node[lab] at (2.0,-0.88) {zero inputs on\\ channels $1,\ldots,K$};
\node[lab] at (4.05,-0.88) {preliminary\\ packet on $J_0$};
\node[lab] at (7.075,-0.88) {exploration packets\\ $1,\ldots,B$};
\node[lab] at (12.0,-0.88) {final packet\\ on $J_{B+1}$};
\end{tikzpicture}
\caption{The VLF-Bandit scheme. With probability $\epsilon_0$, the decoder stops at time zero; otherwise, $\ell_0$ zero inputs on each channel precede a preliminary packet, $B\geq0$ exploration packets, and a final packet. Top: the bits of $W-1$ that each packet carries. Heights: the nominal powers, with the saved allocation $\Omega$ in green. Bottom: the number of channel uses of each phase, for the final packet on the best channel. Not to scale. The quantities are defined in Sections~\ref{sec:scheme_vlf_packets} to~\ref{sec:scheme_vlf_final}.}
\label{fig:vlf_packets}
\end{figure*}
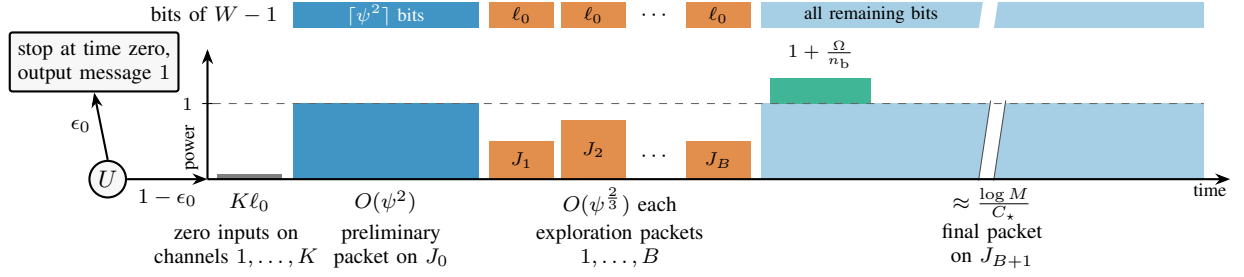

The VLF-Bandit scheme realizes \defnref{def:VLFcode}. It divides the message into successive segments, called packets, and sends each packet by SK feedback transmission~\cite{schalkwijk,schalkwijk1966bandlimited} over one channel, which the receiver selects. After the first symbol of a packet, every use is a refinement of the receiver's message estimate. In a refinement, the transmitter sends a scaled version of the receiver's current estimation error, and the receiver normalizes the output by a factor computed from the earlier outputs of the packet. Both terminals store these normalized outputs and use them to estimate the noise precision within a packet and to compare the channels across packets. After zero-input observations on every channel, the scheme sends a preliminary packet, zero or more short exploration packets, which carry message bits at powers of at most one, and a final packet, which carries the remaining bits with the power saved in exploration (Figure~\ref{fig:vlf_packets}).

\subsubsection{Message packets} \label{sec:scheme_vlf_packets}

Fix $\epsilon\in(0,1)$ and a large enough message size $M$, a power of two, and define the construction scale
\begin{align}
 \psi&\triangleq\log\log M.
 \label{eq:vlf_scale}
\end{align}
Since $\log M$ is of order $N$ for the message sizes of interest, $\psi$ is of order $\log N$, as is the second-order term. As in~\secref{sec:scheme_vlsf}, the decoder stops at time zero and outputs message~1 with probability
\begin{align}
 \epsilon_0&\triangleq\epsilon-\frac{1}{\log M}.
 \label{eq:vlf_stop_zero}
\end{align}
With probability $1-\epsilon_0$, the terminals execute the transmission procedure below, and the margin $\frac{1}{\log M}$ is reserved for decoding errors on this branch.

The message $W-1$ is represented by a binary string of length $\log_2M$. Packet $j=0,1,\ldots$ has alphabet size $M_j$ and carries the next $\log_2M_j$ unused bits, whose binary value plus one defines its message $W_j\in[M_j]$. Its channel is denoted by $J_j$.

\subsubsection{Feedback transmission of a packet}

Fix packet $j$. Its channel $J_j$, alphabet size $M_j$, and nominal powers $P_{j,0},P_{j,1},\ldots$ are prescribed before it starts. The refinement powers $P_{j,n}$, $n\geq1$, are positive, non-increasing, and bounded away from zero, and the initial power $P_{j,0}>0$ is specified separately. With the uses within the packet indexed by $n=0,1,\ldots$, the outputs are $Y_{j,n}=X_{j,n}+Z_{j,n}$ with $Z_{j,n}\sim\mc{N}(0,\sigma_{J_j}^2)$.

The initial symbol uses pulse-amplitude modulation, as in~\cite[Sec.~III]{gallagerNakiboglu2010}. The transmitter sends $X_{j,0}$, and the receiver initializes its real-valued message estimate to $\nn{W}_{j,0}$, where
\begin{align}
 X_{j,0}&\triangleq
 \frac{\sqrt{12P_{j,0}}}{M_j}
 \left(W_j-\frac{M_j+1}{2}\right),
 \qquad
 \nn{W}_{j,0}\triangleq
 \frac{M_j+1}{2}+\frac{M_jY_{j,0}}{\sqrt{12P_{j,0}}}.
 \label{eq:vlf_packet_initial}
\end{align}

Let $F_{j,n}$ denote the accumulated logarithmic error reduction after refinement $n$, updated in~\eqref{eq:vlf_packet_contraction} below, with $F_{j,0}\triangleq0$, and define the normalized estimation error
\begin{align}
 E_{j,n}&\triangleq
 \frac{\sqrt{12P_{j,0}}}{M_j}\exp\{F_{j,n}\}
 \left(W_j-\nn{W}_{j,n}\right),
 \qquad n\geq0,
 \label{eq:vlf_packet_error}
\end{align}
so that $E_{j,0}=-Z_{j,0}$ by~\eqref{eq:vlf_packet_initial}. The transmitter computes $E_{j,n-1}$ from $W_j$ and the feedback. At refinement $n\geq1$, both terminals compute the precision estimate from the normalized outputs $\widetilde Z_{j,1},\ldots,\widetilde Z_{j,n-1}$ of the earlier refinements, defined in~\eqref{eq:vlf_packet_input},
\begin{align}
 \nn{\eta}_{j,n-1}&\triangleq
 \begin{cases}
 \frac{\eta_{\min}}{2},&1\leq n\leq9,\\[2pt]
 \displaystyle\Pi_{\left[\frac{\eta_{\min}}{2},\,2\eta_{\max}\right]}
 \left(\frac{n-3}{\sum_{i=1}^{n-1}\widetilde Z_{j,i}^{2}}\right),&n\geq10,
 \end{cases}
 \label{eq:vlf_packet_precision}
\end{align}
where the precision bounds are given in~\eqref{eq:eta_range}. The transmitter sends $X_{j,n}$, and upon receiving $Y_{j,n}$, the receiver stores the normalized output $\widetilde Z_{j,n}$, where
\begin{align}
 X_{j,n}&\triangleq
 \sqrt{P_{j,n}\nn{\eta}_{j,n-1}}\,E_{j,n-1},
 \qquad
 \widetilde Z_{j,n}\triangleq
 \frac{Y_{j,n}}{\sqrt{1+P_{j,n}\nn{\eta}_{j,n-1}}}.
 \label{eq:vlf_packet_input}
\end{align}
Given the selected channel, the normalized outputs within a packet are i.i.d.\ $\mc{N}(0,\sigma_{J_j}^2)$ samples (\lemref{lem:vlf_packet_gaussian}). The transmitter learns them through feedback, and both terminals update
\begin{align}
 F_{j,n}&\triangleq
 F_{j,n-1}+\CG\left(P_{j,n}\nn{\eta}_{j,n-1}\right),
 \label{eq:vlf_packet_contraction}\\
 \nn{W}_{j,n}&\triangleq
 \nn{W}_{j,n-1}
 +\frac{M_j\exp\{-F_{j,n-1}\}\sqrt{P_{j,n}\nn{\eta}_{j,n-1}}}
 {\sqrt{12P_{j,0}}\left(1+P_{j,n}\nn{\eta}_{j,n-1}\right)}Y_{j,n}.
 \label{eq:vlf_packet_estimate}
\end{align}

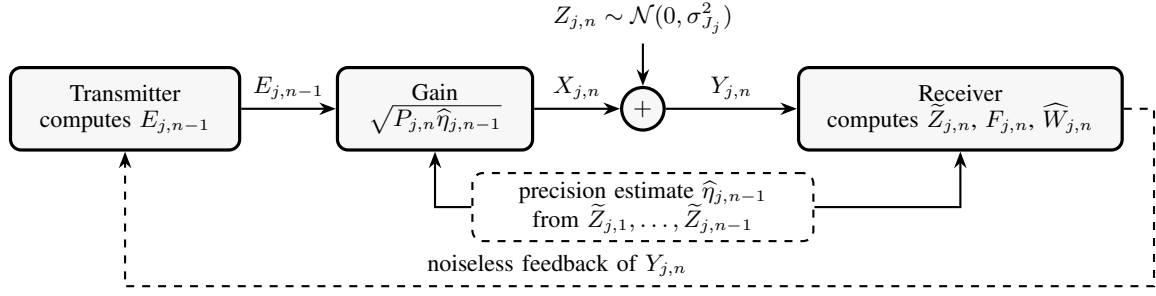
\begin{figure*}[t]
\centering
\begin{tikzpicture}[
  >={Stealth[length=2.2mm]},
  thick,
  font=\small,
  execute at begin node={\linespread{1}\selectfont},
  box/.style  = {rectangle, rounded corners, draw=black, very thick, fill=black!3,
                 align=center, minimum height=3.1em, inner sep=4pt},
  est/.style  = {rectangle, rounded corners, draw=black, thick, dashed, fill=white,
                 align=center, minimum height=2.5em, inner sep=3pt},
  sum/.style  = {circle, draw=black, very thick, fill=black!3, inner sep=0pt, minimum size=1.6em},
  line/.style = {draw, thick, ->}
]
\node[box, text width=2.75cm] (TX) at (0.2,0) {Transmitter\\ computes $E_{j,n-1}$};
\node[box, text width=2.3cm] (G) at (4.3,0) {Gain\\ $\sqrt{P_{j,n}\nn{\eta}_{j,n-1}}$};
\node[sum] (S) at (7.05,0) {$+$};
\node[box, text width=4.0cm] (RX) at (11.25,0) {Receiver\\ computes $\widetilde Z_{j,n}$, $F_{j,n}$, $\nn{W}_{j,n}$};
\node[est, text width=4.3cm] (EST) at (7.05,-1.3) {precision estimate $\nn{\eta}_{j,n-1}$\\ from $\widetilde Z_{j,1},\ldots,\widetilde Z_{j,n-1}$};
\node (NZ) at (7.05,1.2) {$Z_{j,n}\sim\mc{N}(0,\sigma_{J_j}^2)$};

\draw[line] (TX.east) -- node[above] {$E_{j,n-1}$} (G.west);
\draw[line] (G.east) -- node[above] {$X_{j,n}$} (S.west);
\draw[line] (NZ.south) -- (S.north);
\draw[line] (S.east) -- node[above] {$Y_{j,n}$} (RX.west);
\draw[line] (EST.west) -| (G.south);
\draw[line] (EST.east) -| (RX.south);
\draw[line, dashed] (RX.east) -- ++(0.4,0) |- node[pos=0.79, above] {noiseless feedback of $Y_{j,n}$} ($(TX.south)+(0,-1.78)$) -- (TX.south);
\end{tikzpicture}
\caption{Refinement $n\geq1$ of packet $j$ over channel $J_j$. Both terminals compute the precision estimate (dashed box), the constant $\frac{\eta_{\min}}{2}$ for $n\leq9$.}
\label{fig:vlf_refinement}
\end{figure*}

Figure~\ref{fig:vlf_refinement} shows one refinement. If the true precision $\frac{1}{\sigma_{J_j}^2}$ replaced $\nn{\eta}_{j,n-1}$, then~\eqref{eq:vlf_packet_input}--\eqref{eq:vlf_packet_estimate} would be the refinements of the SK scheme~\cite{schalkwijk,gallagerNakiboglu2010} with powers $P_{j,n}$ and a known noise variance. Each packet therefore runs a universal version of the SK scheme. The initial output $Y_{j,0}$ is used only for message estimation.

Each packet has the error target $\exp\{-b\}$, where
\begin{align}
 b&\triangleq\psi+\log\left(\lceil\psi^2\rceil+2\right),
 \label{eq:vlf_packet_budget}
\end{align}
so that the margin $\frac{1}{\log M}$ in~\eqref{eq:vlf_stop_zero} is split equally among $\lceil\psi^2\rceil+2$ packets, the most that the scheme transmits. Packet $j$ stops at
\begin{align}
 \tau^{(j)}&\triangleq
 \inf\left\{n\geq1\colon F_{j,n}\geq\gamma_j\right\},
 \qquad
 \gamma_j\triangleq\log M_j+\frac{1}{2}\log\left(\frac{2\sigma_{\max}^2 b}{3 P_{j,0}}\right).
 \label{eq:vlf_packet_stop}
\end{align}
The receiver rounds $\nn{W}_{j,\tau^{(j)}}$ to the nearest element of $[M_j]$ and expresses that index minus one as a binary string of length $\log_2M_j$, the decoded packet. Packet $j$ takes $\tau^{(j)}+1$ channel uses.

\subsubsection{Initial observations and preliminary packet}

Let $\ell_0\triangleq\left\lceil\psi^{\frac{1}{3}}\right\rceil$. The receiver selects the channels in increasing order, taking $\ell_0$ zero-input observations on each. For channel $k$, the sample count and the empirical variance are initialized to
\begin{align}
 \widetilde L_k&\triangleq\ell_0,\qquad
 \nn{s}_k\triangleq
 \frac{1}{\ell_0}\sum_{n=1}^{\ell_0}Y_{\left(k-1\right)\ell_0+n}^{2}.
 \label{eq:vlf_raw_variance}
\end{align}

Packet zero, the preliminary packet, uses the channel
\begin{align}
 J_0&\triangleq\argmin_{k\in[K]}
 \Pi_{\left[\sigma_{\min}^2,\,\sigma_{\max}^2\right]}\left(\nn{s}_k\right)
 \label{eq:vlf_preliminary_channel}
\end{align}
and carries the first $\lceil\psi^2\rceil$ bits, with
\begin{align}
 M_0&\triangleq2^{\lceil\psi^2\rceil},
 \qquad P_{0,n}\triangleq1,\quad n\geq0.
 \label{eq:vlf_preliminary_payload}
\end{align}
With its order-$\psi^2$ uses, this packet gives an accurate variance estimate of channel~$J_0$, with which the other channels are compared during exploration.

After each completed packet $j$ on channel $k=J_j$, the empirical variance and then the sample count of channel $k$ are updated by
\begin{align}
 \nn{s}_k&\leftarrow
 \frac{\widetilde L_k\nn{s}_k+\sum_{n=1}^{\tau^{(j)}}\widetilde Z_{j,n}^{2}}{\widetilde L_k+\tau^{(j)}},
 \qquad \widetilde L_k\leftarrow \widetilde L_k+\tau^{(j)},
 \label{eq:vlf_sample_update}
\end{align}
and the statistics of the other channels are unchanged. These statistics retain observations across packets, whereas the precision estimate~\eqref{eq:vlf_packet_precision} uses only those of its own packet.

\subsubsection{Exploration packets}

Each exploration packet carries $\ell_0$ bits, so that the channels are compared often.

After updating the statistics from the preliminary packet, the receiver initializes the set $\mc{A}=[K]$ of channels eligible for exploration and the counter $j=0$ of completed exploration packets, and repeats the following steps until exploration ends.

\begin{enumerate}

\item Form the confidence intervals
\begin{align}
  \mc{I}_k&\triangleq
  \left\{\widetilde\sigma^2\in\left[\sigma_{\min}^2,\sigma_{\max}^2\right]\colon
  \widetilde L_k\,\dG\left(\nn{s}_k\|\widetilde\sigma^2\right)\leq b\right\},
  \qquad k\in\mc{A}.
  \label{eq:vlf_confidence}
 \end{align}
If any interval is empty, end exploration and select channel~1 for the final packet.

\item Remove the channels whose entire interval lies above another active channel's interval, by the simultaneous assignment
\begin{align}
  \mc{A}&\leftarrow
  \left\{k\in\mc{A}\colon
  \inf\mc{I}_k\leq\min_{k'\in\mc{A}}\sup\mc{I}_{k'}\right\},
  \label{eq:vlf_elimination}
 \end{align}
which retains a channel attaining the smallest upper endpoint. If one channel remains, end exploration and select it for the final packet. Otherwise, if the completed exploration packets have reached the exploration cap of $\lceil \psi^2\rceil$ channel uses, end exploration and select channel~1 for the final packet.

\item Select the active channel with the fewest retained observations and transmit packet $j+1$ using
\begin{align}
  J_{j+1}&\triangleq\argmin_{k\in\mc{A}}\widetilde L_k,
  \qquad M_{j+1}\triangleq2^{\ell_0},
  \label{eq:vlf_exploration_channel}\\
  P_{j+1,n}&\triangleq
  \Pi_{\left[\frac{1}{\ell_0},\,1\right]}
  \left(1+
  \Pi_{\left[\sigma_{\min}^2,\,\sigma_{\max}^2\right]}\left(\nn{s}_{J_0}\right)
  -\Pi_{\left[\sigma_{\min}^2,\,\sigma_{\max}^2\right]}\left(\nn{s}_{J_{j+1}}\right)\right),
  \qquad n\geq0.
  \label{eq:vlf_exploration_power}
 \end{align}
Complete this packet even if its length crosses the exploration cap. Apply~\eqref{eq:vlf_sample_update}, set $j\leftarrow j+1$, and return to step~1.

\end{enumerate}

At termination, the number of exploration packets is $B=j$, and the final channel is denoted by $J_{B+1}$. The last exploration packet starts before the exploration cap is reached, and each packet takes at least one use, so $B\leq\lceil\psi^2\rceil$ and at most $\lceil\psi^2\rceil+2$ packets are transmitted.

\subsubsection{Final packet and decoding} \label{sec:scheme_vlf_final}

Packet $B+1$ uses channel $J_{B+1}$ and carries all remaining bits, so that
\begin{align}
 \log M_{B+1}&\triangleq\log M-\log M_0-B\ell_0\log2.
 \label{eq:vlf_final_payload}
\end{align}
Exploration saves the allocation $\Omega$, which the final packet spreads over its first $n_{\mr{b}}$ refinements, where
\begin{align}
 \Omega&\triangleq
 \sum_{j=1}^{B}\left(1-P_{j,0}\right)\left(\tau^{(j)}+1\right),
 \qquad n_{\mr{b}}\triangleq\left\lceil1+ \Omega^2\right\rceil,
 \label{eq:vlf_saved_power}\\
 P_{B+1,0}&\triangleq1,\qquad
 P_{B+1,n}\triangleq
 \begin{cases}
 1+\frac{\Omega}{n_{\mr{b}}},&1\leq n\leq n_{\mr{b}},\\[2pt]
 1,&n>n_{\mr{b}}.
 \end{cases}
 \label{eq:vlf_final_power}
\end{align}
The decoder concatenates the decoded binary strings of packets $0,1,\ldots,B+1$ in this order and adds one to the resulting binary value to obtain $\nn{W}\in[M]$. The transmission branch uses $B+2$ packets and has decoding time
\begin{align}
 \tau&=K\ell_0+\sum_{j=0}^{B+1}\left(\tau^{(j)}+1\right).
 \label{eq:vlf_total_length}
\end{align}

Packet parameters and channel selections use only the initial zero-input observations and the normalized outputs of completed packets, which both terminals know through feedback. Decoded packets are used only for final decoding. The parameters of the scheme depend only on $M$, $\epsilon$, $K$, and the known variance bounds.

\section{Main Results} \label{sec:main}

For the schemes in \secref{sec:scheme}, we state second-order achievability bounds for the VLSF-Bandit scheme (\thmref{thm:vlsf_main}) and the VLF-Bandit scheme (\thmref{thm:vlf_main}).

\subsection{VLSF-Bandit Scheme} \label{sec:main_vlsf}

We first state the achievability bound for the single-phase scheme.

\begin{theorem}[Second-order achievability bound for the VLSF-Bandit scheme]
\label{thm:vlsf_main}
Fix $\epsilon\in(0,1)$ and $K\geq1$, and fix the VLSF-Bandit scheme in~\secref{sec:scheme_vlsf} with the stop-at-time-zero strategy at $\epsilon_0=\epsilon-\frac{1}{\log(M-1)}$. For every problem instance
    $\bm{\sigma}^2\in [\sigma_{\min}^2,\sigma_{\max}^2]^K$ with a unique best channel $k^\star$, this scheme is, for all sufficiently large $M$, an $(N,M,\epsilon,K)$-VLSF-Bandit code with $N$ satisfying, as
    $M\to\infty$,
\begin{align}
        \log M
        &\geq
        \frac{N\CG_\star}{1-\epsilon}
        - \left(1+\sum_{k\neq k^\star}\frac{\Delta_k}{D_{k,k^\star}}\right)\log N
        + o(\log N).
        \label{eq:vlsf_main}
    \end{align}
\end{theorem}

\begin{IEEEproof} See \secref{sec:vlsf_proof}.
\end{IEEEproof}

The magnitude of the coefficient of $\log N$ in~\eqref{eq:vlsf_main} decomposes as
\begin{align}
    \underbrace{1}_{\text{VLSF}}
    + \underbrace{\sum_{k\neq k^\star}\frac{\Delta_k}{D_{k,k^\star}}}_{\text{channel exploration}}.
    \label{eq:vlsf_logN_decomposition}
\end{align}
The first term is the single-phase VLSF baseline. It is the magnitude of the coefficient in the achievability part
of~\eqref{eq:pol_upper} and arises from the threshold decoder and the stop-at-time-zero strategy. For the transmitted message, the logarithm of each likelihood mixture in~\eqref{eq:Skn_def} falls below that of the likelihood at the true noise variance by about $\frac{1}{2}\log N_k(n)$, and the $\frac{1}{2}\log N_k(n)$ terms cancel in the ratio. The remaining universal-decoding cost, from score fluctuations and the square-root term in~\eqref{eq:mixture_threshold}, is absorbed in the $o(\log N)$ remainder. For $K=1$, the coefficient is one for every admissible noise variance, without the cost $\frac{1}{2}$ of the universal score in~\cite{yavasTan2025}.

The last term is the time cost of suboptimal channel selections. Under the Gaussian input distribution, each use of channel $k\neq k^\star$ incurs an expected information loss of $\Delta_k$ relative to the best channel. The Lai--Robbins lower bound~\cite{lai1985} states that, for any bandit algorithm whose expected number of suboptimal arm pulls is $o(n^\alpha)$ for every $\alpha>0$ at every admissible problem instance, the expected number of uses of each suboptimal channel $k\neq k^\star$ satisfies
\begin{align}
    \liminf_{n\to\infty}\frac{\E{N_k(n)}}{\log n} \geq \frac{1}{D_{k,k^\star}}.
    \label{eq:lai_robbins_vlsf}
\end{align}
The $\text{kl-UCB}^{++}$ rule attains the right-hand side of~\eqref{eq:lai_robbins_vlsf}~\cite[Th.~2]{menardGarivier2017}. Our bound retains $\frac{1}{D_{k,k^\star}}$ at the random decoding time and weights it by the capacity gap $\Delta_k$.

\subsection{VLF-Bandit Scheme} \label{sec:main_vlf}

We next state the achievability bound for the VLF-Bandit scheme in~\secref{sec:scheme_vlf}. For each suboptimal channel $k\neq k^\star$, define the limiting exploration power
\begin{align}
 P_k^\star&\triangleq\left(v_\star-\sigma_k^2\right)^+,
 \label{eq:vlf_limiting_power}
\end{align}
which the power~\eqref{eq:vlf_exploration_power} of an exploration packet on channel~$k$ approaches as $M\to\infty$.

\begin{theorem}[Second-order achievability bound for the VLF-Bandit scheme]
\label{thm:vlf_main}
Fix $\epsilon\in(0,1)$ and $K\geq1$, and fix the VLF-Bandit scheme in~\secref{sec:scheme_vlf}. For every problem instance
    $\bm{\sigma}^2\in [\sigma_{\min}^2,\sigma_{\max}^2]^K$ with a unique best channel $k^\star$, this scheme is, for all sufficiently large $M$, an $(N,M,\epsilon,K)$-VLF-Bandit code with $N$ satisfying, as
    $M\to\infty$ through powers of two,
\begin{align}
        \log M
        &\geq
        \frac{N\CG_\star}{1-\epsilon}
        -\left(
        \frac{1}{2v_\star^2}
        +\sum_{k\neq k^\star}
        \frac{\CG_\star-\CG\left(\frac{P_k^\star}{\sigma_k^2}\right)-\frac{1-P_k^\star}{2v_\star}}
        {\dG\left(\sigma_k^2\|\sigma_{k^\star}^2\right)}
        \right)\log N
        +o(\log N).
        \label{eq:vlf_main}
    \end{align}
\end{theorem}

\begin{IEEEproof} See \secref{sec:vlf_proof}.
\end{IEEEproof}

The magnitude of the coefficient of $\log N$ in~\eqref{eq:vlf_main} decomposes as
\begin{align}
    \underbrace{\frac{1}{2v_\star^2}}_{\text{universal decoding}}
    \mathbin{+}\underbrace{\sum_{k\neq k^\star}
    \frac{\CG_\star-\CG\left(\frac{P_k^\star}{\sigma_k^2}\right)-\frac{1-P_k^\star}{2v_\star}}
    {\dG\left(\sigma_k^2\|\sigma_{k^\star}^2\right)}}_{\text{channel exploration}}.
    \label{eq:vlf_logN_decomposition}
\end{align}
The decomposition~\eqref{eq:vlf_logN_decomposition} has no baseline, because the packet error targets in~\secref{sec:scheme_vlf} affect only the $o(\log N)$ remainder.

The universal-decoding term is the cost of estimating the noise precision within the final packet. After $m$ refinements, the precision estimate is nearly unbiased with variance of order $\frac{1}{m}$, so the concavity of the error reduction~\eqref{eq:vlf_packet_contraction} gives an expected loss of the same order at the next refinement. These losses accumulate over the order-$\log M$ refinements of the packet to a cost of order $\log\log M$. For a single AWGN channel with unknown noise variance ($K=1$), the bound improves the magnitude $\frac{1}{2}+\frac{\CG_\star}{C_{1,\star}}$ in~\cite[Th.~5]{yavasTan2025} to $\frac{1}{2v_\star^2}$. With known noise variance, the SK scheme can be shown to achieve the second-order term $-O(\log\log N)$, improving the coefficient of $\log N$ in~\cite[Th.~4]{yavasTan2025} from $-\frac{\CG_\star}{C_{1,\star}}$ to zero.

The channel-exploration term accounts for both communication during exploration and reuse of the saved allocation. At the limiting power $P_k^\star$, an exploration use of channel $k$ carries $\CG\left(\frac{P_k^\star}{\sigma_k^2}\right)$ nats instead of $\CG_\star$, and the final packet converts the saved allocation $1-P_k^\star$ of the use into $\frac{1-P_k^\star}{2v_\star}$ nats, since $\frac{1}{2v_\star}$ is the derivative of the capacity of channel $k^\star$ with respect to the input power at power one. The numerator is the remaining information loss of the use, and $P_k^\star$ minimizes it among powers in $[0,1]$.

Every VLSF-Bandit code is a VLF-Bandit code, and the magnitude of the coefficient of $\log N$ is strictly smaller in \thmref{thm:vlf_main} than in \thmref{thm:vlsf_main}. Indeed, $\frac{1}{2v_\star^2}<1$, and each exploration summand is smaller than $\frac{\Delta_k}{D_{k,k^\star}}$, since its numerator is at most $\Delta_k$ and its denominator exceeds $D_{k,k^\star}$.

The bounds in Theorems~\ref{thm:vlsf_main} and~\ref{thm:vlf_main} both attain the optimal first-order term $\frac{N\CG_\star}{1-\epsilon}$. To see this, reveal the noise variances to both terminals and simulate each selected channel on channel $k^\star$ by adding independent Gaussian noise of variance $\sigma_{A_t}^2-\sigma_{k^\star}^2$ from common randomness, which preserves the error probability, expected decoding time, and expected energy. The converse argument in~\cite[Sec.~II-E]{truong2016gaussian} extends to the simulated full-feedback code by summing conditional AWGN information over the uses up to~$\tau$ and applying concavity under~\eqref{eq:powerconstraint}, and gives $(1-\epsilon)\log M\leq\CG_\star\E{\tau}+h(\epsilon)$ for every VLF-Bandit code, which is the upper bound in~\eqref{eq:pol_upper} with $C=\CG_\star$.

\subsection{Numerical Illustration} \label{sec:main_numerical}

For the known variance interval $[0.5,2.5]$ and a single channel with $\sigma_1^2=1$, the magnitudes of the $\log N$ coefficients in the VLSF-Bandit and VLF-Bandit bounds are $1$ and $0.125$, respectively. For $K=2$ with $\bm{\sigma}^2=(1,2)$, these magnitudes are approximately $4.043$ and $0.754$, respectively.

\suppressfloats[t]

\begin{figure*}[t]
\centering
\includegraphics[width=0.88\textwidth]{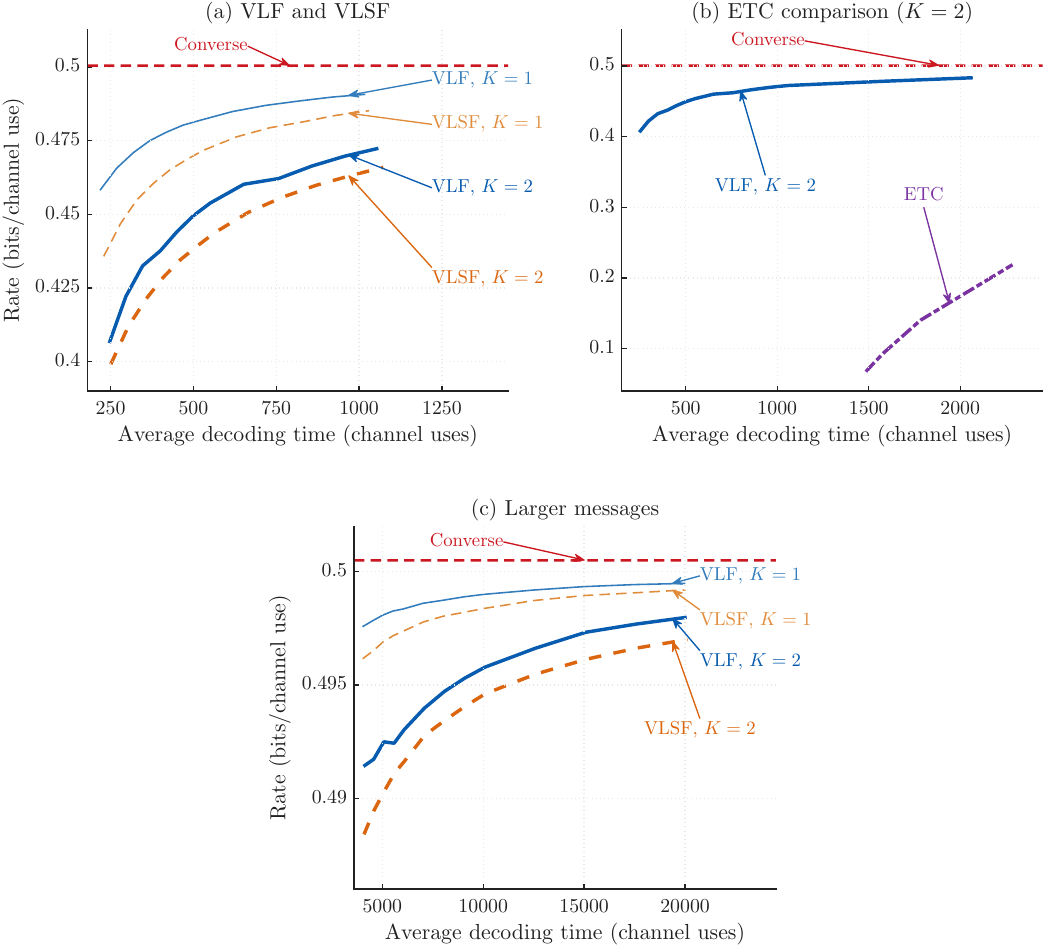}
\caption{Rate evaluations for the known variance interval $[0.5,2.5]$, $\epsilon=10^{-3}$, and $\epsilon_0=0$. The instances are $\sigma_1^2=1$ for $K=1$ and $\bm\sigma^2=(1,2)$ for $K=2$. Thin and thick curves denote $K=1$ and $K=2$, respectively; ETC uses $K=2$. The VLSF-Bandit curves report simulated true-message threshold-crossing rates. Each VLF-Bandit and VLSF-Bandit rate evaluation uses at least 100\,000 independent trials; each ETC evaluation uses 50\,000. The dashed red curve is the converse.}
\label{fig:finite_rates}
\end{figure*}

Figure~\ref{fig:finite_rates} compares numerical rate evaluations of the VLF-Bandit, VLSF-Bandit, and ETC schemes at $\epsilon=10^{-3}$ and $\epsilon_0=0$, together with the converse in~\cite{truong2016gaussian}. We choose the free parameters of VLF-Bandit numerically. The plotted rate is $\frac{\log_2 M}{\widehat N}$, where $\widehat N$ is the sample average duration. For VLSF-Bandit, this duration is the first time the transmitted message reaches the threshold $\gamma$, which upper-bounds the decoder's stopping time.

Reading the curves at $N=500$, the VLF-Bandit and VLSF-Bandit rates are approximately $0.481$ and $0.470$ bits per channel use for $K=1$, and $0.450$ and $0.438$ for $K=2$, respectively. At $N=2000$ and $K=2$, VLF-Bandit achieves about $0.483$ bits per channel use, compared with $0.179$ for ETC.

\section{Proof of the VLSF-Bandit Results}

\label{sec:vlsf_proof}

This section proves \thmref{thm:vlsf_main}. Wrong-message threshold crossings and the tail of the true-message stopping time give the error bound. A comparison with the true information density, an overshoot bound, and the suboptimal channel-use counts give the expected decoding time. We then apply the stop-at-time-zero strategy and expand the bound as $M\to\infty$.

For the analysis, we continue the channel-selection rule and the scores beyond $\tau$ using the infinite codebook and fresh channel noises. Let
\begin{align}
 \mc{F}_t&\triangleq\sigma(A^t,X^t,Y^t),\qquad t\geq0.
 \label{eq:vlsf_Ft_filtration}
\end{align}
The channel choice $A_t$ is $\mc{F}_{t-1}$-measurable. On $\{A_t=k\}$, the current transmitted symbol and channel noise, conditioned on $\mc{F}_{t-1}$, are independent zero-mean Gaussian variables with variances $1$ and $\sigma_k^2$, respectively.

\subsection{Supporting Lemmas}

\label{sec:vlsf_supporting_lemmas}

We first present three tools for the VLSF-Bandit proof.

\begin{lemma}[Doob's maximal inequalities~{\cite[Th.~1.1]{acciaio2013doob}}]\label{lem:ville}
Let $\{X_n\}_{n\geq0}$ be a non-negative submartingale. For every integer $H\geq0$,
\begin{align}
 \E{\left(\max_{0\leq n\leq H}X_n\right)^4}
 &\leq\left(\frac43\right)^4\E{X_H^4}.
 \label{eq:doob_fourth}
\end{align}
If $\{X_n\}_{n\geq0}$ is a martingale with $X_0=1$, then
\begin{align}
 \E{\max_{0\leq n\leq H}X_n}
 &\leq\frac{e}{e-1}\bigl(1+\E{X_H\log X_H}\bigr).
 \label{eq:doob_entropy}
\end{align}
\end{lemma}

\begin{lemma}[Overshoot bound]\label{lem:overshoot} In this lemma and its proof, all notation is local. Let $\{ X_n\}_{n\geq1}$ be an integrable process adapted to a filtration
$\{\mc{F}_n\}_{n\geq0}$. Fix $c\in[0,\infty)$ such that, for every $n\geq1$ and every non-negative
$\mc{F}_{n-1}$-measurable random variable $a$,
\begin{align}
    \E{(X_n-a)^+\mid\mc{F}_{n-1}}
    &\leq c\,\Prob{X_n>a\mid\mc{F}_{n-1}}
    \quad\text{a.s.}
    \label{eq:overshoot_residual}
\end{align}
Fix $\omega\in\mathbb{R}$. Let $S_0$ be an integrable
$\mc{F}_0$-measurable random variable satisfying $S_0<\omega$ almost
surely, and define
\begin{align}
    S_n &\triangleq S_0+\sum_{t=1}^nX_t,
    \qquad
    \tau\triangleq\inf\{n\geq1\colon S_n\geq \omega\}.
    \label{eq:overshoot_tau_b}
\end{align}
Then, for every $\ntr\in\mathbb{N}$,
\begin{align}
    \E{S_{\tau\wedge\ntr}}
    &\leq \omega+c.
    \label{eq:overshoot_bound}
\end{align}
\end{lemma}

\begin{IEEEproof} For $n\geq1$, let $a_n=( \omega- S_{n-1})^+$. On
$\{\tau\geq n\}$, we have $S_{n-1}<\omega$ and $a_n=\omega-S_{n-1}$.
Therefore,
\begin{align}
    \E{(S_{\tau}-\omega)\1{\tau\leq\ntr}}
    &=
    \sum_{n=1}^{\ntr}
    \E{\1{\tau\geq n}(X_n-a_n)^+}
    \leq
    c\sum_{n=1}^{\ntr}
    \Prob{\tau\geq n,\ X_n>a_n}
    \leq c,
    \label{eq:overshoot_excess_bound}
\end{align}
where the first inequality applies~\eqref{eq:overshoot_residual} with $a=a_n$ on the event $\{\tau\geq n\}\in\mc{F}_{n-1}$, and the last inequality follows because the events in the sum are disjoint. On $\{\tau>\ntr\}$, we have $S_{\ntr}<\omega$. Splitting $\E{S_{\tau\wedge\ntr}}$ over the events $\{\tau\leq\ntr\}$ and $\{\tau>\ntr\}$ and applying~\eqref{eq:overshoot_excess_bound} proves~\eqref{eq:overshoot_bound}.
\end{IEEEproof}

\begin{lemma}[Fixed-horizon $\text{kl-UCB}^{++}$ pull count, adapted from {\cite[Th.~2]{menardGarivier2017}}]\label{lem:klucbpp_fixed_horizon}
Consider a $K$-armed bandit whose rewards belong to a one-parameter canonical exponential family parameterized by its mean, with KL divergence $d(\cdot\|\cdot)$, whose raw empirical means lie in the mean domain almost surely. Fix a closed interval $[\mu^-,\mu^+]$ inside the mean domain, containing all arm means, on which the family variance is at most $V<\infty$. Suppose that the best arm $k^\star$ is unique, with mean $\mu_\star=\max_{j\in[K]}\mu_j$. Run $\text{kl-UCB}^{++}$ with horizon $T\geq K$ and exploration budget $g_T$ in~\eqref{eq:budget_def}, projecting each empirical mean onto $[\mu^-,\mu^+]$ and restricting the index candidates to that interval. If $k\neq k^\star$ and
\begin{align}
 \sqrt{\frac{22VK}{T}}\leq\delta\leq\frac{\mu_\star-\mu_k}{3},
 \label{eq:generic_klucbpp_delta}
\end{align}
then
\begin{align}
 \E{N_k(T)}
 &\leq\frac{G_K(T)}{d(\mu_k+\delta\|\mu_\star-\delta)}
 +\frac{(16e^2+2)2VK}{\delta^2}+1.
 \label{eq:generic_klucbpp_pull_bound}
\end{align}
\end{lemma}

\begin{IEEEproof}
Apply the decomposition in~\cite[eqs.~(24)--(30)]{menardGarivier2017}. Projection toward an interval containing $\mu_\star$ can only reduce the lower-tail KL deviation of the best arm's empirical mean, and once $n>\frac{G_K(T)}{d(\mu_k+\delta\|\mu_\star-\delta)}$, an index of arm $k$ above $\mu_\star-\delta$ requires its raw empirical mean to exceed $\mu_k+\delta$, so the same tail bounds apply. The variance bound is needed only between the cutoff means and the true means, all in $[\mu^-,\mu^+]$, and the maximal inequality~\cite[Lemma~4]{menardGarivier2017} needs no variance bound outside this interval. The estimates in~\cite[eqs.~(27) and~(30)]{menardGarivier2017}, together with the initial pull, give~\eqref{eq:generic_klucbpp_pull_bound}.
\end{IEEEproof}

\subsection{Wrong-Message Error Control}
\label{sec:vlsf_wrong}

The following lemma bounds the probability that the score of a wrong message reaches a given level within a given number of uses.

\begin{lemma}[Finite-horizon wrong-message control]
\label{lem:wrong_anytime_yi}
Let $\beta\triangleq\frac{\sigma_{\max}^2}{\sigma_{\min}^2}$ and
\begin{align}
 c_0&\triangleq\frac{e}{e-1}
       \left(\frac54+K\log\bigl(2\beta\log(4\beta)\bigr)+\frac K2\right).
 \label{eq:mixture_c0}
\end{align}
For every $u\geq0$ and $H\in\mathbb N$, the VLSF-Bandit process satisfies
\begin{align}
 \Prob{\exists m\neq W:\max_{1\leq n\leq H}S_n(m)\geq u}
 &\leq(M-1)c_0(1+\log H)\exp\{-u\}.
 \label{eq:wrong_anytime_yi}
\end{align}
\end{lemma}
\begin{IEEEproof} See \appref{app:awgn_wrong_anytime}.
\end{IEEEproof}

\subsection{True-Message Score}
\label{sec:vlsf_true_score}

We compare the score of the transmitted message with the information density at the true noise variances. Define
\begin{align}
 I_n&\triangleq\sum_{t=1}^n\log\frac{\phi_{\sigma_{A_t}^2}(Y_t-X_t)}{\phi_{v_{A_t}}(Y_t)},\qquad I_0=0,
 \label{eq:In_def}\\
 \tau_W&\triangleq\inf\{n\geq1\colon S_n(W)\geq\gamma\}.
 \label{eq:vlsf_tauW_def}
\end{align}
Conditioned on $\mc{F}_{t-1}$, the increment $I_t-I_{t-1}$ has mean $C^{(A_t)}$, and $\tau\leq\tau_W$ by the stopping rule. The next lemma bounds, at the time $\tau_W\wedge H$, the expected gap between $I_n$ and $S_n(W)$ and the expected score itself.

\begin{lemma}[True-message score]\label{lem:vlsf_plugin_loss}
Let
\begin{align}
 c_1&\triangleq2\beta+\frac12\log(4\beta+2\eta_{\max}),
 \label{eq:mixture_c1}\\
 c_2&\triangleq4K(c_1+8\beta^2),
 \label{eq:mixture_c2}
\end{align}
where $\beta=\frac{\sigma_{\max}^2}{\sigma_{\min}^2}$ and $\eta_{\max}=\frac{1}{\sigma_{\min}^2}$.
For every $\gamma>0$ and $H\in\mathbb N$,
\begin{align}
 \E{I_{\tau_W\wedge H}-S_{\tau_W\wedge H}(W)}
 &\leq c_2\sqrt{1+\log_2 H},
 \label{eq:mixture_stopped_loss}\\
 \E{S_{\tau_W\wedge H}(W)}&\leq\gamma+c_1.
 \label{eq:mixture_overshoot}
\end{align}
\end{lemma}
\begin{IEEEproof}
The two bounds are proved in \appref{app:awgn_plugin_loss} and \appref{app:awgn_bov}, respectively.
\end{IEEEproof}

\subsection{Output-Power $\text{kl-UCB}^{++}$ Pull Counts}

\label{sec:vlsf_pulls}

The expected decoding time also depends on how often each suboptimal channel is used. The channel-selection rule acts on the output powers. Their negatives form the exponential-family rewards to which we apply \lemref{lem:klucbpp_fixed_horizon}.

\begin{lemma}[Exponential-family reduction]\label{lem:awgn_op_exp_family} For channel $k\in[ K]$, under the Gaussian random-coding ensemble, the reward $- Y^2$, with $Y\sim\mc{N}(0, v_k)$
and $v_k=1+\sigma_k^2$, has a one-parameter canonical exponential-family law with mean $-v_k$. The variance
of this family is at most $V_{\mr{sq}}\triangleq2(1+\sigma_{\max}^2)^2$ over the output-variance interval
$[1+\sigma_{\min}^2,1+\sigma_{\max}^2]$, and the KL divergence between the laws with output variances $u$ and
$v$ is $\dG(u\|v)$.
\end{lemma}

\begin{IEEEproof} See \appref{app:awgn_sqres_exp_family}.
\end{IEEEproof}

By \lemref{lem:awgn_op_exp_family}, the projected empirical reward mean is
$-\nn{v}_k(t)$. Maximizing a plausible reward mean $-(1+s)$ is equivalent
to minimizing $s$, so the selection rule~\eqref{eq:klucb_rule} is the
$\text{kl-UCB}^{++}$ rule for the rewards $-Y_t^2$, with
$\mu_k=-v_k$ and $\mu_\star=-v_\star$. The reward-family KL divergence
satisfies
\begin{align}
    d(\mu_k+\delta\|\mu_\star-\delta)
    &=
    \dG(v_k-\delta\|v_\star+\delta).
    \label{eq:vlsf_reward_kl_identification}
\end{align}

The following tail bound lets us use the fixed-horizon estimates without restricting the decoding time.

\begin{lemma}[Tail of the true-message stopping time]
\label{lem:vlsf_residual_tail}
For every $\gamma>0$, the true-message stopping time has finite expectation. Let
\begin{align}
 c_3&\triangleq\frac{288}{\CG(\eta_{\min})^4}
             +\frac{384c_2^2}{\CG(\eta_{\min})^2}.
 \label{eq:mixture_tail_constant}
\end{align}
If $\log\gamma\geq1\vee\frac{2}{\sqrt{\CG(\eta_{\min})}}$, then, at the design horizon $T=K\vee\lceil\gamma\log^2\gamma\rceil$,
\begin{align}
 \Prob{\tau_W>T}&\leq\frac{c_3}{\gamma^2\log^3\gamma},
 \label{eq:mixture_true_tail_gamma_probability}\\
 \E{(\tau_W-T)^+}&\leq\frac{c_3}{\gamma\log\gamma}.
 \label{eq:vlsf_tau_tail_mean}
\end{align}
\end{lemma}
\begin{IEEEproof} See \appref{app:awgn_score_tail}.
\end{IEEEproof}

The next lemma bounds the expected number of uses of each suboptimal channel up to the design horizon~$T$.

\begin{lemma}[$\text{kl-UCB}^{++}$ pull count]\label{lem:awgn_klucbpp_pull} Under the VLSF-Bandit scheme in \secref{sec:scheme_vlsf}, fix a suboptimal channel $k\neq k^\star$. Then, as $\gamma\to\infty$,
\begin{align}
    \E{N_k(T)}
    &\leq
    \frac{\log\gamma}{D_{k,k^\star}}
    +O\bigl((\log\gamma)^{\frac{3}{4}}\bigr).
    \label{eq:awgn_klucbpp_pull_asymp}
\end{align}
\end{lemma}

\begin{IEEEproof} See \appref{app:awgn_klucbpp_pull_bound}.
\end{IEEEproof}

\subsection{Proof of \thmref{thm:vlsf_main}}
\label{sec:vlsf_proof_main}

We use the threshold and design horizon in \secref{sec:scheme_vlsf}. Until the stop-at-time-zero strategy is applied below, all probabilities and expectations are conditioned on the transmission branch.

\subsubsection{Error probability bound}

By \lemref{lem:vlsf_residual_tail}, $\E{\tau_W}<\infty$, so $\tau\leq\tau_W<\infty$ almost surely and the decoder stops. An error by time $T$ requires a wrong-message threshold crossing by then, and an error after time $T$ requires $\tau_W>T$. For all sufficiently large $M$, \lemref{lem:wrong_anytime_yi} and~\eqref{eq:mixture_true_tail_gamma_probability} give
\begin{align}
 \Prob{\nn{W}\neq W}
 &\leq c_0(M-1)(1+\log T)\exp\{-\gamma\}
       +\frac{c_3}{\gamma^2\log^3\gamma}.
 \label{eq:mixture_error_split}
\end{align}
With $g\triangleq\log\frac{M-1}{\epsilon_{\mr{dec}}}$, the threshold~\eqref{eq:mixture_threshold} is $\gamma=g+\sqrt{\log(1+g)}$, and the choice of $\epsilon_0$ in \thmref{thm:vlsf_main} gives
\begin{align}
 \frac1{\epsilon_{\mr{dec}}}&=(1-\epsilon_0)\log(M-1)\leq\gamma.
 \label{eq:vlsf_main_eps_dec}
\end{align}
Dividing \eqref{eq:mixture_error_split} by $\epsilon_{\mr{dec}}$ and using \eqref{eq:vlsf_main_eps_dec} in its last term yields
\begin{align}
 \frac{\Prob{\nn{W}\neq W}}{\epsilon_{\mr{dec}}}
 &\leq c_0(1+\log T)\exp\{-\sqrt{\log(1+g)}\}
       +\frac{c_3}{\gamma\log^3\gamma}.
 \label{eq:mixture_error_ratio}
\end{align}
The right-hand side of~\eqref{eq:mixture_error_ratio} decays to zero as $\gamma\to\infty$. Hence, for all sufficiently large $M$,
\begin{align}
 \Prob{\nn{W}\neq W}&\leq\epsilon_{\mr{dec}},
 \label{eq:vlsf_error_verified}
\end{align}
and the condition in \lemref{lem:vlsf_residual_tail} and $\epsilon_0\in[0,\epsilon)$ hold as well. The value of $M$ beyond which these hold depends only on $\epsilon$, $K$, and the known variance bounds, and is therefore the same at every admissible instance.

\subsubsection{Average decoding time bound}
\label{sec:vlsf_ave_dec_time}

Since $I_n-\sum_{t=1}^n C^{(A_t)}$ is a martingale, the optional stopping theorem for bounded stopping times~\cite[Th.~10.10]{williams1991} and $C^{(k)}=C_\star-\Delta_k$ for $k\neq k^\star$ give
\begin{align}
 C_\star\E{\tau_W\wedge T}
 &=\E{I_{\tau_W\wedge T}}+\sum_{k\neq k^\star}\Delta_k\E{N_k(\tau_W\wedge T)}.
 \label{eq:vlsf_optional_stopping_info}
\end{align}
Bounding $\E{I_{\tau_W\wedge T}}$ by \lemref{lem:vlsf_plugin_loss}, each $N_k(\tau_W\wedge T)$ by $N_k(T)$, and $\E{\tau_W}-\E{\tau_W\wedge T}=\E{(\tau_W-T)^+}$ by~\eqref{eq:vlsf_tau_tail_mean} gives
\begin{align}
 C_\star\E{\tau_W}
 &\leq\gamma+c_1+c_2\sqrt{1+\log_2 T}
       +\sum_{k\neq k^\star}\Delta_k\E{N_k(T)}+\frac{C_\star c_3}{\gamma\log\gamma}.
 \label{eq:vlsf_master_raw}
\end{align}
Since $\log T=\log\gamma+O(\log\log\gamma)$, the term $c_2\sqrt{1+\log_2 T}$ is $O(\sqrt{\log\gamma})$, and substituting \eqref{eq:awgn_klucbpp_pull_asymp} gives
\begin{align}
 \E{\tau}
 &\leq\E{\tau_W}
 \leq\frac1{C_\star}\left(\gamma+\sum_{k\neq k^\star}\frac{\Delta_k}{D_{k,k^\star}}\log\gamma\right)
 +O\bigl((\log\gamma)^{\frac34}\bigr).
 \label{eq:vlsf_tau_expected_bound}
\end{align}

\subsubsection{Power constraint}

It remains to verify~\eqref{eq:powerconstraint}. For $t\geq1$, conditioned on $W$, the outputs $Y^{t-1}$, and the codeword symbols $\{\mb{c}_s(m)\colon s\in[t-1],\,m\in[M]\}$, the input $X_t$ has distribution $\mc{N}(0,1)$, while $\{\tau\geq t\}$ is determined by these variables. Thus, $\E{X_t^2\1{\tau\geq t}}=\Prob{\tau\geq t}$. Interchanging the non-negative sum and expectation gives
\begin{align}
 \E{\sum_{t=1}^{\tau}X_t^2}
 &=\sum_{t=1}^{\infty}\E{X_t^2\1{\tau\geq t}}
 =\sum_{t=1}^{\infty}\Prob{\tau\geq t}=\E{\tau}.
 \label{eq:vlsf_power_verified}
\end{align}
Hence, the power constraint~\eqref{eq:powerconstraint} holds with equality. For all sufficiently large $M$, the error and power bounds hold at every variance vector in $[\sigma_{\min}^2,\sigma_{\max}^2]^K$.

\subsubsection{Second-order expansion}

By \eqref{eq:mixture_threshold} and~\eqref{eq:vlsf_main_eps_dec}, the threshold satisfies
\begin{align}
 \gamma&=\log M+\log\log M+O(\sqrt{\log\log M}).
 \label{eq:vlsf_main_gamma}
\end{align}
The independent stop at time zero keeps the error probability at most $\epsilon_0+(1-\epsilon_0)\epsilon_{\mr{dec}}=\epsilon$ and multiplies the expected decoding time and energy by $1-\epsilon_0$. Thus, the scheme is an $(N,M,\epsilon,K)$-VLSF-Bandit code with $N=(1-\epsilon_0)\E{\tau}$. Substituting \eqref{eq:vlsf_main_gamma} into \eqref{eq:vlsf_tau_expected_bound} gives
\begin{align}
 N&\leq\frac{1-\epsilon}{C_\star}\log M
 +\frac{1-\epsilon}{C_\star}\left(1+\sum_{k\neq k^\star}\frac{\Delta_k}{D_{k,k^\star}}\right)\log\log M
 +O\bigl((\log\log M)^{\frac34}\bigr).
 \label{eq:vlsf_NM_bound_main}
\end{align}
The converse in \secref{sec:main_vlf} and \eqref{eq:vlsf_NM_bound_main} imply $\log N=\log\log M+O(1)$. Replacing $\log\log M$ by $\log N$ costs $O(1)$, and inversion yields
\begin{align}
 \log M&\geq\frac{NC_\star}{1-\epsilon}
 -\left(1+\sum_{k\neq k^\star}\frac{\Delta_k}{D_{k,k^\star}}\right)\log N+o(\log N),
 \label{eq:vlsf_main_inversion}
\end{align}
where the remainder is $O((\log N)^{\frac34})$. This proves \thmref{thm:vlsf_main}.\hfill\IEEEQED

\section{Proof of the VLF-Bandit Results}

\label{sec:vlf_proof}

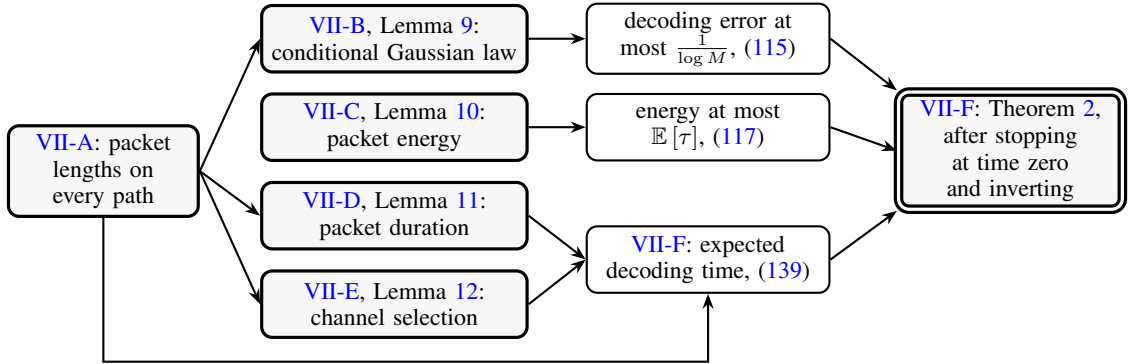
\begin{figure*}[!htbp]
\centering
\begin{tikzpicture}[
  >={Stealth[length=2mm]},
  thick,
  font=\small,
  execute at begin node={\linespread{1}\selectfont},
  pstep/.style = {rectangle, rounded corners, draw=black, very thick, fill=black!3,
                 align=center, inner sep=3pt, minimum height=2.5em},
  res/.style  = {rectangle, rounded corners, draw=black, thick, fill=white,
                 align=center, inner sep=3pt, minimum height=2.5em},
  line/.style = {draw, thick, ->}
]
\node[pstep, text width=2.3cm] (A) at (0,-1.3) {\ref{sec:vlf_deterministic}: packet lengths on every path};
\node[pstep, text width=3.3cm] (B) at (3.85,0.45) {\ref{sec:vlf_packet_gaussian}, \lemref{lem:vlf_packet_gaussian}:\\ conditional Gaussian law};
\node[pstep, text width=3.3cm] (C) at (3.85,-0.72) {\ref{sec:vlf_power}, \lemref{lem:vlf_packet_energy}:\\ packet energy};
\node[pstep, text width=3.3cm] (D) at (3.85,-1.89) {\ref{sec:vlf_packet_duration}, \lemref{lem:vlf_packet_duration}:\\ packet duration};
\node[pstep, text width=3.3cm] (E) at (3.85,-3.06) {\ref{sec:vlf_channel_selection}, \lemref{lem:vlf_channel_selection}:\\ channel selection};
\node[res, text width=3.0cm] (B2) at (8.0,0.45) {decoding error at most $\frac{1}{\log M}$, \eqref{eq:vlf_active_error}};
\node[res, text width=3.0cm] (C2) at (8.0,-0.72) {energy at most $\E{\tau}$, \eqref{eq:vlf_total_energy}};
\node[res, text width=3.0cm] (F) at (8.0,-2.47) {\ref{sec:vlf_theorem_proof}: expected decoding time, \eqref{eq:vlf_active_length}};
\node[pstep, text width=2.75cm, double, double distance=1.2pt] (G) at (12.0,-1.03) {\ref{sec:vlf_theorem_proof}: \thmref{thm:vlf_main}, after stopping at time zero and inverting};
\draw[line] (A.east) -- (B.west);
\draw[line] (A.east) -- (D.west);
\draw[line] (A.east) -- (E.west);
\draw[line] (B) -- (B2);
\draw[line] (C) -- (C2);
\draw[line] (D.east) -- (F.west);
\draw[line] (E.east) -- (F.west);
\draw[line] (A.south) |- ($(E.south)+(0,-0.3)$) -| (F.south);
\draw[line] (B2.east) -- (G.north west);
\draw[line] (C2.east) -- (G.west);
\draw[line] (F.east) -- (G.south west);
\end{tikzpicture}
\caption{Steps of the proof of \thmref{thm:vlf_main} and the subsections that contain them. An arrow points from a result to a step that uses it, and \lemref{lem:vlf_packet_gaussian} also enters the proofs of Lemmas~\ref{lem:vlf_packet_energy}--\ref{lem:vlf_channel_selection}.}
\label{fig:vlf_proof_map}
\end{figure*}

This section proves \thmref{thm:vlf_main}. The conditional Gaussian law of the normalized outputs gives the error and energy bounds. We then derive the packet durations, combine them with the channel-selection bounds, and apply the stop-at-time-zero strategy. Figure~\ref{fig:vlf_proof_map} shows how these steps depend on each other.

Fix the problem instance and let $M\to\infty$ through powers of two. Until the stop-at-time-zero strategy is applied in~\secref{sec:vlf_theorem_proof}, all probabilities and expectations are conditioned on the transmission branch.

\subsection{Packet Lengths}

\label{sec:vlf_deterministic}

Every precision estimate in~\eqref{eq:vlf_packet_precision} lies in $\left[\frac{\eta_{\min}}{2},2\eta_{\max}\right]$. Hence, for a packet with constant power $P_{j,n}=P_{j,0}$, the stopping rule~\eqref{eq:vlf_packet_stop} gives, for $\gamma_j>0$, the pathwise bounds
\begin{align}
 \frac{\gamma_j}{\CG\left(2P_{j,0}\eta_{\max}\right)}
 &\leq\tau^{(j)}\leq\left\lceil\frac{\gamma_j}{\CG\left(\frac{P_{j,0}\eta_{\min}}{2}\right)}\right\rceil.
 \label{eq:vlf_constant_crossing_bounds}
\end{align}

For the preliminary and exploration packets, $\frac{1}{\ell_0}\leq P_{j,0}\leq1$, so that
\begin{align}
 0\leq\gamma_j-\log M_j
 &\leq\frac{1}{2}\log\left(\frac{2\sigma_{\max}^2b\ell_0}{3}\right)= O\left(\log \psi\right)
 \label{eq:vlf_threshold_excess}
\end{align}
for sufficiently large $M$. Since $\log \psi= o\left(\ell_0\right)$, it follows that $\log M_j\leq\gamma_j\leq2\log M_j$ for $0\leq j\leq B$. Substituting these bounds into~\eqref{eq:vlf_constant_crossing_bounds}, and using $\CG(pa)\geq p\CG(a)$ for $0\leq p\leq1$, $\log M_j=\ell_0\log2$ for $1\leq j\leq B$, and $P_{j,0}\geq\frac{1}{\ell_0}$ for the upper bound on $\tau^{(j)}$, we obtain
\begin{align}
 \frac{\log M_0}{\CG\left(2\eta_{\max}\right)}
 &\leq\tau^{(0)}\leq\frac{2\log M_0}{\CG\left(\frac{\eta_{\min}}{2}\right)}+1,
 \label{eq:vlf_preliminary_bounds}\\
 \frac{\ell_0\log2}{\CG\left(2\eta_{\max}\right)}
 &\leq\tau^{(j)},\label{eq:vlf_exploration_lower}\\
 \tau^{(j)}+1&\leq\frac{2\log2}{\CG\left(\frac{\eta_{\min}}{2}\right)}\ell_0^2+2= O\left(\psi^{\frac{2}{3}}\right),\qquad 1\leq j\leq B.
 \label{eq:vlf_exploration_bounds}
\end{align}

The last exploration packet starts before the completed exploration packets reach $\lceil\psi^2\rceil$ channel uses, so \eqref{eq:vlf_exploration_lower}--\eqref{eq:vlf_exploration_bounds} give
\begin{align}
 \sum_{j=1}^{B}\left(\tau^{(j)}+1\right)
 &\leq\lceil \psi^2\rceil+\frac{2\log2}{\CG\left(\frac{\eta_{\min}}{2}\right)}\ell_0^2+2= O\left(\psi^2\right),\label{eq:vlf_exploration_cap_length}\\
 B\ell_0\log2&\leq\CG\left(2\eta_{\max}\right)\sum_{j=1}^{B}\tau^{(j)}=O\left(\psi^2\right).
 \label{eq:vlf_exploration_cap}
\end{align}

By~\eqref{eq:vlf_final_payload} and~\eqref{eq:vlf_saved_power}, the final packet therefore satisfies
\begin{align}
 0\leq\Omega&=O\left(\psi^2\right),\label{eq:vlf_saved_allocation_bound}\\
 \log M_{B+1}&=\log M-O\left(\psi^2\right)>0,
 \label{eq:vlf_final_admissible}\\
 n_{\mr{b}}&= O\left(1+\psi^4\right)= o\left(\log M\right),\label{eq:vlf_boost_length_bound}\\
 0\leq\frac{\Omega}{n_{\mr{b}}}&\leq\frac{\Omega}{1+\Omega^2}\leq\frac{1}{2}.
 \label{eq:vlf_final_power_bounds}
\end{align}

Thus, the final packet has a positive integer number of bits and refinement powers in $\left[1,\frac{3}{2}\right]$, its threshold is $\gamma_{B+1}=\log M_{B+1}+O\left(\log \psi\right)$, and, since each increment of $F_{B+1,n}$ is at least $\CG\left(\frac{\eta_{\min}}{2}\right)$,
\begin{align}
 \tau^{(B+1)}+1&=O\left(\log M\right).
 \label{eq:vlf_final_length_pathwise}
\end{align}

Finally, every threshold is at least $\ell_0\log2$ for sufficiently large $M$, whereas the first nine refinements of every packet use the precision $\frac{\eta_{\min}}{2}$ and powers at most $\frac{3}{2}$, so that $F_{j,9}\leq9\CG\left(\frac{3\eta_{\min}}{4}\right)<\ell_0\log2$. Hence, $\tau^{(j)}\geq10$, and the first nine refinements of every packet are transmitted.\footnote{Nine is the smallest number of refinements at the precision $\frac{\eta_{\min}}{2}$ for which every later unprojected precision estimate has a finite fourth moment, as \appref{app:vlf_packet_duration} requires; a chi-square variable $\chi_m^2$ has $\E{(\chi_m^2)^{-4}}<\infty$ if and only if $m>8$. These refinements also form an energy reserve against the upward bias of the projected estimates (\appref{app:vlf_packet_energy}).}

\subsection{Normalized Outputs and Decoding Error}

\label{sec:vlf_packet_gaussian}

The analysis of a packet conditions on the information retained at its start. For $j\geq0$, define the retained history
\begin{align}
 \mc{G}_{j,0}&\triangleq\sigma\left(U,Y^{K\ell_0},\left(\tau^{(i)},\left(\widetilde Z_{i,n}\right)_{n=1}^{\tau^{(i)}}\right)_{i=0}^{j-1}\right),
 \label{eq:vlf_packet_start_history}
\end{align}
where $\tau^{(i)}\triangleq0$ for a packet that is not transmitted. This history excludes the initial outputs $Y_{i,0}$ and the decoded packets, which depend on the message and enter only the message estimates. It determines whether packet $j$ is transmitted and, if so, its channel, alphabet size, threshold, powers, and first use. For the analysis, continue its recursion beyond $\tau^{(j)}$ using fresh $\mc{N}(0,\sigma_{J_j}^2)$ noises, independent of the actual transmission and of each other given $\mc{G}_{j,0}$. Let
\begin{align}
 \mc{G}_{j,n}&\triangleq\mc{G}_{j,0}\vee\sigma\left(\widetilde Z_{j,1},\ldots,\widetilde Z_{j,n}\right),\qquad n\geq1,
 \label{eq:vlf_packet_history}
\end{align}
where $\vee$ denotes the join of sigma-fields, and let $\mc{G}_{j,\tau^{(j)}}$ be the stopped sigma-field. Since $\{B+1=j\}\in\mc{G}_{j,0}$ and $\mc{G}_{j,0}\subseteq\mc{G}_{j+1,0}$, the index $B+1$ of the final packet is a stopping time for $\left(\mc{G}_{j,0}\right)_{j\geq0}$; $\mc{G}_{B+1,0}$ denotes its stopped sigma-field.

The next lemma gives the conditional law of the normalized outputs and of the normalized error.

\begin{lemma}[Conditional Gaussian law]\label{lem:vlf_packet_gaussian} Fix a transmitted packet $j$. For every $n\geq0$, conditioned on $\mc{G}_{j,0}$ and $W$, the variables $\widetilde Z_{j,1},\ldots,\widetilde Z_{j,n},\allowbreak E_{j,n}$ are independent with distribution $\mc{N}\left(0,\sigma_{J_j}^2\right)$, and
\begin{align}
 \Prob{E_{j,\tau^{(j)}}\leq z\mid\mc{G}_{j,\tau^{(j)}},W}
 &=\int_{-\infty}^{z}\phi_{\sigma_{J_j}^2}(y)\,dy,\qquad z\in\mathbb{R}.
 \label{eq:vlf_stopped_gaussian}
\end{align}
Moreover, $\mc{G}_{j,\tau^{(j)}}$ is independent of $W$, and, conditioned on $\mc{G}_{j,0}$, the packet message $W_j$ is uniform on $[M_j]$.
\end{lemma}

\begin{IEEEproof} Write $s=\sigma_{J_j}^2$. On $\{\tau^{(j)}\geq n\}\in\mc{G}_{j,n-1}$, the next refinement uses a channel and a time determined by the retained history. Since $\mc{G}_{j,n-1}$, $W$, and $E_{j,n-1}$ depend only on the common randomness, the message, and earlier channel noises, memorylessness gives $Z_{j,n}\sim\mc{N}(0,s)$ conditionally on these variables. The continuation noises have the same conditional law on $\{\tau^{(j)}<n\}$. The initial noise also has this law given $\mc{G}_{j,0}$ and $W$, so $E_{j,0}=-Z_{j,0}$ proves the case $n=0$.

Suppose that $E_{j,n-1}$ has distribution $\mc{N}(0,s)$ conditioned on $\mc{G}_{j,n-1}$ and $W$. By~\eqref{eq:vlf_packet_error}--\eqref{eq:vlf_packet_estimate}, whose squared gain $P_{j,n}\nn{\eta}_{j,n-1}$ is $\mc{G}_{j,n-1}$-measurable,
\begin{align}
 \begin{pmatrix}E_{j,n}\\\widetilde Z_{j,n}\end{pmatrix}
 &=\frac{1}{\sqrt{1+P_{j,n}\nn{\eta}_{j,n-1}}}
 \begin{pmatrix}1&-\sqrt{P_{j,n}\nn{\eta}_{j,n-1}}\\\sqrt{P_{j,n}\nn{\eta}_{j,n-1}}&1\end{pmatrix}
 \begin{pmatrix}E_{j,n-1}\\Z_{j,n}\end{pmatrix}.
 \label{eq:vlf_gaussian_rotation}
\end{align}

Since the matrix is orthogonal, $E_{j,n}$ and $\widetilde Z_{j,n}$ are independent $\mc{N}(0,s)$ variables conditioned on $\mc{G}_{j,n-1}$ and $W$. In particular, $E_{j,n}$ has distribution $\mc{N}(0,s)$ conditioned on $\mc{G}_{j,n}=\mc{G}_{j,n-1}\vee\sigma\left(\widetilde Z_{j,n}\right)$ and $W$. This completes an induction over $n$ and proves the first assertion.

The packet length is bounded given $\mc{G}_{j,0}$ by~\secref{sec:vlf_deterministic}. Partitioning over $\{\tau^{(j)}=n\}$ therefore gives~\eqref{eq:vlf_stopped_gaussian}.

Since $E_{j,0}=-Z_{j,0}$, the rotation~\eqref{eq:vlf_gaussian_rotation} makes the normalized outputs functions of the retained history and fresh packet noises alone. Starting with the zero-input observations, induction over packets therefore makes every retained history independent of $W$. Given $\mc{G}_{j,0}$, the packet's bit positions are fixed, so the independent uniform bits of $W-1$ make $W_j$ uniform on $[M_j]$.
\end{IEEEproof}

An incorrect packet decision implies $\left|W_j-\nn{W}_{j,\tau^{(j)}}\right|\geq\frac{1}{2}$. Since $F_{j,\tau^{(j)}}\geq\gamma_j$, the definitions~\eqref{eq:vlf_packet_error} and~\eqref{eq:vlf_packet_stop} give
\begin{align}
 \{\text{packet }j\text{ is decoded incorrectly}\}
 &\subseteq\left\{\left|E_{j,\tau^{(j)}}\right|\geq\frac{\sqrt{3P_{j,0}}\exp\{\gamma_j\}}{M_j}\right\}
 =\left\{E_{j,\tau^{(j)}}^2\geq2\sigma_{\max}^2b\right\}.
 \label{eq:vlf_packet_error_event}
\end{align}

The bound $\Prob{|Z|\geq a}\leq\exp\left\{-\frac{a^2}{2\sigma^2}\right\}$ for $Z\sim\mc{N}(0,\sigma^2)$ and $a\geq0$, together with~\eqref{eq:vlf_stopped_gaussian} and $\sigma_{J_j}^2\leq\sigma_{\max}^2$, gives
\begin{align}
 \Prob{\text{packet }j\text{ is decoded incorrectly}\mid\mc{G}_{j,0}}
 &\leq\exp\left\{-\frac{\sigma_{\max}^2b}{\sigma_{J_j}^2}\right\}
 \leq\exp\{-b\}.
 \label{eq:vlf_packet_error_bound}
\end{align}

Since whether packet $j$ is transmitted is determined by $\mc{G}_{j,0}$ and at most $\lceil \psi^2\rceil+2$ packets are transmitted, the union bound and~\eqref{eq:vlf_packet_budget} give
\begin{align}
 \Prob{\nn{W}\ne W}
 &\leq\left(\lceil\psi^2\rceil+2\right)\exp\{-b\}=\frac{1}{\log M}.
 \label{eq:vlf_active_error}
\end{align}

\subsection{Power Constraint}

\label{sec:vlf_power}

The energy of a packet is random, because its gains depend on the precision estimates. The next lemma bounds its expected energy by its nominal allocation.

\begin{lemma}[Packet energy]\label{lem:vlf_packet_energy} For every transmitted packet $j$,
\begin{align}
 \E{\sum_{n=0}^{\tau^{(j)}}X_{j,n}^2\,\middle|\,\mc{G}_{j,0}}
 &\leq P_{j,0}+\E{\sum_{n=1}^{\tau^{(j)}}P_{j,n}\,\middle|\,\mc{G}_{j,0}}.
 \label{eq:vlf_packet_energy}
\end{align}
\end{lemma}

\begin{IEEEproof} See \appref{app:vlf_packet_energy}.
\end{IEEEproof}

The preliminary and exploration packets have total nominal allocation $\sum_{j=0}^{B}(\tau^{(j)}+1)-\Omega$ by~\eqref{eq:vlf_saved_power}, and the final packet has allocation at most $\tau^{(B+1)}+1+\Omega$ by~\eqref{eq:vlf_final_power}. Summing~\eqref{eq:vlf_packet_energy} by the tower property, and noting that the $K\ell_0$ initial uses carry no energy, gives
\begin{align}
 \E{\sum_{t=1}^{\tau}X_t^2}
 &\leq\E{\sum_{j=0}^{B}\left(\tau^{(j)}+1\right)}-\E{\Omega}+\E{\tau^{(B+1)}+1}+\E{\Omega}
 =\E{\tau}-K\ell_0\leq\E{\tau}.
 \label{eq:vlf_total_energy}
\end{align}

\subsection{Expected Packet Duration}

\label{sec:vlf_packet_duration}

The duration of a packet is governed by the true-precision reduction
\begin{align}
 \overline F_{j,n}&\triangleq\sum_{i=1}^{n}\CG\left(\frac{P_{j,i}}{\sigma_{J_j}^2}\right),\qquad n\geq0,
 \label{eq:vlf_true_capacity_sum}
\end{align}
which the packet would accumulate if the precision were known. We compare $\overline F_{j,n}$ with $F_{j,n}$ for a transmitted packet $j$ whose refinement powers satisfy
\begin{align}
 q\leq P_{j,n}\leq2q,\qquad n\geq1,
 \label{eq:vlf_duration_power_range}
\end{align}
for a $\mc{G}_{j,0}$-measurable $q\in(0,1]$. By~\secref{sec:vlf_deterministic}, this holds with $q=P_{j,0}$ for the preliminary and exploration packets and with $q=1$ for the final packet. Write $s=\sigma_{J_j}^2$ and condition on $\mc{G}_{j,0}$; by \lemref{lem:vlf_packet_gaussian}, the normalized outputs are then i.i.d.\ $\mc{N}(0,s)$, and the alphabet size, the threshold, and the powers are fixed. For $m\geq9$, refinement $m+1$ uses the estimate $\nn{\eta}_{j,m}$ in place of $\frac{1}{s}$, and concavity of $\CG$ gives the non-negative Taylor remainder
\begin{align}
 R_{j,m}&\triangleq\CG\left(\frac{P_{j,m+1}}{s}\right)
 +\frac{P_{j,m+1}s}{2\left(P_{j,m+1}+s\right)}\left(\nn{\eta}_{j,m}-\frac{1}{s}\right)
 -\CG\left(P_{j,m+1}\nn{\eta}_{j,m}\right)\geq0.
 \label{eq:vlf_duration_deficit}
\end{align}

Since the first nine refinements use the precision $\frac{\eta_{\min}}{2}$ and $\tau^{(j)}\geq10$, summing~\eqref{eq:vlf_duration_deficit} over the refinements of the packet gives
\begin{align}
 \overline F_{j,\tau^{(j)}}
 &=F_{j,\tau^{(j)}}
 +\sum_{n=1}^{9}\left[\CG\left(\frac{P_{j,n}}{s}\right)-\CG\left(\frac{P_{j,n}\eta_{\min}}{2}\right)\right]-\sum_{m=9}^{\tau^{(j)}-1}\frac{P_{j,m+1}s}{2\left(P_{j,m+1}+s\right)}\left(\nn{\eta}_{j,m}-\frac{1}{s}\right)
 +\sum_{m=9}^{\tau^{(j)}-1}R_{j,m}.
 \label{eq:vlf_duration_identity}
\end{align}

In \appref{app:vlf_packet_duration}, we calculate
\begin{align}
 \E{R_{j,m}}&=\frac{P_{j,m+1}^2}{2\left(P_{j,m+1}+s\right)^2m}+O\left(q^2m^{-\frac{3}{2}}\right),\qquad m\geq9.
 \label{eq:vlf_duration_coefficient}
\end{align}

The linear sum stops at $\tau^{(j)}$, and the event $\{\tau^{(j)}>m\}$ is determined by $\widetilde Z_{j,1},\ldots,\widetilde Z_{j,m-1}$, which also enter $\nn{\eta}_{j,m}$. We show in \appref{app:vlf_packet_duration} that its expectation nevertheless satisfies
\begin{align}
 \left|\E{\sum_{m=9}^{\tau^{(j)}-1}\frac{P_{j,m+1}s}{2\left(P_{j,m+1}+s\right)}
 \left(\nn{\eta}_{j,m}-\frac{1}{s}\right)}\right|&=O\left(q\right).
 \label{eq:vlf_stopped_precision_bias}
\end{align}

In~\eqref{eq:vlf_duration_identity}, the first sum is at most $9q\eta_{\max}$, the overshoot $F_{j,\tau^{(j)}}-\gamma_j$ is at most $2q\eta_{\max}$, and $\tau^{(j)}\leq\left\lceil\frac{\gamma_j}{\CG\left(\frac{q\eta_{\min}}{2}\right)}\right\rceil$ because every increment of $F_{j,n}$ is at least $\CG\left(\frac{q\eta_{\min}}{2}\right)$. As $R_{j,m}\geq0$, its stopped sum is at most its sum up to this deterministic bound. Taking expectations in~\eqref{eq:vlf_duration_identity} and applying~\eqref{eq:vlf_duration_coefficient} and~\eqref{eq:vlf_stopped_precision_bias} yields
\begin{align}
 \E{\overline F_{j,\tau^{(j)}}\,\middle|\,\mc{G}_{j,0}}
 &\leq\gamma_j+
 \sum_{m=9}^{\left\lceil\frac{\gamma_j}{\CG\left(\frac{q\eta_{\min}}{2}\right)}\right\rceil-1}
 \frac{P_{j,m+1}^2}{2\left(P_{j,m+1}+s\right)^2m}+O\left(q\right).
 \label{eq:vlf_capacity_sum_expectation}
\end{align}

The summation in~\eqref{eq:vlf_capacity_sum_expectation} produces the universal-decoding term. We now convert the capacity-sum bound into packet durations.

\begin{lemma}[Expected packet duration]\label{lem:vlf_packet_duration} For every transmitted preliminary or exploration packet $j$, as $M\to\infty$ through powers of two,
\begin{align}
 \CG\left(\frac{P_{j,0}}{\sigma_{J_j}^2}\right)\E{\tau^{(j)}+1\mid\mc{G}_{j,0}}
 &\leq\log M_j+O\left(\log \psi\right).
 \label{eq:vlf_short_duration}
\end{align}
For the final packet,
\begin{align}
 \CG\left(\frac{1}{\sigma_{J_{B+1}}^2}\right)\E{\tau^{(B+1)}+1\mid\mc{G}_{B+1,0}}
 &\leq\log M_{B+1}-\frac{\Omega}{2\left(1+\sigma_{J_{B+1}}^2\right)}
 +\frac{\psi}{2\left(1+\sigma_{J_{B+1}}^2\right)^2}+o\left(\psi\right).
 \label{eq:vlf_final_duration}
\end{align}
The remainders are uniform over the retained histories and the variances in $\left[\sigma_{\min}^2,\sigma_{\max}^2\right]$, and in~\eqref{eq:vlf_short_duration}, also over the exploration powers in $\left[\frac{1}{\ell_0},1\right]$.
\end{lemma}

\begin{IEEEproof} See \appref{app:vlf_packet_duration_lemma}.
\end{IEEEproof}

\subsection{Channel Selection}

\label{sec:vlf_channel_selection}

The expected decoding time also depends on how often each channel is explored. Define the number of exploration uses of channel $k$, including the initial use of each packet,
\begin{align}
 T_k&\triangleq\sum_{j=1}^{B}\left(\tau^{(j)}+1\right)\1{J_j=k},
 \qquad k\in[K].
 \label{eq:vlf_exploration_uses}
\end{align}

We use the following bounds on the exploration uses, powers, and channel-selection errors.

\begin{lemma}[Channel-selection bounds]\label{lem:vlf_channel_selection} As $M\to\infty$ through powers of two,
\begin{align}
 \E{T_k}&\leq
 \frac{\psi}{\dG\left(\sigma_k^2\|\sigma_{k^\star}^2\right)}+o(\psi),
 \qquad k\neq k^\star,
 \label{eq:vlf_suboptimal_uses}\\
 \E{T_{k^\star}}&=o(\psi),\label{eq:vlf_best_exploration_uses}\\
 \E{B}&=O\left(\psi^{\frac{2}{3}}\right),
 \label{eq:vlf_best_uses_packets}\\
 \E{\sum_{j=1}^{B}\left|P_{j,0}-P_k^\star\right|
       \left(\tau^{(j)}+1\right)\1{J_j=k}}
 &=o(\psi),
 \qquad k\neq k^\star.
 \label{eq:vlf_weighted_power}
\end{align}
Moreover,
\begin{align}
 \Prob{J_0\neq k^\star}
 &=o\left(\psi^{-1}\right),\label{eq:vlf_preliminary_channel_error}\\
 \Prob{J_{B+1}\neq k^\star}
 &= O\left(\frac{1}{\log M}\right).
 \label{eq:vlf_channel_errors}
\end{align}
\end{lemma}

\begin{IEEEproof} See \appref{app:vlf_channel_selection}.
\end{IEEEproof}

\subsection{Proof of \thmref{thm:vlf_main}}

\label{sec:vlf_theorem_proof}

By~\eqref{eq:vlf_total_length}, the decoding time consists of the $K\ell_0=o(\psi)$ initial uses and the packet durations. We bound $\CG_\star$ times each expected packet duration by the number of nats that the packet carries plus its excess cost.

For the preliminary packet, $\CG_\star-\CG\left(\frac{1}{\sigma_{J_0}^2}\right)$ vanishes on $\{J_0=k^\star\}$ and is bounded otherwise. Averaging~\eqref{eq:vlf_short_duration} and using~\eqref{eq:vlf_preliminary_bounds} and~\eqref{eq:vlf_preliminary_channel_error} gives
\begin{align}
 \CG_\star\E{\tau^{(0)}+1}
 &\leq \log M_0+O\left(\log \psi\right)+O\left(\psi^2\right)\Prob{J_0\ne k^\star}
 =\log M_0+o(\psi).
 \label{eq:vlf_preliminary_information}
\end{align}

Summing~\eqref{eq:vlf_short_duration} over the exploration packets with the tower property, and using $\E{B}\,O(\log\psi)=o(\psi)$ by~\eqref{eq:vlf_best_uses_packets}, gives
\begin{align}
 \CG_\star\E{\sum_{j=1}^{B}\left(\tau^{(j)}+1\right)}
 &\leq\E{B}\,\ell_0\log2
 +\E{\sum_{j=1}^{B}\left[\CG_\star-\CG\left(\frac{P_{j,0}}{\sigma_{J_j}^2}\right)\right]\left(\tau^{(j)}+1\right)}+o(\psi).
 \label{eq:vlf_exploration_common_scale}
\end{align}

For the final packet, replacing $\sigma_{J_{B+1}}^2$ by $\sigma_{k^\star}^2$ in~\eqref{eq:vlf_final_duration} costs $O(1)$ in expectation: the affected terms are at most $O(\log M)$ by~\eqref{eq:vlf_saved_allocation_bound} and~\eqref{eq:vlf_final_length_pathwise}, and channel misselection has probability $O\left(\frac{1}{\log M}\right)$ by~\eqref{eq:vlf_channel_errors}. Hence,
\begin{align}
 \CG_\star\E{\tau^{(B+1)}+1}
 &\leq\E{\log M_{B+1}}-\frac{\E{\Omega}}{2v_\star}
 +\frac{\psi}{2v_\star^2}+o(\psi).
 \label{eq:vlf_final_common_scale}
\end{align}

The payloads satisfy $\log M_0+B\ell_0\log2+\log M_{B+1}=\log M$ by~\eqref{eq:vlf_final_payload}. Adding~\eqref{eq:vlf_preliminary_information}--\eqref{eq:vlf_final_common_scale} and substituting $\Omega$ from~\eqref{eq:vlf_saved_power} yields
\begin{align}
 \CG_\star\E{\tau}
 &\leq\log M+\frac{\psi}{2v_\star^2}
 +\E{\sum_{j=1}^{B}\left[\CG_\star-\CG\left(\frac{P_{j,0}}{\sigma_{J_j}^2}\right)-\frac{1-P_{j,0}}{2v_\star}\right]\left(\tau^{(j)}+1\right)}+o(\psi).
 \label{eq:vlf_duration_with_power_credit}
\end{align}

The bracket is uniformly bounded and Lipschitz in $P_{j,0}\in[0,1]$. Hence, packets on the best channel contribute $o(\psi)$ by~\eqref{eq:vlf_best_exploration_uses}, and~\eqref{eq:vlf_weighted_power} replaces the suboptimal-channel powers by $P_k^\star$ at a cost $o(\psi)$. Using~\eqref{eq:vlf_exploration_uses} gives
\begin{align}
 \CG_\star\E{\tau}
 &\leq\log M+\frac{\psi}{2v_\star^2}
 +\sum_{k\ne k^\star}\left[\CG_\star-\CG\left(\frac{P_k^\star}{\sigma_k^2}\right)-\frac{1-P_k^\star}{2v_\star}\right]\E{T_k}+o(\psi).
 \label{eq:vlf_duration_limiting_powers}
\end{align}

The channel ordering and the tangent bound for $\CG$ give $\CG\left(\frac{p}{\sigma_k^2}\right)\leq\CG\left(\frac{p}{\sigma_{k^\star}^2}\right)\leq\CG_\star-\frac{1-p}{2v_\star}$ for $0\leq p\leq1$. The multipliers of $\E{T_k}$ in~\eqref{eq:vlf_duration_limiting_powers} are therefore non-negative, and~\eqref{eq:vlf_suboptimal_uses} yields
\begin{align}
 \CG_\star\E{\tau}
 &\leq\log M+
 \left(\frac{1}{2v_\star^2}
 +\sum_{k\ne k^\star}
 \frac{\CG_\star-\CG\left(\frac{P_k^\star}{\sigma_k^2}\right)-\frac{1-P_k^\star}{2v_\star}}
 {\dG\left(\sigma_k^2\|\sigma_{k^\star}^2\right)}\right)\psi+o(\psi).
 \label{eq:vlf_active_length}
\end{align}

\emph{Stop-at-time-zero strategy:} We now include the initial randomization, so probabilities and expectations below are unconditional. For $\epsilon_0=\epsilon-\frac{1}{\log M}$ in~\eqref{eq:vlf_stop_zero}, the bounds~\eqref{eq:vlf_active_error} and~\eqref{eq:vlf_total_energy} give
\begin{align}
 \Prob{\nn{W}\ne W}
 &\leq\epsilon_0
 +\frac{1-\epsilon_0}{\log M}
 \leq\epsilon,
 \label{eq:vlf_mixed_error}\\
 \E{\sum_{t=1}^{\tau}X_t^2}
 &=\left(1-\epsilon_0\right)\E{\sum_{t=1}^{\tau}X_t^2\,\middle|\,\tau>0}
 \leq\left(1-\epsilon_0\right)\E{\tau\mid\tau>0}
 =\E{\tau}.
 \label{eq:vlf_mixed_energy}
\end{align}

The error and power bounds~\eqref{eq:vlf_mixed_error}--\eqref{eq:vlf_mixed_energy} hold throughout the prescribed variance interval.

The duration in~\eqref{eq:vlf_active_length} is multiplied by $1-\epsilon_0=1-\epsilon+\frac{1}{\log M}$. With $N=\E{\tau}$, the converse in~\secref{sec:main_vlf} and this bound give $\psi=\log N+O(1)$. Substituting this relation into~\eqref{eq:vlf_active_length} and inverting after the randomization yields, as $M\to\infty$ through powers of two at every instance,
\begin{align}
 \log M
 &\geq\frac{N\CG_\star}{1-\epsilon}
 -\left(\frac{1}{2v_\star^2}
 +\sum_{k\neq k^\star}
 \frac{\CG_\star-\CG\left(\frac{P_k^\star}{\sigma_k^2}\right)-\frac{1-P_k^\star}{2v_\star}}
 {\dG\left(\sigma_k^2\|\sigma_{k^\star}^2\right)}\right)\log N
 +o(\log N).
 \label{eq:vlf_main_inversion}
\end{align}
This proves \thmref{thm:vlf_main}.\hfill\IEEEQED

\section{Comparison With the Explore-then-Commit Strategy}

\label{sec:etc_comparison}

An ETC strategy separates the two tasks that the VLF-Bandit scheme performs jointly. It first identifies a channel from zero-input pilot observations and then sends the entire message over the declared channel as one packet (\secref{sec:scheme_vlf}), which spends the energy saved during identification. The ETC strategy is a VLF-Bandit code.

Let $K\geq2$, and use the stop-at-time-zero randomization~\eqref{eq:vlf_stop_zero}. Until it is applied below, all probabilities and expectations in this section are conditioned on the transmission branch. During identification, the transmitter sends zero, so a pilot observation on channel $k$ satisfies $\frac{Y_t^2}{\sigma_k^2}\sim\chi_1^2$. The computation in~\appref{app:awgn_sqres_exp_family}, with $v_k$ replaced by $\sigma_k^2$, shows that the rewards $-Y_t^2$ on channel $k$ form a one-parameter exponential family with mean $-\sigma_k^2$, in which the divergence between the members with means $-u$ and $-v$ is $\dG(u\|v)$. The results of Kaufmann and Koolen~\cite{kaufmann2021mixture} on best-arm identification in such families therefore apply. The characteristic time of the problem is $\Upsilon_{\mr{ETC}}\left(\bm{\sigma}^2\right)$, where
\begin{align}
 \left(\Upsilon_{\mr{ETC}}\left(\bm{\sigma}^2\right)\right)^{-1}
 &\triangleq\sup_{\substack{\bm{\lambda}\in\left[0,1\right]^K:\\ \sum_{i=1}^K\lambda_i=1}}
 \inf_{\substack{\widetilde{\bm{\sigma}}^2\in\left(0,\infty\right)^K:\\ \exists k\ne k^\star:\ \widetilde\sigma_k^2<\widetilde\sigma_{k^\star}^2}}
 \sum_{i=1}^K\lambda_i\dG\left(\sigma_i^2\|\widetilde\sigma_i^2\right).
 \label{eq:etc_upsilon_awgn}
\end{align}

We use the asymptotically optimal Track-and-Stop rule in~\cite[Sec.~5.1]{kaufmann2021mixture} at error level $\delta_{\mr p}=\frac{1}{\log M}$. Let $\tau_{\mr p}\geq K$ be the number of pilot uses and $J_0$ the declared channel, so that $\Prob{J_0\ne k^\star}\leq\delta_{\mr p}$. The best channel is unique, and the optimal allocation is unique and continuous~\cite[Th.~5 and Prop.~6]{garivier2016}. The expected-time argument in~\cite[Th.~17]{kaufmann2021mixture} and its tail bound~\cite[App.~D]{kaufmann2021mixture} give, for $\psi=\log\log M$,
\begin{align}
 \E{\tau_{\mr p}}&\leq\Upsilon_{\mr{ETC}}(\bm\sigma^2)\psi+o(\psi),
 \label{eq:etc_identification_mean}\\
 \E{\tau_{\mr p}^2}&=O(\psi^2).
 \label{eq:etc_identification_bound}
\end{align}

After identification, the transmitter sends the entire message as packet zero on channel $J_0$, with $M_0=M$, the threshold in~\eqref{eq:vlf_packet_stop} at $b=\psi$, $P_{0,0}=1$, and
\begin{align}
 P_{0,n}&\triangleq
 \begin{cases}
 1+\frac{1}{\tau_{\mr{p}}},&1\leq n\leq\tau_{\mr{p}}^2,\\[2pt]
 1,&n>\tau_{\mr{p}}^2.
 \end{cases}
 \label{eq:etc_reused_power}
\end{align}

The extra allocation is at most $\tau_{\mr p}$, the energy saved by the zero pilots. Since the pilot history is independent of the message, Lemmas~\ref{lem:vlf_packet_gaussian} and~\ref{lem:vlf_packet_energy} apply to the subsequent packet. Its error probability is at most $\exp\{-\psi\}=\frac{1}{\log M}$, and the saved allocation covers its excess energy.

The packet-duration calculation in~\eqref{eq:vlf_capacity_sum_expectation}, with $q=1$, gives a capacity credit of $\frac{\E{\tau_{\mr p}}}{2v_\star}+o(\psi)$ for the saved allocation. The second-moment bound in~\eqref{eq:etc_identification_bound} makes the loss from unused allocation negligible, while a wrongly declared channel contributes only $O(1)$ to the expected duration. Adding the pilot duration yields
\begin{align}
 \CG_\star\E{\tau_{\mr{p}}+\tau^{(0)}+1}
 &\leq\log M+\frac{\psi}{2v_\star^2}
 +\left(\CG_\star-\frac{1}{2v_\star}\right)\E{\tau_{\mr{p}}}+ o\left(\psi\right)
 \leq\log M+\rho_{\mr{ETC}}\psi+o(\psi),
 \label{eq:etc_active_coefficient}\\*
 \rho_{\mr{ETC}}&\triangleq\frac{1}{2v_\star^2}
 +\left(\CG_\star-\frac{1}{2v_\star}\right)\Upsilon_{\mr{ETC}}\left(\bm{\sigma}^2\right).
 \label{eq:rho_etc_awgn}
\end{align}

The stop-at-time-zero strategy and the inversion in~\secref{sec:vlf_theorem_proof} then show that this ETC strategy achieves the expansion~\eqref{eq:vlf_main} with the coefficient of $\log N$ replaced by $-\rho_{\mr{ETC}}$.

Denote the magnitude of the coefficient of $\log N$ in~\eqref{eq:vlf_main} by
\begin{align}
 \rho_{\mr{joint}}&\triangleq\frac{1}{2v_\star^2}
 +\sum_{k\ne k^\star}\frac{\CG_\star-\CG\left(\frac{P_k^\star}{\sigma_k^2}\right)-\frac{1-P_k^\star}{2v_\star}}
 {\dG\left(\sigma_k^2\|\sigma_{k^\star}^2\right)}.
 \label{eq:rho_joint_awgn}
\end{align}

We show that $\rho_{\mr{joint}}<\rho_{\mr{ETC}}$ for every instance with $K\geq2$. Let $\sigma_{(2)}^2\triangleq\min_{k\ne k^\star}\sigma_k^2$, and fix weights $\bm\lambda$ in~\eqref{eq:etc_upsilon_awgn}. Alternatives that make channels $k$ and $k^\star$ nearly tied at variance $\sigma_{k^\star}^2$ bound the inner infimum by $\lambda_k\dG(\sigma_k^2\|\sigma_{k^\star}^2)$. A near tie between the best and second-best channels at variance $\sigma_{(2)}^2$ bounds it by $\lambda_{k^\star}\dG(\sigma_{k^\star}^2\|\sigma_{(2)}^2)$. Summing the resulting lower bounds on the weights gives
\begin{align}
 \Upsilon_{\mr{ETC}}\left(\bm{\sigma}^2\right)
 &\geq\frac{1}{\dG\left(\sigma_{k^\star}^2\|\sigma_{(2)}^2\right)}
 +\sum_{k\ne k^\star}\frac{1}{\dG\left(\sigma_k^2\|\sigma_{k^\star}^2\right)}.
 \label{eq:etc_characteristic_lower}
\end{align}

Also, $P_k^\star$ maximizes $\CG\left(\frac{p}{\sigma_k^2}\right)-\frac{p}{2v_\star}$ over $0\leq p\leq1$, and the maximum is non-negative because $p=0$ is feasible. Subtracting~\eqref{eq:rho_joint_awgn} from~\eqref{eq:rho_etc_awgn} now gives
\begin{align}
 \rho_{\mr{ETC}}-\rho_{\mr{joint}}
 &\geq\frac{\CG_\star-\frac{1}{2v_\star}}{\dG\left(\sigma_{k^\star}^2\|\sigma_{(2)}^2\right)}
 +\sum_{k\ne k^\star}\frac{\CG\left(\frac{P_k^\star}{\sigma_k^2}\right)-\frac{P_k^\star}{2v_\star}}
 {\dG\left(\sigma_k^2\|\sigma_{k^\star}^2\right)}>0.
 \label{eq:etc_joint_difference_awgn}
\end{align}

The first term is strictly positive, and every summand is non-negative. ETC must spend zero-input observations on the best channel itself, whereas the joint scheme learns its noise variance while transmitting message bits. Communication on suboptimal channels with $P_k^\star>0$ provides the additional non-negative terms in~\eqref{eq:etc_joint_difference_awgn}.

\begin{figure}[t]
\centering
\includegraphics[width=0.68\columnwidth]{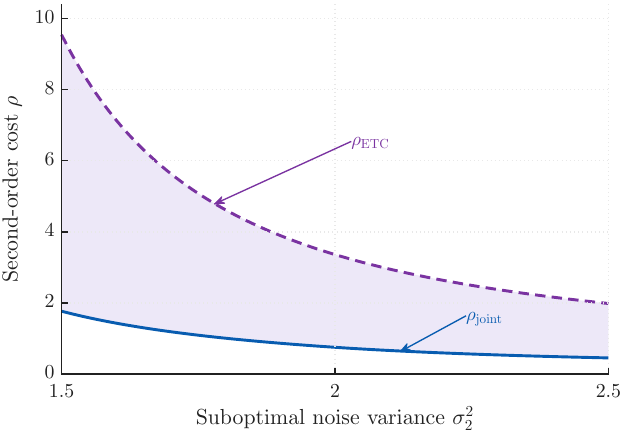}
\caption{Second-order coefficients for $K=2$, unit average power, $\sigma_1^2=1$, and $\sigma_2^2\in[1.5,2.5]$. The shaded region shows $\rho_{\mr{ETC}}-\rho_{\mr{joint}}$, evaluated from~\eqref{eq:rho_etc_awgn} and~\eqref{eq:rho_joint_awgn}.}
\label{fig:etc_gap}
\end{figure}

Figure~\ref{fig:etc_gap} illustrates the coefficient comparison for $\sigma_1^2=1$ and $\sigma_2^2\in[1.5,2.5]$. At $(\sigma_1^2,\sigma_2^2)=(1,2)$, the costs in the two achievability bounds are $\rho_{\mr{ETC}}\approx3.362$ and $\rho_{\mr{joint}}\approx0.754$, a difference of about $2.608$.

\section{Conclusion}
\label{sec:conclusion}

This paper couples VLF coding with bandit channel selection over $K$ AWGN channels whose noise variances are unknown. The VLSF-Bandit bound has a baseline of one and a channel-exploration term at second order. The VLF-Bandit bound has no baseline and a smaller magnitude of the total coefficient, consisting of a universal-decoding term and a smaller channel-exploration term. It improves on the ETC bound in every instance with $K\geq2$.

Several problems remain open. The most important is a tight second-order converse. It is unknown whether every sequence of VLF-Bandit codes whose expected number of uses of each suboptimal channel is $o(N^\alpha)$ for every $\alpha>0$ at every admissible problem instance must pay a channel-exploration cost of the Lai--Robbins order, since the change-of-measure arguments behind~\cite{lai1985} constrain bandit policies at deterministic horizons and do not directly apply to codes, whose error criterion involves only the message estimate. A second direction is a joint scheme for channel families whose outputs do not reveal the unknown parameter under the capacity-achieving input, such as the BSC.

\appendices

\renewcommand{\thesubsection}{\thesection.\Roman{subsection}}
\renewcommand{\thesubsectiondis}{\Roman{subsection}.}

\renewcommand{\thesubsubsection}{\thesubsection.\arabic{subsubsection}}
\renewcommand{\thesubsubsectiondis}{\arabic{subsubsection})}

\renewcommand{\theHsubsection}{\thesection.\Roman{subsection}}
\renewcommand{\theHsubsubsection}{\theHsubsection.\arabic{subsubsection}}

\section{Proof of the VLSF-Bandit Lemmas}

\label{app:Gaussian}

\subsection{Proof of \lemref{lem:wrong_anytime_yi}}
\label{app:awgn_wrong_anytime}

Fix $w_0\in[M]$ and $m\neq w_0$, and condition on $W=w_0$. The output density at a variance vector $\bm{s}$ and its average over the product of the priors in~\eqref{eq:Skn_def} are
\begin{align}
 p_{\bm{s},n}(y^n)&\triangleq\prod_{t=1}^n\phi_{1+s_{A_t}}(y_t),
 \label{eq:mixture_output_density}\\
 q_n(y^n)&\triangleq\frac{1}{[\log(4\beta)]^K}
       \int_{[\frac{\sigma_{\min}^2}{2},\,2\sigma_{\max}^2]^K}
          p_{\bm{s},n}(y^n)\prod_{k=1}^K\frac{ds_k}{s_k},
 \label{eq:mixture_output_density_average}
\end{align}
where the selections $A_t$ are determined by the past of the output path. Under the true law, the candidate codeword $\mb{c}(m)$ is independent of the outputs, which have the density $p_{\bm{\sigma}^2,n}$ once the transmitted codeword is integrated out. Let $\mathbb Q$ be the auxiliary law under which the candidate $\mb{c}(m)$ is transmitted and the noise variance of each channel is drawn independently from the prior before transmission; by Gaussian convolution, the outputs have the density $q_n$ under $\mathbb Q$. Write $\mathbb P_n$ and $\mathbb Q_n$ for the joint laws of the candidate prefix $(\mb{c}_1(m),\ldots,\mb{c}_n(m))$ and $Y^n$ under the true law and under $\mathbb Q$. The candidate prefix has the same Gaussian density under both laws. Its remaining density factor is $p_{\bm{\sigma}^2,n}$ under $\mathbb P_n$ and the product of the normalized residual mixtures under $\mathbb Q_n$. Since $q_n$ is the product of the output mixtures, \eqref{eq:Skn_def} gives
\begin{align}
 L_n&\triangleq\frac{p_{\bm{\sigma}^2,n}(Y^n)}{q_n(Y^n)},
 \label{eq:mixture_output_lr}\\
 \frac{d\mathbb P_n}{d\mathbb Q_n}&=L_n \exp\{-S_n(m)\}.
 \label{eq:mixture_joint_lr}
\end{align}
As the likelihood ratio of the two output laws, $\{L_n\}_{n\geq0}$ is a martingale under $\mathbb Q$ in the output filtration, with $L_0=1$.

Let $\tau'$ be the first $n\geq1$ with $S_n(m)\geq u$. Then
\begin{align}
 \Prob{\tau'\leq H}
 &=\sum_{n=1}^H\mathbb E_{\mathbb Q}\!\left[L_n\exp\{-S_n(m)\}\1{\tau'=n}\right]
 \leq \exp\{-u\}\mathbb E_{\mathbb Q}\!\left[\max_{0\leq n\leq H}L_n\right]
 \label{eq:mixture_wrong_maximum}\\
 &\leq\frac{e}{e-1}\left(1+D(\mathbb P_H^Y\|\mathbb Q_H^Y)\right)\exp\{-u\},
 \label{eq:mixture_entropy_maximum}
\end{align}
where $\mathbb P_H^Y$ and $\mathbb Q_H^Y$ are the laws of $Y^H$. The equality in~\eqref{eq:mixture_wrong_maximum} sums~\eqref{eq:mixture_joint_lr} over the disjoint events $\{\tau'=n\}$, each of which is determined by the candidate prefix and $Y^n$, and the inequality uses $S_n(m)\geq u$ on $\{\tau'=n\}$. The maximum depends only on the outputs, so~\eqref{eq:doob_entropy} applies in the output filtration and, with $\mathbb E_{\mathbb Q}[L_H\log L_H]=D(\mathbb P_H^Y\|\mathbb Q_H^Y)$, gives~\eqref{eq:mixture_entropy_maximum}.

It remains to bound this divergence. We restrict the prior to a box around the true variances,
\begin{align}
 r_H\triangleq\frac{\sigma_{\min}^2}{2}\wedge\frac{1+\frac{\sigma_{\min}^2}{2}}{\sqrt H},
 \qquad
 \mc{B}\triangleq\prod_{k=1}^K[\sigma_k^2-r_H,\sigma_k^2+r_H],
 \label{eq:mixture_prior_box}
\end{align}
and let $\pi_{\mc{B}}$ be the product of the priors restricted to $\mc{B}$ and normalized. Each factor of $\mc{B}$ lies inside the enlarged prior interval and has prior mass at least $\frac{r_H}{\sigma_{\max}^2\log(4\beta)}$. Restricting the integral in~\eqref{eq:mixture_output_density_average} to $\mc{B}$, applying Jensen's inequality to $\pi_{\mc{B}}$, and conditioning each log-likelihood increment on the selected channel and the past, under which the output has the law $\mc{N}(0,v_{A_t})$, we obtain
\begin{align}
 D(\mathbb P_H^Y\|\mathbb Q_H^Y)
 &=\E{\log\frac{p_{\bm{\sigma}^2,H}(Y^H)}{q_H(Y^H)}}
 \label{eq:mixture_entropy_expectation}\\
 &\leq K\log\frac{\sigma_{\max}^2\log(4\beta)}{r_H}
       -\E{\log\int\frac{p_{\bm{s},H}(Y^H)}{p_{\bm{\sigma}^2,H}(Y^H)}\,\pi_{\mc{B}}(d\bm{s})}
 \label{eq:mixture_entropy_restriction}\\
 &\leq K\log\frac{\sigma_{\max}^2\log(4\beta)}{r_H}
       +\int\E{\log\frac{p_{\bm{\sigma}^2,H}(Y^H)}{p_{\bm{s},H}(Y^H)}}\,\pi_{\mc{B}}(d\bm{s})
 \label{eq:mixture_entropy_jensen}\\
 &\leq K\log\frac{\sigma_{\max}^2\log(4\beta)}{r_H}
       +\sup_{\bm{s}\in\mc{B}}\sum_{k=1}^K
             \E{N_k(H)}\dG(v_k\|1+s_k)
 \label{eq:mixture_entropy_localization}\\
 &\leq K\log\bigl(2\beta\log(4\beta)\sqrt H\bigr)
       +\frac{Hr_H^2}{4\left(1+\frac{\sigma_{\min}^2}{2}\right)^2}
 \label{eq:mixture_entropy_quadratic}\\
 &\leq\frac K2\log H+K\log\bigl(2\beta\log(4\beta)\bigr)+\frac14.
 \label{eq:mixture_entropy_bound}
\end{align}
In~\eqref{eq:mixture_entropy_quadratic}, the first term comes from $r_H\geq\frac{\sigma_{\min}^2}{2\sqrt H}$, and the second from the fact that the counts add up to $H$ and from the bound $\dG(v_k\|1+s_k)\leq\frac{(v_k-1-s_k)^2}{4(v_k\wedge(1+s_k))^2}$. For $\bm{s}\in\mc{B}$, these two variances are both at least $1+\frac{\sigma_{\min}^2}{2}$ and differ by at most $r_H$.

Since $\frac54+K\log(2\beta\log(4\beta))\geq0$, substituting~\eqref{eq:mixture_entropy_bound} into~\eqref{eq:mixture_entropy_maximum} and using the definition of $c_0$ give
\begin{align}
 \Prob{\tau'\leq H}
 &\leq c_0(1+\log H)\exp\{-u\}.
 \label{eq:mixture_wrong_single}
\end{align}
As $\{\tau'\leq H\}=\{\max_{1\leq n\leq H}S_n(m)\geq u\}$, a union bound over $m\neq w_0$ and an average over $W$ prove~\eqref{eq:wrong_anytime_yi}.

\subsection{Proof of \eqref{eq:mixture_stopped_loss}}
\label{app:awgn_plugin_loss}

We use the reward-stack representation in~\cite[Sec.~4.6]{lattimore2020}. For each channel $k$, let $\breve Y_{k,1},\breve Y_{k,2},\ldots$ be i.i.d.\ $\mc{N}(0,v_k)$, independently over $k$, and let $\breve Y_{k,j}$ be the output of the $j$-th use of channel $k$. Since the selection rule uses only past outputs, and the next output has the law $\mc{N}(0,v_k)$ conditionally on the past and on the selection of channel $k$, the selections and outputs have the same law as in the continued process. Define
\begin{align}
 \chi_{k,j}^2&\triangleq\sum_{i=1}^j\frac{\breve Y_{k,i}^2}{v_k}.
 \label{eq:mixture_output_square_stack}
\end{align}
For the first $j$ uses of channel $k$, let $\Lambda_{k,j}^{\mr{num}}$ be the logarithm of the likelihood of the residuals at the true noise variance divided by the residual mixture in~\eqref{eq:Skn_def}, and let $\Lambda_{k,j}^{\mr{den}}$ be the same quantity for the outputs and the output mixture, both mixtures with their normalizing factors. These redundancies vanish at $j=0$. Grouping the factors of $I_n$ in~\eqref{eq:In_def} and of $S_n(W)$ by channel gives, for every $n$,
\begin{align}
 I_n-S_n(W)
 &=\sum_{k=1}^K\left(\Lambda_{k,N_k(n)}^{\mr{num}}-\Lambda_{k,N_k(n)}^{\mr{den}}\right).
 \label{eq:mixture_true_redundancy_identity}
\end{align}
We show that the first redundancy is at most $\frac12\log j$ plus a constant and that the second is at least $\frac12\log j$ minus a constant and a chi-square fluctuation, so that the leading logarithms cancel in~\eqref{eq:mixture_true_redundancy_identity}.

For the residual mixture, we write the noise variance through its precision $\eta=\frac1s$, so that $\frac{ds}{s}=\frac{d\eta}{\eta}$, the enlarged interval becomes $[\frac{\eta_{\min}}{2},2\eta_{\max}]$, and the true precision is $\eta_k=\frac{1}{\sigma_k^2}$. Fix $n$ with $j=N_k(n)\geq1$, so that the first $j$ residuals of channel $k$ have the squared sum $\SE_{k,W}(n)$. We keep only the precisions in the segment $|\eta-\eta_k|\leq\frac{\eta_{\min}}{2\sqrt j}$, which lies in the enlarged interval because $\eta_{\min}\leq\eta_k\leq\eta_{\max}$. On this segment, $\frac1\eta\geq\frac{\sigma_{\min}^2}{2}$ and the second derivative of $\frac j2\log\eta$ is at least $-2j\sigma_{\max}^4$. Taylor's theorem gives
\begin{align}
 \exp\{-\Lambda_{k,j}^{\mr{num}}\}
 &=\frac{1}{\log(4\beta)}\int_{\frac{\eta_{\min}}{2}}^{2\eta_{\max}}
   \left(\frac{\eta}{\eta_k}\right)^{\frac j2}
   \exp\left\{-\frac{(\eta-\eta_k)\SE_{k,W}(n)}{2}\right\}\frac{d\eta}{\eta}
 \label{eq:mixture_numerator_precision}\\
 &\geq\frac{\sigma_{\min}^2}{2\log(4\beta)}
   \int_{\eta_k-\frac{\eta_{\min}}{2\sqrt j}}^{\eta_k+\frac{\eta_{\min}}{2\sqrt j}}
   \exp\left\{\frac{(\eta-\eta_k)\bigl(j\sigma_k^2-\SE_{k,W}(n)\bigr)}{2}
   -j\sigma_{\max}^4(\eta-\eta_k)^2\right\}d\eta
 \label{eq:mixture_numerator_segment}\\
 &\geq\frac{\sigma_{\min}^2e^{-\frac14}}{\log(4\beta)}
   \int_0^{\frac{\eta_{\min}}{2\sqrt j}}
   \cosh\left(\frac{x\bigl(j\sigma_k^2-\SE_{k,W}(n)\bigr)}{2}\right)dx
 \label{eq:mixture_numerator_fold}\\
 &\geq\frac{\sigma_{\min}^2\eta_{\min}e^{-\frac14}}{2\log(4\beta)\sqrt j}
 =\frac{e^{-\frac14}}{2\beta\log(4\beta)\sqrt j}.
 \label{eq:mixture_numerator_integral}
\end{align}
Pairing the points $\eta_k\pm x$ gives the hyperbolic cosine in~\eqref{eq:mixture_numerator_fold}, which is at least one. Since $j\sigma_{\max}^4(\eta-\eta_k)^2\leq\frac14$ on the segment, the last two rows follow. Taking logarithms gives, for every residual sequence,
\begin{align}
 \Lambda_{k,j}^{\mr{num}}
 &\leq\frac12\log j+\frac14+\log\bigl(2\beta\log(4\beta)\bigr),
 \qquad j\geq1.
 \label{eq:mixture_numerator_bound}
\end{align}

The output mixture is bounded from above by a similar argument, in the natural parameter $\theta=\frac{1}{2(1+s)}$ of the output family. Then $\frac{ds}{s}=\frac{2(1+s)^2}{s}\,d\theta$, the enlarged interval becomes $[\frac{1}{2+4\sigma_{\max}^2},\frac{1}{2+\sigma_{\min}^2}]$, and the true output variance corresponds to $\theta_k=\frac{1}{2v_k}$. Since $\frac{(1+s)^2}{s}$ is convex, the prior density in $\theta$ is largest at an endpoint of the enlarged interval, and both endpoints give
\begin{align}
 \frac{2(1+s)^2}{s\log(4\beta)}
 &\leq\Bigl(1+\frac{\sigma_{\min}^2}{2}\Bigr)\frac{8\beta+4\eta_{\max}}{\log(4\beta)},
 \qquad s\in\Bigl[\frac{\sigma_{\min}^2}{2},2\sigma_{\max}^2\Bigr].
 \label{eq:mixture_output_prior_bound}
\end{align}
In the reward-stack representation, the first $j$ outputs of channel $k$ have the squared sum $v_k\chi_{k,j}^2$, and, for $j\geq1$,
\begin{align}
 \exp\{-\Lambda_{k,j}^{\mr{den}}\}
 &=\frac{1}{\log(4\beta)}\int_{\frac{1}{2+4\sigma_{\max}^2}}^{\frac{1}{2+\sigma_{\min}^2}}
   \left(\frac{\theta}{\theta_k}\right)^{\frac j2}
   \exp\left\{-(\theta-\theta_k)v_k\chi_{k,j}^2\right\}\frac{2(1+s)^2}{s}\,d\theta
 \label{eq:mixture_denominator_natural}\\
 &\leq\Bigl(1+\frac{\sigma_{\min}^2}{2}\Bigr)\frac{8\beta+4\eta_{\max}}{\log(4\beta)}
   \int_{-\infty}^{\infty}\exp\left\{(\theta-\theta_k)v_k\bigl(j-\chi_{k,j}^2\bigr)
   -j\Bigl(1+\frac{\sigma_{\min}^2}{2}\Bigr)^2(\theta-\theta_k)^2\right\}d\theta
 \label{eq:mixture_denominator_gaussian}\\
 &=\frac{(8\beta+4\eta_{\max})\sqrt\pi}{\log(4\beta)\sqrt j}
   \exp\left\{\frac{v_k^2\bigl(\chi_{k,j}^2-j\bigr)^2}{4j\left(1+\frac{\sigma_{\min}^2}{2}\right)^2}\right\}
 \label{eq:mixture_denominator_integral}\\
 &\leq\frac{4\sqrt\pi(2\beta+\eta_{\max})}{\log(4\beta)\sqrt j}
   \exp\left\{\frac{\beta^2\bigl(\chi_{k,j}^2-j\bigr)^2}{j}\right\},
 \label{eq:mixture_denominator_simple}\\
 \Lambda_{k,j}^{\mr{den}}
 &\geq\frac12\log j
 -\log\frac{4\sqrt\pi(2\beta+\eta_{\max})}{\log(4\beta)}
 -\frac{\beta^2\left(\chi_{k,j}^2-j\right)^2}{j}.
 \label{eq:mixture_denominator_bound}
\end{align}
Inequality~\eqref{eq:mixture_denominator_gaussian} uses~\eqref{eq:mixture_output_prior_bound} and Taylor's theorem with second derivative at most $-2j(1+\frac{\sigma_{\min}^2}{2})^2$, then extends the integral to the real line. The remaining rows use $v_k\leq(2+\sigma_{\min}^2)\beta$ and take logarithms.

Subtracting~\eqref{eq:mixture_denominator_bound} from~\eqref{eq:mixture_numerator_bound} cancels the leading logarithms and gives
\begin{align}
 \Lambda_{k,j}^{\mr{num}}-\Lambda_{k,j}^{\mr{den}}
 &\leq\frac14+\log\bigl(8\sqrt\pi\beta(2\beta+\eta_{\max})\bigr)
       +\frac{\beta^2\left(\chi_{k,j}^2-j\right)^2}{j}
 \label{eq:mixture_loss_prefix}\\
 &\leq4(c_1+\beta^2)+\frac{\beta^2\left(\chi_{k,j}^2-j\right)^2}{j},
 \label{eq:mixture_loss_prefix_simple}
\end{align}
where~\eqref{eq:mixture_loss_prefix_simple} follows from the definition of $c_1$ in \lemref{lem:vlsf_plugin_loss}, $\beta\geq1$, and $\frac14+\frac12\log\pi<1$.

Since the redundancies of an unused channel vanish, \eqref{eq:mixture_true_redundancy_identity} and~\eqref{eq:mixture_loss_prefix_simple} give, for every integer $H\geq1$ and on every path,
\begin{align}
 \max_{0\leq n\leq H}(I_n-S_n(W))^+
 &\leq4K(c_1+\beta^2)+\beta^2\sum_{k=1}^K\max_{1\leq j\leq H}\frac{\left(\chi_{k,j}^2-j\right)^2}{j}.
 \label{eq:mixture_loss_pathwise}
\end{align}
The right-hand side depends on the outputs alone, whose law is the same in the reward-stack representation. There, $\chi_{k,j}^2-j$ is a sum of $j$ i.i.d.\ centered variables, so $|\chi_{k,j}^2-j|$ is a non-negative submartingale in $j$, whose fourth moments are those of a centered chi-square variable,
\begin{align}
 \E{\left(\chi_{k,n}^2-n\right)^4}&=12n^2+48n\leq60n^2,\qquad n\geq1.
 \label{eq:mixture_stack_fourth}
\end{align}
We apply Minkowski's inequality to~\eqref{eq:mixture_loss_pathwise}, split the indices into the dyadic blocks $2^\ell\leq j<2^{\ell+1}$, and apply~\eqref{eq:doob_fourth} and~\eqref{eq:mixture_stack_fourth} on each block. This gives
\begin{align}
 \left(\E{\left[\max_{0\leq n\leq H}(I_n-S_n(W))^+\right]^2}\right)^{\frac12}
 &\leq4K(c_1+\beta^2)
 +\beta^2\sum_{k=1}^K\left(\E{\max_{1\leq j\leq H}\frac{\left(\chi_{k,j}^2-j\right)^4}{j^2}}\right)^{\frac12}
 \label{eq:mixture_loss_minkowski}\\
 &\leq4K(c_1+\beta^2)
 +\beta^2\sum_{k=1}^K\left(\sum_{\ell=0}^{\lfloor\log_2H\rfloor}2^{-2\ell}
       \E{\max_{1\leq j\leq2^{\ell+1}}\left(\chi_{k,j}^2-j\right)^4}\right)^{\frac12}
 \label{eq:mixture_stack_dyadic}\\
 &\leq4K(c_1+\beta^2)+28K\beta^2\sqrt{1+\log_2H}
 \label{eq:mixture_stack_maximum}\\
 &\leq c_2\sqrt{1+\log_2H},
 \label{eq:mixture_loss_max_second}
\end{align}
since $240(\frac43)^4<28^2$, $\sqrt{1+\log_2H}\geq1$, and $c_2=4K(c_1+8\beta^2)$. Finally, $I_{\tau_W\wedge H}-S_{\tau_W\wedge H}(W)$ is at most the maximum on the left-hand side of~\eqref{eq:mixture_loss_minkowski}, so the Cauchy--Schwarz inequality gives
\begin{align}
 \E{I_{\tau_W\wedge H}-S_{\tau_W\wedge H}(W)}
 &\leq c_2\sqrt{1+\log_2H},
 \label{eq:mixture_loss_stopped}
\end{align}
which proves~\eqref{eq:mixture_stopped_loss}.

\subsection{Proof of~\eqref{eq:mixture_overshoot}}
\label{app:awgn_bov}

We verify the mean-excess condition of \lemref{lem:overshoot}, with $c=c_1$, for the score increments
\begin{align}
 \xi_t&\triangleq S_t(W)-S_{t-1}(W).
 \label{eq:vlsf_xi_def}
\end{align}
Condition on $\mc{F}_{t-1}$ and $A_t=k$. Then $\xi_t$ is the logarithm of the ratio of two predictive densities, that of the residual $Y_t-X_t$ under the residual mixture and that of the output $Y_t$ under the output mixture. These are mixtures of $\mc{N}(0,s)$ and $\mc{N}(0,1+s)$ densities, respectively, over $s\in[\frac{\sigma_{\min}^2}{2},2\sigma_{\max}^2]$, with mixing weights fixed by $\mc{F}_{t-1}$. Write the independent standard Gaussian variables $X_t$ and $\frac{Y_t-X_t}{\sigma_k}$ in polar coordinates as $\sqrt{2\rho}\cos\varphi$ and $\sqrt{2\rho}\sin\varphi$. In these coordinates, $\rho$ is exponential with mean one and independent of $\varphi$, and the residual and the output are $\sigma_k\sqrt{2\rho}\sin\varphi$ and $\sqrt{2\rho}(\cos\varphi+\sigma_k\sin\varphi)$.

For a fixed angle, regard the increment as a function $\xi_t(\rho)$ of $\rho$. Each Gaussian log-density in the two mixtures is affine in $\rho$, with slope $-\frac{\sigma_k^2\sin^2\varphi}{s}$ for the residual and $-\frac{(\cos\varphi+\sigma_k\sin\varphi)^2}{1+s}$ for the output, so $\xi_t'(\rho)$ is a weighted average of the first slopes minus a weighted average of the second. On the prior interval, $\sigma_k^2\leq2\beta s$ and, by the Cauchy--Schwarz inequality, $(\cos\varphi+\sigma_k\sin\varphi)^2\leq v_k\leq2\beta(1+s)$, so all these slopes lie in $[-2\beta,0]$. At $\rho=0$, the two predictive densities are averages of $(2\pi s)^{-\frac12}$ and of $(2\pi(1+s))^{-\frac12}$ over the prior interval, and the extreme values of these two functions bound their ratio. Therefore,
\begin{align}
 |\xi_t'(\rho)|&\leq2\beta,\qquad \rho\geq0,
 \label{eq:mixture_radial_derivative}\\
 |\xi_t(0)|&\leq\frac12\log(4\beta+2\eta_{\max})=c_1-2\beta.
 \label{eq:mixture_radial_origin}
\end{align}

Fix a non-negative $\mc{F}_{t-1}$-measurable threshold $u$. The set of values $r\geq0$ with $\xi_t(r)>u$ is open in $[0,\infty)$ and is a disjoint union of intervals with endpoints $r_-,r_+$. On such an interval, $\xi_t(r)-u\leq2\beta(r-r_-)+c_1-2\beta$, by~\eqref{eq:mixture_radial_derivative} if $r_->0$, since then $\xi_t(r_-)=u$, and by~\eqref{eq:mixture_radial_derivative} and~\eqref{eq:mixture_radial_origin} if $r_-=0$. Integration by parts applied to this bound gives
\begin{align}
 \int_{r_-}^{r_+}\bigl(\xi_t(r)-u\bigr)\exp\{-r\}\,dr
 &\leq c_1\int_{r_-}^{r_+}\exp\{-r\}\,dr,
 \label{eq:mixture_radial_integral}
\end{align}
since the upper boundary term is non-positive and vanishes when $r_+=\infty$. Summing over the intervals and averaging over the angle, we obtain
\begin{align}
 \E{(\xi_t-u)^+\mid\mc{F}_{t-1}}
 &\leq c_1\Prob{\xi_t>u\mid\mc{F}_{t-1}}.
 \label{eq:mixture_excess}
\end{align}
The increments are also integrable, since $|\xi_t(\rho)|\leq|\xi_t(0)|+2\beta\rho$ and $\E{\rho}=1$ give $\E{|\xi_t|\mid\mc{F}_{t-1}}\leq c_1$. Since $S_0(W)=0<\gamma$, \lemref{lem:overshoot} with $\omega=\gamma$ now gives~\eqref{eq:mixture_overshoot}.

\subsection{Proof of \lemref{lem:awgn_op_exp_family}}

\label{app:awgn_sqres_exp_family}

Conditioned on selecting channel $k$, the output $Y$ has distribution $\mc{N}(0,v_k)$ with $v_k=1+\sigma_k^2$ under the random-coding ensemble. A change of variables gives the density of the reward $-Y^2$, and, with the natural parameter $\theta=\frac{1}{2v_k}>0$, its density with respect to a dominating measure $\mu_{\mr{sq}}$ takes the canonical form,
\begin{align}
    \frac{dP_{-Y^2}}{ds}(s)
    &=
    \frac{1}{\sqrt{2\pi v_k(-s)}}
    \exp\!\left\{\frac{s}{2v_k}\right\}
    \1{s<0},
    \label{eq:app_op_density}\\
    \mu_{\mr{sq}}(dr)
    &\triangleq
    \frac{\1{r<0}}{\sqrt{2\pi(-r)}}\,dr,
    \label{eq:app_op_base}\\
    \frac{d\nu_\theta}{d\mu_{\mr{sq}}}(s)
    &=\exp\{\theta s-\kappa(\theta)\},
    \label{eq:app_op_density2}\\
    \kappa(\theta)&\triangleq-\frac{1}{2}\log(2\theta).
    \label{eq:app_op_cumulant}
\end{align}
Thus, $\{\nu_\theta\}_{\theta>0}$ is a one-parameter canonical exponential family. Its mean, its variance, and the divergence between two of its members follow from the cumulant~\eqref{eq:app_op_cumulant},
\begin{align}
    \kappa'(\theta)&=-\frac{1}{2\theta}=-v_k,
    \label{eq:app_op_mean} \\
    \kappa''(\theta)&=\frac{1}{2\theta^2}=2v_k^2
    \leq 2(1+\sigma_{\max}^2)^2=V_{\mr{sq}},
    \label{eq:app_op_var}\\
 D\left(\nu_{\frac{1}{2u}}\middle\|\nu_{\frac{1}{2v}}\right)
 &=\kappa\left(\frac{1}{2v}\right)-\kappa\left(\frac{1}{2u}\right)
 -\kappa'\left(\frac{1}{2u}\right)\left(\frac{1}{2v}-\frac{1}{2u}\right)
 =\frac{1}{2}\left(\frac{u}{v}-1-\log\frac{u}{v}\right)=\dG(u\|v).
 \label{eq:app_op_divergence}
\end{align}
This proves the lemma.

\subsection{Proof of \lemref{lem:vlsf_residual_tail}}
\label{app:awgn_score_tail}

Define the centered information density
\begin{align}
 \widetilde I_n&\triangleq I_n-\sum_{t=1}^n C^{(A_t)},\qquad n\geq0.
 \label{eq:mixture_centered_information}
\end{align}
By~\eqref{eq:In_def}, on $\{A_t=k\}$ and conditioned on $\mc{F}_{t-1}$, the increment $\widetilde I_t-\widetilde I_{t-1}$ is a quadratic form in the independent standard Gaussian variables $X_t$ and $\frac{Y_t-X_t}{\sigma_k}$, with eigenvalues $\pm\frac{1}{2\sqrt{v_k}}$. An orthogonal rotation gives
\begin{align}
 \widetilde I_t-\widetilde I_{t-1}
 &\overset{\mr{d}}=\frac{G_1G_2}{\sqrt{v_{A_t}}},
 \label{eq:mixture_information_product_normal}
\end{align}
where $G_1$ and $G_2$ are independent standard Gaussian variables under this conditional law. The conditional second and fourth moments of the increment are thus at most one and nine, respectively, and its conditional odd moments vanish. Orthogonality gives $\E{\widetilde I_{t-1}^2}\leq t-1$. Expanding the fourth power conditionally and summing the resulting bounds gives
\begin{align}
 \E{\widetilde I_n^4}
 &\leq\sum_{t=1}^n\bigl(6(t-1)+9\bigr)
 =3n^2+6n\leq9n^2,\qquad n\geq1.
 \label{eq:mixture_information_fourth}
\end{align}

Let $n\geq\frac{4\gamma}{\CG(\eta_{\min})}$ be an integer. As every selected-channel capacity is at least $\CG(\eta_{\min})$, the score satisfies $S_n(W)\geq\widetilde I_n+n\CG(\eta_{\min})-(I_n-S_n(W))^+$, and therefore
\begin{align}
 \{\tau_W>n\}
 &\subseteq\left\{\widetilde I_n<-\frac{n\CG(\eta_{\min})}{2}\right\}
 \cup\left\{(I_n-S_n(W))^+>\frac{n\CG(\eta_{\min})}{4}\right\}.
 \label{eq:mixture_tail_split}
\end{align}
The union bound and Markov's inequality for the fourth and the second powers, followed by~\eqref{eq:mixture_information_fourth} and~\eqref{eq:mixture_loss_max_second}, give
\begin{align}
 \Prob{\tau_W>n}
 &\leq\frac{16\E{\widetilde I_n^4}}{\CG(\eta_{\min})^4n^4}
       +\frac{16\E{\left[(I_n-S_n(W))^+\right]^2}}{\CG(\eta_{\min})^2n^2}
 \label{eq:mixture_tail_markov}\\
 &\leq\frac{144}{\CG(\eta_{\min})^4n^2}
       +\frac{16c_2^2(1+\log_2n)}{\CG(\eta_{\min})^2n^2}.
 \label{eq:mixture_true_tail_finite}
\end{align}
This bound is summable over $n$, so $\E{\tau_W}<\infty$. For every integer $H\geq\frac{4\gamma}{\CG(\eta_{\min})}$, it also gives
\begin{align}
 \E{(\tau_W-H)^+}
 &=\sum_{i=H}^{\infty}\Prob{\tau_W>i}
 \leq\frac{288}{\CG(\eta_{\min})^4H}
       +\frac{64c_2^2(1+\log_2H)}{\CG(\eta_{\min})^2H},
 \label{eq:mixture_true_tail_mean_finite}
\end{align}
since $\sum_{i=H}^{\infty}\frac{1}{i^2}\leq\frac{2}{H}$ and, by comparison with an integral, $\sum_{i=H}^{\infty}\frac{1+\log_2i}{i^2}\leq\frac{4(1+\log_2H)}{H}$.

Under the condition on $\gamma$ in \lemref{lem:vlsf_residual_tail}, the design horizon satisfies $T\geq\gamma\log^2\gamma\geq e\vee\frac{4\gamma}{\CG(\eta_{\min})}$, and the right-hand sides of~\eqref{eq:mixture_true_tail_finite} and~\eqref{eq:mixture_true_tail_mean_finite} decrease in $n$ and $H$ beyond $e$. Taking $n=H=T$, replacing $T$ by $\gamma\log^2\gamma$, and using $1+\log_2(\gamma\log^2\gamma)\leq6\log\gamma$, we arrive at
\begin{align}
 \Prob{\tau_W>T}
 &\leq\frac{144}{\CG(\eta_{\min})^4\gamma^2\log^4\gamma}
       +\frac{96c_2^2}{\CG(\eta_{\min})^2\gamma^2\log^3\gamma}
 \leq\frac{c_3}{\gamma^2\log^3\gamma},
 \label{eq:mixture_gamma_probability_proof}\\
 \E{(\tau_W-T)^+}
 &\leq\frac{288}{\CG(\eta_{\min})^4\gamma\log^2\gamma}
       +\frac{384c_2^2}{\CG(\eta_{\min})^2\gamma\log\gamma}
 \leq\frac{c_3}{\gamma\log\gamma},
 \label{eq:mixture_gamma_residual_proof}
\end{align}
where both final inequalities use $\log\gamma\geq1$ and the definition of $c_3$. Together with the finiteness of $\E{\tau_W}$, this is \lemref{lem:vlsf_residual_tail}.

\subsection{Proof of \lemref{lem:awgn_klucbpp_pull}}

\label{app:awgn_klucbpp_pull_bound}

We apply \lemref{lem:klucbpp_fixed_horizon} at the horizon~$T$, in the reward-stack representation in \appref{app:awgn_plugin_loss}. By \lemref{lem:awgn_op_exp_family}, the rewards $-\breve Y_{k,s}^2$ are negative almost surely and have means $\mu_k=-v_k$ and variance bounded by $V_{\mr{sq}}$ on $[-(1+\sigma_{\max}^2),-(1+\sigma_{\min}^2)]$, an interval inside the mean domain $(-\infty,0)$. The projected means and restricted indices are those in \lemref{lem:klucbpp_fixed_horizon}. Let $\delta=(\log T)^{-\frac{1}{4}}$. For all sufficiently large $\gamma$, this $\delta$ satisfies~\eqref{eq:generic_klucbpp_delta} with $V=V_{\mr{sq}}$. Hence, \lemref{lem:klucbpp_fixed_horizon} and~\eqref{eq:vlsf_reward_kl_identification} give
\begin{align}
 \E{N_k(T)}&\leq\frac{G_K(T)}{\dG(v_k-\delta\|v_\star+\delta)}+2(16e^2+2)V_{\mr{sq}}K\sqrt{\log T}+1
 \label{eq:app_pull_fixed}\\
 &\leq\frac{\log T+\log(1+\log^2T)}{D_{k,k^\star}-O(\delta)}+O\bigl(\sqrt{\log T}\bigr)
 \label{eq:app_pull_substituted}\\
 &=\frac{\log\gamma}{D_{k,k^\star}}+O\bigl(\delta\log\gamma+\log\log\gamma+\sqrt{\log\gamma}\bigr)
 \label{eq:app_pull_expanded}\\
 &\leq\frac{\log\gamma}{D_{k,k^\star}}+O\bigl((\log\gamma)^{\frac{3}{4}}\bigr),
 \label{eq:app_pull_asymp}
\end{align}
where~\eqref{eq:app_pull_substituted} uses $G_K(T)\leq\log T+\log(1+\log^2T)$, which follows from~\eqref{eq:budget_def}, and $\dG(v_k-\delta\|v_\star+\delta)\geq D_{k,k^\star}-O(\delta)$, a consequence of the continuous differentiability of $\dG$ and of $\dG(v_k\|v_\star)=D_{k,k^\star}>0$; \eqref{eq:app_pull_expanded} follows from~\eqref{eq:mixture_horizon}, which gives $\log T=\log\gamma+O(\log\log\gamma)$. The last row is~\eqref{eq:awgn_klucbpp_pull_asymp}.

\section{Proof of the VLF-Bandit Lemmas}

\label{app:vlf_lemmas}

All probabilities and expectations retain the transmission-branch convention in~\secref{sec:vlf_proof}. Throughout this appendix, $\chi_n^2$ denotes a variable with the chi-square distribution with $n$ degrees of freedom.

\subsection{Proof of \lemref{lem:vlf_packet_energy}}

\label{app:vlf_packet_energy}

Fix a transmitted packet $j$, condition on $\mc{G}_{j,0}$, and write $s=\sigma_{J_j}^2$; all expectations and probabilities below use this conditional law. By \lemref{lem:vlf_packet_gaussian}, $W_j$ is uniform on $[M_j]$, so the initial symbol~\eqref{eq:vlf_packet_initial} has energy
\begin{align}
 \E{X_{j,0}^2}&=\frac{12P_{j,0}}{M_j^2}\cdot\frac{M_j^2-1}{12}\leq P_{j,0}.
 \label{eq:vlf_initial_energy}
\end{align}
For $n\geq1$, the estimate $\nn{\eta}_{j,n-1}$ and the indicator $\1{\tau^{(j)}\geq n}$ are $\mc{G}_{j,n-1}$-measurable, and $E_{j,n-1}$ has conditional second moment $s$ given $\mc{G}_{j,n-1}$ by \lemref{lem:vlf_packet_gaussian}. Hence, by~\eqref{eq:vlf_packet_input},
\begin{align}
 \E{X_{j,n}^2\1{\tau^{(j)}\geq n}}
 &=sP_{j,n}\E{\nn{\eta}_{j,n-1}\1{\tau^{(j)}\geq n}},\qquad n\geq1,
 \label{eq:vlf_refinement_energy_identity}\\
 &\leq\frac{1}{2}P_{j,n}\Prob{\tau^{(j)}\geq n},\qquad 1\leq n\leq9,
 \label{eq:vlf_reserve_refinements}
\end{align}
where the inequality holds because the first nine refinements use the estimate $\frac{\eta_{\min}}{2}$, which is at most $\frac{1}{2s}$ since $s\leq\sigma_{\max}^2$. These refinements spend at most half of their nominal energy, and the other half is a reserve.

For $n\geq10$, refinement $n$ uses the estimate $\nn{\eta}_{j,m}$ with $m=n-1\geq9$, which by~\eqref{eq:vlf_packet_precision} is the projection onto $\left[\frac{\eta_{\min}}{2},2\eta_{\max}\right]$ of the unprojected estimate $\eta_{j,m}^{\circ}=\frac{m-2}{\sum_{i=1}^{m}\widetilde Z_{j,i}^2}$. By \lemref{lem:vlf_packet_gaussian}, $\frac{1}{s}\sum_{i=1}^{m}\widetilde Z_{j,i}^2$ has the law of $\chi_m^2$. Projection at the upper end decreases the estimate, whereas projection at the lower end increases it by at most $\frac{\eta_{\min}}{2}$ and happens only if $\eta_{j,m}^{\circ}<\frac{\eta_{\min}}{2}$, so that
\begin{align}
 s\E{\nn{\eta}_{j,m}}
 &\leq s\E{\eta_{j,m}^{\circ}}+\frac{s\eta_{\min}}{2}\Prob{\eta_{j,m}^{\circ}<\frac{\eta_{\min}}{2}}
 \label{eq:vlf_clipping_split}\\
 &\leq1+\frac{1}{2}\Prob{\chi_m^2>2\left(m-2\right)}
 \label{eq:vlf_clipping_chi}\\
 &\leq1+\frac{1}{2}\exp\left\{1-\frac{m}{2}\left(1-\log2\right)\right\},
 \label{eq:vlf_precision_clipping_bias}
\end{align}
where~\eqref{eq:vlf_clipping_chi} uses $\E{\left(\chi_m^2\right)^{-1}}=\frac{1}{m-2}$ and $s\eta_{\min}\leq1$, which also shows that $\eta_{j,m}^{\circ}<\frac{\eta_{\min}}{2}$ forces $\frac{1}{s}\sum_{i=1}^{m}\widetilde Z_{j,i}^2>2(m-2)$. Markov's inequality, applied to $\exp\{\frac{\chi_m^2}{4}\}$, whose mean is $2^{\frac{m}{2}}$, gives~\eqref{eq:vlf_precision_clipping_bias}.

By~\eqref{eq:vlf_packet_precision} and~\eqref{eq:vlf_packet_contraction}, the quantities $\nn{\eta}_{j,n-1}$ and $F_{j,n-1}$ are non-increasing in the squared normalized outputs $\widetilde Z_{j,1}^2,\allowbreak\ldots,\allowbreak\widetilde Z_{j,n-1}^2$, which are independent by \lemref{lem:vlf_packet_gaussian}. Hence, $\1{\tau^{(j)}\geq n}=\1{F_{j,n-1}<\gamma_j}$ is non-decreasing, and association~\cite[Th.~2.1]{esaryProschanWalkup1967} gives non-positive covariance with $\nn{\eta}_{j,n-1}$, so
\begin{align}
 \E{\nn{\eta}_{j,n-1}\1{\tau^{(j)}\geq n}}
 &\leq\E{\nn{\eta}_{j,n-1}}\Prob{\tau^{(j)}\geq n},\qquad n\geq10.
 \label{eq:vlf_refinement_association}
\end{align}
Summing~\eqref{eq:vlf_refinement_energy_identity} over $n\geq1$, and bounding its terms by~\eqref{eq:vlf_reserve_refinements} for $n\leq9$ and by~\eqref{eq:vlf_refinement_association} and~\eqref{eq:vlf_precision_clipping_bias} for $n\geq10$, we get
\begin{align}
 \E{\sum_{n=1}^{\tau^{(j)}}X_{j,n}^2}
 &=\sum_{n=1}^{\infty}sP_{j,n}\E{\nn{\eta}_{j,n-1}\1{\tau^{(j)}\geq n}}
 \label{eq:vlf_refinement_energy_sum}\\
 &\leq\E{\sum_{n=1}^{\tau^{(j)}}P_{j,n}}
 -\frac{1}{2}\sum_{n=1}^{9}P_{j,n}\Prob{\tau^{(j)}\geq n}
 \notag\\
 &\qquad+\frac{1}{2}\sum_{n=10}^{\infty}P_{j,n}\Prob{\tau^{(j)}\geq n}\exp\left\{1-\frac{n-1}{2}\left(1-\log2\right)\right\}
 \label{eq:vlf_refinement_energy_reserve}\\
 &\leq\E{\sum_{n=1}^{\tau^{(j)}}P_{j,n}}
 -\left(\frac{9}{2}-\frac{\exp\left\{1-\frac{9}{2}\left(1-\log2\right)\right\}}{2\left(1-\exp\left\{-\frac{1}{2}\left(1-\log2\right)\right\}\right)}\right)P_{j,10}\Prob{\tau^{(j)}\geq10}
 \label{eq:vlf_reserve_comparison}\\
 &\leq\E{\sum_{n=1}^{\tau^{(j)}}P_{j,n}}.
 \label{eq:vlf_reserve_covers}
\end{align}
Since $P_{j,n}$ and $\Prob{\tau^{(j)}\geq n}$ are non-increasing in $n$, the subtracted sum in~\eqref{eq:vlf_refinement_energy_reserve} is at least $\frac{9}{2}P_{j,10}\Prob{\tau^{(j)}\geq10}$, whereas the last sum is at most $P_{j,10}\Prob{\tau^{(j)}\geq10}$ times a geometric series; this gives~\eqref{eq:vlf_reserve_comparison}. The coefficient in~\eqref{eq:vlf_reserve_comparison} is positive, giving~\eqref{eq:vlf_reserve_covers}. The reserve of the first nine refinements thus covers the bias of the later estimates, and adding~\eqref{eq:vlf_initial_energy} proves~\eqref{eq:vlf_packet_energy}.

\subsection{Proof of~\eqref{eq:vlf_duration_coefficient} and~\eqref{eq:vlf_stopped_precision_bias}}

\label{app:vlf_packet_duration}

Let $j$ be a transmitted packet that satisfies~\eqref{eq:vlf_duration_power_range}. We condition on $\mc{G}_{j,0}$ and write $s=\sigma_{J_j}^2$; by \lemref{lem:vlf_packet_gaussian} and the continued recursion, the normalized outputs are then i.i.d.\ $\mc{N}(0,s)$ at every deterministic index. The constants below depend only on the known variance interval.

\subsubsection{Moments of the precision estimate}

For $m\geq9$, the unprojected estimate $\eta_{j,m}^{\circ}$ has the law of $\frac{m-2}{s\chi_m^2}$, and the inverse moments of $\chi_m^2$ give
\begin{align}
 \E{\left(\eta_{j,m}^{\circ}-\frac{1}{s}\right)^2}
 &=\frac{2}{s^2\left(m-4\right)},\label{eq:vlf_duration_raw_second_moment}\\
 \E{\left(\eta_{j,m}^{\circ}-\frac{1}{s}\right)^4}
 &=\frac{12\left(m+10\right)}{s^4\left(m-4\right)\left(m-6\right)\left(m-8\right)}=O\left(m^{-2}\right).
 \label{eq:vlf_duration_raw_moments}
\end{align}
The projection interval $\left[\frac{\eta_{\min}}{2},2\eta_{\max}\right]$ contains $\frac{1}{s}$, so projection cannot increase the distance to $\frac{1}{s}$. It is active only if $\chi_m^2<\frac{m}{2}$ or $\chi_m^2>2(m-2)\geq\frac{3m}{2}$, an event whose probability is at most $2\exp\left\{-\frac{m}{24}\right\}$ by the Chernoff bound. The Cauchy--Schwarz inequality, applied on this event with~\eqref{eq:vlf_duration_raw_second_moment} and~\eqref{eq:vlf_duration_raw_moments}, gives the first two of the bounds
\begin{align}
 \left|\E{\nn{\eta}_{j,m}}-\frac{1}{s}\right|&=O\left(\exp\left\{-\frac{m}{48}\right\}\right),\label{eq:vlf_duration_projected_bias}\\
 \E{\left(\nn{\eta}_{j,m}-\frac{1}{s}\right)^2}
 &=\frac{2}{s^2m}+O\left(m^{-2}\right),\label{eq:vlf_duration_projected_variance}\\
 \E{\left|\nn{\eta}_{j,m}-\frac{1}{s}\right|^4}&=O\left(m^{-2}\right),
 \label{eq:vlf_duration_projected_moments}
\end{align}
and~\eqref{eq:vlf_duration_raw_moments} gives the third directly.

\subsubsection{Proof of~\eqref{eq:vlf_duration_coefficient}}

Write $p=P_{j,m+1}\leq2q$. Taylor's theorem at $\eta=\frac{1}{s}$, with the third derivative of $\CG(p\eta)$ bounded by $p^3$, gives
\begin{align}
 \left|R_{j,m}-\frac{p^2s^2}{4\left(p+s\right)^2}
 \left(\nn{\eta}_{j,m}-\frac{1}{s}\right)^2\right|
 &\leq\frac{p^3}{6}\left|\nn{\eta}_{j,m}-\frac{1}{s}\right|^3.
 \label{eq:vlf_duration_taylor_error}
\end{align}
We take expectations, using~\eqref{eq:vlf_duration_projected_variance} for the quadratic term and~\eqref{eq:vlf_duration_projected_moments} with Lyapunov's inequality for the cubic term. Since $q\leq1$, this yields
\begin{align}
 \E{R_{j,m}}
 &=\frac{p^2}{2\left(p+s\right)^2m}+O\left(q^2m^{-2}\right)+O\left(q^3m^{-\frac{3}{2}}\right)
 =\frac{p^2}{2\left(p+s\right)^2m}+O\left(q^2m^{-\frac{3}{2}}\right),
 \label{eq:vlf_duration_coefficient_expanded}
\end{align}
which is~\eqref{eq:vlf_duration_coefficient}.

\subsubsection{Proof of~\eqref{eq:vlf_stopped_precision_bias}}

We compare the stopped sum in~\eqref{eq:vlf_stopped_precision_bias} with the sum up to the deterministic time at which the true-precision reduction crosses the threshold,
\begin{align}
 \overline{\tau}^{(j)}&\triangleq\inf\left\{n\geq0\colon\overline F_{j,n}\geq\gamma_j\right\}.
 \label{eq:vlf_true_crossing}
\end{align}
Since $\gamma_j>0$ for sufficiently large $M$ by~\eqref{eq:vlf_packet_stop}, $\overline{\tau}^{(j)}\geq1$. Every increment of $F_{j,n}$ and of $\overline F_{j,n}$ is at least $\CG\left(\frac{q\eta_{\min}}{2}\right)$, so both $\tau^{(j)}$ and $\overline{\tau}^{(j)}$ are at most $\left\lceil\frac{\gamma_j}{\CG\left(\frac{q\eta_{\min}}{2}\right)}\right\rceil$. As the two reductions increase in $n$, $\tau^{(j)}>m$ if and only if $F_{j,m}<\gamma_j$, and $\overline{\tau}^{(j)}>m$ if and only if $\overline F_{j,m}<\gamma_j$.

The two reductions stay close. The derivative of $\CG\left(P_{j,n}\eta\right)$ in $\eta>0$ is at most $\frac{P_{j,n}}{2}\leq q$, and the first nine refinements use the precision $\frac{\eta_{\min}}{2}$. Writing $\left\|\cdot\right\|_4$ for the fourth-moment norm, we get from Minkowski's inequality and~\eqref{eq:vlf_duration_projected_moments}, for $m\geq9$,
\begin{align}
 \left\|F_{j,m}-\overline F_{j,m}\right\|_4
 &\leq9q\eta_{\max}
 +q\sum_{n=9}^{m-1}\left\|\nn{\eta}_{j,n}-\frac{1}{s}\right\|_4
 =O\left(q\sqrt m\right).
 \label{eq:vlf_duration_fourth_deviation}
\end{align}
As a consequence, the two stopping indicators rarely differ. Every increment of $\overline F_{j,n}$ is at least $\CG\left(q\eta_{\min}\right)\geq q\CG\left(\eta_{\min}\right)$, so $\overline F_{j,m}$ is at distance at least $\frac{q\CG\left(\eta_{\min}\right)}{2}\left|m-\overline{\tau}^{(j)}\right|$ from $\gamma_j$ whenever $\left|m-\overline{\tau}^{(j)}\right|\geq2$, and for such $m$,
\begin{align}
 \left\{\1{\tau^{(j)}>m}\ne\1{\overline{\tau}^{(j)}>m}\right\}
 &\subseteq\left\{\left|F_{j,m}-\overline F_{j,m}\right|\geq\frac{q\CG\left(\eta_{\min}\right)}{2}\left|m-\overline{\tau}^{(j)}\right|\right\}.
 \label{eq:vlf_duration_mismatch_event}
\end{align}
Markov's inequality for the fourth power and~\eqref{eq:vlf_duration_fourth_deviation} bound the probability of the right-hand side, and the bound one covers $\left|m-\overline{\tau}^{(j)}\right|\leq1$, so that
\begin{align}
 \Prob{\1{\tau^{(j)}>m}\ne\1{\overline{\tau}^{(j)}>m}}
 &=O\left(\min\left\{1,\frac{m^2}{\left(1+\left|m-\overline{\tau}^{(j)}\right|\right)^4}\right\}\right),\qquad m\geq9.
 \label{eq:vlf_duration_indicator_mismatch}
\end{align}

The coefficient $\frac{P_{j,m+1}s}{2\left(P_{j,m+1}+s\right)}$ in~\eqref{eq:vlf_stopped_precision_bias} is at most $q$. The stopped sum and the sum up to $\overline{\tau}^{(j)}-1$ differ only in the terms with $\1{\tau^{(j)}>m}\ne\1{\overline{\tau}^{(j)}>m}$, so
\begin{align}
 &\left|\E{\sum_{m=9}^{\tau^{(j)}-1}\frac{P_{j,m+1}s}{2\left(P_{j,m+1}+s\right)}\left(\nn{\eta}_{j,m}-\frac{1}{s}\right)}
 -\sum_{m=9}^{\overline{\tau}^{(j)}-1}\frac{P_{j,m+1}s}{2\left(P_{j,m+1}+s\right)}\E{\nn{\eta}_{j,m}-\frac{1}{s}}\right|
 \notag\\
 &\qquad\leq q\sum_{m=9}^{\infty}
 \E{\left|\nn{\eta}_{j,m}-\frac{1}{s}\right|
 \left|\1{\tau^{(j)}>m}-\1{\overline{\tau}^{(j)}>m}\right|}
 \label{eq:vlf_duration_stopped_prefix_comparison}\\
 &\qquad=O\left(q\sum_{m=9}^{\infty}m^{-\frac{1}{2}}
 \min\left\{1,\frac{m}{\left(1+\left|m-\overline{\tau}^{(j)}\right|\right)^2}\right\}\right)
 \label{eq:vlf_duration_summand}\\
 &\qquad=O\left(q\left(\frac{1}{\left(\overline{\tau}^{(j)}\right)^2}\sum_{m\leq\frac{\overline{\tau}^{(j)}}{2}}\sqrt m
 +\sum_{m\geq2\overline{\tau}^{(j)}}m^{-\frac{3}{2}}
 +\frac{1}{\sqrt{\overline{\tau}^{(j)}}}\sum_{i=0}^{\infty}\min\left\{1,\frac{2\overline{\tau}^{(j)}}{(1+i)^2}\right\}\right)\right)
 \label{eq:vlf_duration_ranges}\\
 &\qquad=O(q).
 \label{eq:vlf_duration_ranges_sum}
\end{align}
The Cauchy--Schwarz inequality with~\eqref{eq:vlf_duration_projected_variance} and~\eqref{eq:vlf_duration_indicator_mismatch} gives~\eqref{eq:vlf_duration_summand}. For~\eqref{eq:vlf_duration_ranges}, split at $m=\frac{\overline{\tau}^{(j)}}{2}$ and $m=2\overline{\tau}^{(j)}$. The two outer sums contribute $O\left(\left(\overline{\tau}^{(j)}\right)^{-\frac{1}{2}}\right)$. In the middle range, each distance $i=\left|m-\overline{\tau}^{(j)}\right|$ occurs at most twice; splitting the last sum at $i=\sqrt{\overline{\tau}^{(j)}}$ gives $O(1)$, proving~\eqref{eq:vlf_duration_ranges_sum}. The deterministic sum in~\eqref{eq:vlf_duration_stopped_prefix_comparison} is $O(q)$ because $\frac{P_{j,m+1}s}{2\left(P_{j,m+1}+s\right)}\leq q$ and~\eqref{eq:vlf_duration_projected_bias} is summable. The triangle inequality then gives~\eqref{eq:vlf_stopped_precision_bias}.

\subsection{Proof of \lemref{lem:vlf_packet_duration}}

\label{app:vlf_packet_duration_lemma}

Both bounds follow from~\eqref{eq:vlf_capacity_sum_expectation}. As in~\secref{sec:vlf_packet_duration}, write $s=\sigma_{J_j}^2$ and condition on $\mc{G}_{j,0}$.

\subsubsection{Preliminary and exploration packets, proof of~\eqref{eq:vlf_short_duration}}

For a preliminary or exploration packet, $P_{j,n}=q=P_{j,0}$ for every $n$, so $\overline F_{j,\tau^{(j)}}=\tau^{(j)}\CG\left(\frac{q}{s}\right)$. Concavity of $\CG$ and $\frac{1}{\ell_0}\leq q\leq1$ give
\begin{align}
 \frac{\gamma_j}{\CG\left(\frac{q\eta_{\min}}{2}\right)}
 &\leq\frac{\ell_0\gamma_j}{\CG\left(\frac{\eta_{\min}}{2}\right)}
 =O\left(\psi^2\ell_0\right),
 \label{eq:vlf_short_duration_limit}
\end{align}
where $\gamma_j=O(\psi^2)$ by~\secref{sec:vlf_deterministic}. The summation in~\eqref{eq:vlf_capacity_sum_expectation} is therefore $O(\log\psi)$, since its coefficient is at most $\frac12$ and $\ell_0=\left\lceil\psi^{\frac13}\right\rceil$. Counting the initial use and applying~\eqref{eq:vlf_capacity_sum_expectation}, \eqref{eq:vlf_threshold_excess}, and $\CG\left(\frac{q}{s}\right)\leq\frac{\eta_{\max}}2$ gives
\begin{align}
 \CG\left(\frac{q}{s}\right)\E{\tau^{(j)}+1\mid\mc{G}_{j,0}}
 &=\E{\overline F_{j,\tau^{(j)}}\,\middle|\,\mc{G}_{j,0}}+\CG\left(\frac{q}{s}\right)
 \leq\log M_j+O\left(\log\psi\right),
 \label{eq:vlf_short_duration_raw}
\end{align}
which proves~\eqref{eq:vlf_short_duration}.

\subsubsection{Final packet, proof of~\eqref{eq:vlf_final_duration}}

Set $j=B+1$ and $q=1$. The first $n_{\mr{b}}$ refinements use the power $1+\frac{\Omega}{n_{\mr{b}}}$, and the later ones use power one. By~\eqref{eq:vlf_boost_length_bound}--\eqref{eq:vlf_final_power_bounds}, each increment of $F_{j,n}$ is at most $\CG\left(3\eta_{\max}\right)$ and $n_{\mr{b}}\CG\left(3\eta_{\max}\right)=o(\log M)<\gamma_j$, so $\tau^{(j)}>n_{\mr{b}}$. The true-precision reduction therefore splits into the reduction at power one and the gain from the larger powers,
\begin{align}
 \overline F_{j,\tau^{(j)}}
 &=\tau^{(j)}\CG\left(\frac{1}{s}\right)
 +n_{\mr{b}}\left[\CG\left(\frac{1+\frac{\Omega}{n_{\mr{b}}}}{s}\right)-\CG\left(\frac{1}{s}\right)\right].
 \label{eq:vlf_final_capacity_identity}
\end{align}
The gain satisfies
\begin{align}
 n_{\mr{b}}\left[\CG\left(\frac{1+\frac{\Omega}{n_{\mr{b}}}}{s}\right)-\CG\left(\frac{1}{s}\right)\right]
 &=\frac{n_{\mr{b}}}{2}\log\left(1+\frac{\Omega}{n_{\mr{b}}\left(1+s\right)}\right)
 \label{eq:vlf_final_gain_identity}\\
 &\geq\frac{\Omega}{2\left(1+s\right)}-\frac{\Omega^2}{4n_{\mr{b}}\left(1+s\right)^2}
 \geq\frac{\Omega}{2\left(1+s\right)}-\frac{1}{4\left(1+s\right)^2},
 \label{eq:vlf_final_saved_credit}
\end{align}
by the definition of $\CG$, $\log\left(1+x\right)\geq x-\frac{x^2}{2}$ for $x\geq0$, and $n_{\mr{b}}\geq1+\Omega^2$ from~\eqref{eq:vlf_saved_power}. For the true-precision reduction, \eqref{eq:vlf_capacity_sum_expectation} with $q=1$ gives
\begin{align}
 \E{\overline F_{j,\tau^{(j)}}\,\middle|\,\mc{G}_{j,0}}
 &\leq\gamma_j
 +\frac{1}{2\left(1+s\right)^2}\sum_{m=9}^{\left\lceil\frac{\gamma_j}{\CG\left(\frac{\eta_{\min}}{2}\right)}\right\rceil-1}\frac{1}{m}
 +\frac{\Omega}{4n_{\mr{b}}}\sum_{m=1}^{n_{\mr{b}}}\frac{1}{m}+O(1)
 \label{eq:vlf_final_capacity_coefficient}\\
 &\leq\log M_j+\frac{\psi}{2\left(1+s\right)^2}+O\left(\log \psi\right).
 \label{eq:vlf_final_capacity_bound}
\end{align}
For~\eqref{eq:vlf_final_capacity_coefficient}, only $m<n_{\mr{b}}$ has power larger than one. The derivative of $\frac{p^2}{2\left(p+s\right)^2}$ is at most $\frac{4}{27p}<\frac{1}{4}$ for $p\geq1$, so the coefficient increases by at most $\frac{\Omega}{4n_{\mr{b}}}$. In~\eqref{eq:vlf_final_capacity_bound}, the first harmonic sum is at most $\psi+O(1)$ because its upper limit is $O\left(\log M\right)$. The second contributes at most $\frac{\Omega}{2\sqrt{n_{\mr{b}}}}\leq\frac{1}{2}$ by $\sum_{m=1}^{n}\frac{1}{m}\leq2\sqrt n$, and $\gamma_j=\log M_j+O\left(\log\psi\right)$. Taking expectations in~\eqref{eq:vlf_final_capacity_identity}, using~\eqref{eq:vlf_final_saved_credit} and~\eqref{eq:vlf_final_capacity_bound}, and adding $\CG\left(\frac{1}{s}\right)\leq\frac{\eta_{\max}}{2}$ for the initial use gives
\begin{align}
 \CG\left(\frac{1}{s}\right)\E{\tau^{(j)}+1\mid\mc{G}_{j,0}}
 &\leq\log M_j-\frac{\Omega}{2\left(1+s\right)}+\frac{\psi}{2\left(1+s\right)^2}+O\left(\log\psi\right),
 \label{eq:vlf_final_duration_proved}
\end{align}
which gives~\eqref{eq:vlf_final_duration} and completes the proof of \lemref{lem:vlf_packet_duration}.

\subsection{Proof of \lemref{lem:vlf_channel_selection}}

\label{app:vlf_channel_selection}

If $K=1$, then $B=0$ and $J_0=J_{B+1}=1$, and the lemma holds. Assume that $K\geq2$. The implicit constants below may depend on $K$, the instance, the variance bounds, and the parameters $u$, $\delta$, and $\delta'$ introduced below, but not on $M$ or on the history. Part~1 shows that the variance estimates are accurate with high probability, and parts~2 to~5 prove~\eqref{eq:vlf_preliminary_channel_error}, \eqref{eq:vlf_suboptimal_uses}--\eqref{eq:vlf_best_uses_packets}, \eqref{eq:vlf_channel_errors}, and~\eqref{eq:vlf_weighted_power}, in this order.

\subsubsection{Variance estimates}

The retained observations consist of the initial zero-input outputs and the normalized outputs of packets $0,1,\ldots,B$. By \lemref{lem:vlf_packet_gaussian}, each next observation on channel $k$ has conditional law $\mc{N}(0,\sigma_k^2)$, and its channel depends only on $U$ and preceding retained observations. The reward-stack representation in \appref{app:awgn_plugin_loss} therefore applies, with stacks independent of one another and of $U$. Let $\nn{s}_{k,n}$ be the empirical second moment of channel $k$'s first $n$ stack entries, so $\nn{s}_k=\nn{s}_{k,\widetilde L_k}$ at packet boundaries. For every integer $n\geq1$, $x>0$, and $0<\delta\leq\frac{\sigma_{\min}^2}{2}$,
\begin{align}
 \Prob{n\,\dG\left(\nn{s}_{k,n}\|\sigma_k^2\right)\geq x}
 &\leq2\exp\{-x\},
 \label{eq:vlf_variance_confidence}\\
 \Prob{\exists\,k\in[K],\ m\geq n\colon\left|\nn{s}_{k,m}-\sigma_k^2\right|>\delta}
 &\leq\frac{2K\exp\left\{-\frac{n\delta^2}{6\sigma_{\max}^4}\right\}}
 {1-\exp\left\{-\frac{\delta^2}{6\sigma_{\max}^4}\right\}}.
 \label{eq:vlf_prefix_absolute_uniform}
\end{align}
Equation~\eqref{eq:vlf_variance_confidence} follows from~\cite[Lemma~4]{menardGarivier2017}, applied to both tails at fixed $n$. The bound $\dG\left(\sigma_k^2\pm\delta\|\sigma_k^2\right)\geq\frac{\delta^2}{6\sigma_{\max}^4}$ together with~\eqref{eq:vlf_variance_confidence} at $x=\frac{m\delta^2}{6\sigma_{\max}^4}$ and a union bound over $k$ and $m\geq n$ then gives~\eqref{eq:vlf_prefix_absolute_uniform}.

Before the final packet, every count $\widetilde L_k$ is at most the deterministic number $\overline L=O\left(\psi^2\right)$ obtained by adding $K\ell_0$ and the upper bounds in~\eqref{eq:vlf_preliminary_bounds} and~\eqref{eq:vlf_exploration_cap_length}. Let $\mc{E}$ be the event that $n\,\dG\left(\nn{s}_{k,n}\|\sigma_k^2\right)\leq b$ for every $k\in[K]$ and $1\leq n\leq\overline L$. On $\mc{E}$, every queried interval~\eqref{eq:vlf_confidence} contains its channel's true variance. If $k^\star$ is active, then $\inf\mc{I}_{k^\star}\leq\sigma_{k^\star}^2\leq\min_{k\in\mc{A}}\sigma_k^2\leq\min_{k\in\mc{A}}\sup\mc{I}_k$, so the elimination rule~\eqref{eq:vlf_elimination} retains $k^\star$. The best channel therefore stays active throughout exploration on $\mc{E}$, and a singleton active set consists of $k^\star$.

Since the best channel is unique, we can fix $u$ and $\delta$ with
\begin{align}
 0<\delta&\leq\frac{\sigma_{\min}^2}{2},
 \qquad
 \sigma_{k^\star}^2+\delta<u<
 \min_{k\neq k^\star}\left(\sigma_k^2-\delta\right).
 \label{eq:vlf_comparison_variance}
\end{align}
Let $\mc{V}_\delta$ be the event that $\left|\nn{s}_{k,m}-\sigma_k^2\right|\leq\delta$ for every $k\in[K]$ and $m\geq\ell_0$. For the simultaneous confidence event $\mc{E}$, the union bound with~\eqref{eq:vlf_packet_budget} and~\eqref{eq:vlf_variance_confidence} at $x=b$, and then~\eqref{eq:vlf_prefix_absolute_uniform} with $n=\ell_0$ and $\ell_0\geq\psi^{\frac{1}{3}}$, give
\begin{align}
 \Prob{\mc{E}^{\mr{c}}}
 &\leq2K\overline L\exp\{-b\}
 =O\left(\frac{1}{\log M}\right),
 \label{eq:vlf_confidence_failure}\\
 \Prob{\left(\mc{E}\cap\mc{V}_\delta\right)^{\mr{c}}}
 &=O\left(\frac{1}{\log M}\right)+O\left(\exp\left\{-\frac{\ell_0\delta^2}{6\sigma_{\max}^4}\right\}\right)
 =o\left(\psi^{-1}\right).
 \label{eq:vlf_count_exception_probability}
\end{align}

\subsubsection{Preliminary channel, proof of~\eqref{eq:vlf_preliminary_channel_error}}

On $\mc{V}_\delta$, the initial estimates $\nn{s}_{k,\ell_0}$ are within $\delta$ of the true variances, and projection onto $\left[\sigma_{\min}^2,\sigma_{\max}^2\right]$ preserves this accuracy. By~\eqref{eq:vlf_comparison_variance}, the projected estimate of the best channel is then the smallest, so $J_0=k^\star$ by~\eqref{eq:vlf_preliminary_channel}. Consequently, by~\eqref{eq:vlf_prefix_absolute_uniform} with $n=\ell_0$ and $\ell_0\geq\psi^{\frac{1}{3}}$,
\begin{align}
 \Prob{J_0\neq k^\star}
 &\leq\Prob{\mc{V}_\delta^{\mr{c}}}
 =O\left(\exp\left\{-\frac{\delta^2}{6\sigma_{\max}^4}\psi^{\frac{1}{3}}\right\}\right),
 \label{eq:vlf_preliminary_error_proved}
\end{align}
Since $\psi\exp\left\{-\frac{\delta^2}{6\sigma_{\max}^4}\psi^{\frac13}\right\}\to0$, this proves~\eqref{eq:vlf_preliminary_channel_error}.

\subsubsection{Exploration uses, proof of~\eqref{eq:vlf_suboptimal_uses}--\eqref{eq:vlf_best_uses_packets}}

For $k\neq k^\star$, let $n_k\triangleq1+\left\lceil\frac{b}{\dG\left(\sigma_k^2-\delta\|u\right)}\right\rceil$. Since $\frac{b}{\psi}\to1$, each $n_k$ is of order $\psi$ and exceeds $\ell_0$ for sufficiently large $M$. We first show that, on $\mc{E}\cap\mc{V}_\delta$, the best channel is never explored, and a suboptimal channel $k$ is explored only until its count reaches $n_k$.

Consider a history in $\mc{E}\cap\mc{V}_\delta$. After the preliminary packet, which uses the channel $J_0=k^\star$ by part~2, the best channel has the count $\widetilde L_{k^\star}\geq\tau^{(0)}\geq\frac{\log M_0}{\CG\left(2\eta_{\max}\right)}$ by~\eqref{eq:vlf_preliminary_bounds}, and its estimate satisfies $\nn{s}_{k^\star}\leq\sigma_{k^\star}^2+\delta<u$. A suboptimal channel $k$ with $\widetilde L_k\geq n_k$ has the estimate $\nn{s}_k\geq\sigma_k^2-\delta>u$. Since $\dG\left(a\|u\right)$ decreases in $a<u$ and increases in $a>u$, for sufficiently large $M$,
\begin{align}
 \widetilde L_{k^\star}\dG\left(\nn{s}_{k^\star}\|u\right)
 &\geq\frac{\log M_0}{\CG\left(2\eta_{\max}\right)}\dG\left(\sigma_{k^\star}^2+\delta\|u\right)>b,
 \label{eq:vlf_best_upper_endpoint}\\
 \widetilde L_k\dG\left(\nn{s}_k\|u\right)
 &\geq n_k\dG\left(\sigma_k^2-\delta\|u\right)>b,
 \qquad \widetilde L_k\geq n_k.
 \label{eq:vlf_suboptimal_lower_endpoint}
\end{align}
The divergence $\dG\left(a\|\widetilde\sigma^2\right)$ increases when $\widetilde\sigma^2$ moves away from $a$. Thus, \eqref{eq:vlf_best_upper_endpoint}--\eqref{eq:vlf_suboptimal_lower_endpoint} give $\sup\mc{I}_{k^\star}<u<\inf\mc{I}_k$, and the elimination rule~\eqref{eq:vlf_elimination} removes channel $k$ at the next decision.

Two consequences follow. First, a suboptimal channel starts an exploration packet only while $\widetilde L_k<n_k$, and its last packet adds $O\left(\ell_0^2\right)$ observations by~\eqref{eq:vlf_exploration_bounds}. Thus, its exploration packets contain at most $n_k+O\left(\ell_0^2\right)$ refinements. Each packet has at least $\frac{\ell_0\log2}{\CG\left(2\eta_{\max}\right)}$ refinements by~\eqref{eq:vlf_exploration_lower}, so including one initial use per packet gives
\begin{align}
 T_k
 &\leq\left(1+\frac{\CG\left(2\eta_{\max}\right)}{\ell_0\log2}\right)\left(n_k+O\left(\ell_0^2\right)\right)
 =n_k+O\left(\psi^{\frac{2}{3}}\right),
 \qquad k\neq k^\star,
 \label{eq:vlf_physical_count_on_event}
\end{align}
where $n_k=O(\psi)$ and $\ell_0=\left\lceil\psi^{\frac13}\right\rceil$.
Second, at every decision, an active suboptimal channel has fewer than $n_k=O(\psi)$ observations, whereas $\widetilde L_{k^\star}$ is of order $\psi^2$, so the minimum-count rule never selects the best channel, and $T_{k^\star}=0$.

The event $\mc{E}^{\mr{c}}\cup\mc{V}_\delta^{\mr{c}}$ contributes $o(\psi)$ to $\E{T_k}$ by~\eqref{eq:vlf_exploration_cap_length} and~\eqref{eq:vlf_count_exception_probability}. Taking expectations in~\eqref{eq:vlf_physical_count_on_event} and using $\frac{b}{\psi}\to1$, we get
\begin{align}
 \limsup_{M\to\infty}\frac{\E{T_k}}{\psi}
 &\leq\frac{1}{\dG\left(\sigma_k^2-\delta\|u\right)},
 \qquad k\neq k^\star.
 \label{eq:vlf_suboptimal_uses_delta}
\end{align}
Letting $\delta\downarrow0$ at fixed $u$, and then $u\downarrow\sigma_{k^\star}^2$, replaces the right-hand side by $\frac{1}{\dG\left(\sigma_k^2\|\sigma_{k^\star}^2\right)}$, by continuity of $\dG$; this proves~\eqref{eq:vlf_suboptimal_uses}. With $T_{k^\star}=0$ on $\mc{E}\cap\mc{V}_\delta$, the same bound on $\mc{E}^{\mr{c}}\cup\mc{V}_\delta^{\mr{c}}$ gives~\eqref{eq:vlf_best_exploration_uses}. Finally, $B\leq\frac{\CG\left(2\eta_{\max}\right)}{\ell_0\log2}\sum_{k=1}^KT_k$ by~\eqref{eq:vlf_exploration_lower}, and $\sum_{k=1}^K\E{T_k}=O(\psi)$ by~\eqref{eq:vlf_suboptimal_uses}--\eqref{eq:vlf_best_exploration_uses}. These two bounds and $\ell_0\geq\psi^{\frac{1}{3}}$ give~\eqref{eq:vlf_best_uses_packets}.

\subsubsection{Final channel, proof of~\eqref{eq:vlf_channel_errors}}

On $\mc{E}$, neither an empty interval nor an incorrect singleton selects the final channel, so the final channel can be incorrect on $\mc{E}$ only if exploration reaches its cap with at least two active channels. On such a history, evaluate $\mc{A}$ and the counts $\widetilde L_k$ at the cap decision. Then, for sufficiently large $M$,
\begin{align}
 \min_{k\in\mc{A}}\widetilde L_k
 &\geq\frac{1}{K}\sum_{j=1}^{B}\tau^{(j)}-O\left(\ell_0^2\right)
 \geq\frac{\psi^2}{2K}-O\left(\ell_0^2\right)
 \geq\frac{\psi^2}{3K}.
 \label{eq:vlf_survivor_count_lower}
\end{align}
To see the first inequality, consider a channel of $\mc{A}$ with the smallest count. Since it stays active, the minimum-count rule bounds each explored channel's count before its last packet by this final count. By~\eqref{eq:vlf_exploration_lower}--\eqref{eq:vlf_exploration_bounds}, each packet adds $O\left(\ell_0^2\right)$ observations, so summing these bounds gives the first inequality. At the cap, exploration uses at least $\lceil\psi^2\rceil$ channel uses, with an initial-use fraction at most $\frac{\CG\left(2\eta_{\max}\right)}{\ell_0\log2}$, so $\sum_{j=1}^{B}\tau^{(j)}\geq\frac{\psi^2}{2}$.

By~\eqref{eq:vlf_prefix_absolute_uniform} with $n=\left\lceil\frac{\psi^2}{3K}\right\rceil$, except on an event of probability $O\left(\exp\left\{-\frac{\delta^2\psi^2}{18K\sigma_{\max}^4}\right\}\right)$, every channel in $\mc{A}$ has an estimate within $\delta$ of its true variance. With all counts at least $\frac{\psi^2}{3K}$, the fixed positive divergences in~\eqref{eq:vlf_best_upper_endpoint}--\eqref{eq:vlf_suboptimal_lower_endpoint} and $b=O(\psi)$ give $\sup\mc{I}_{k^\star}<u<\inf\mc{I}_k$ for every $k\in\mc{A}\setminus\{k^\star\}$ and sufficiently large $M$. Only $k^\star$ remains after elimination, contradicting two active channels. The exceptional probability is $o\left(\frac{1}{\log M}\right)$, so~\eqref{eq:vlf_confidence_failure} gives
\begin{align}
 \Prob{J_{B+1}\neq k^\star}
 &\leq\Prob{\mc{E}^{\mr{c}}}+O\left(\exp\left\{-\frac{\delta^2\psi^2}{18K\sigma_{\max}^4}\right\}\right)
 =O\left(\frac{1}{\log M}\right),
 \label{eq:vlf_final_channel_proved}
\end{align}
which proves~\eqref{eq:vlf_channel_errors}.

\subsubsection{Exploration powers, proof of~\eqref{eq:vlf_weighted_power}}

Fix $0<\delta'\leq\frac{\sigma_{\min}^2}{2}$ with $2\delta'<\min_{k\neq k^\star}\left(\sigma_k^2-\sigma_{k^\star}^2\right)$, and let $\mc{V}_{\delta'}$ be $\mc{V}_\delta$ with $\delta'$ in place of $\delta$. On $\mc{V}_{\delta'}$, part~2 gives $J_0=k^\star$, and all estimates at packet starts are within $\delta'$ of their true variances. For $k\neq k^\star$, $P_k^\star=\Pi_{[0,1]}\left(1+\sigma_{k^\star}^2-\sigma_k^2\right)$. The $1$-Lipschitz property of projection and the lower-endpoint change of $\frac{1}{\ell_0}$ in~\eqref{eq:vlf_exploration_power} give $\left|P_{j,0}-P_k^\star\right|\leq2\delta'+\frac{1}{\ell_0}$ on $\mc{V}_{\delta'}$ for every exploration packet on channel $k$. On the complement, $\left|P_{j,0}-P_k^\star\right|\leq1$. Splitting the expectation and applying~\eqref{eq:vlf_exploration_cap_length} and~\eqref{eq:vlf_prefix_absolute_uniform} gives
\begin{align}
 \E{\sum_{j=1}^{B}\left|P_{j,0}-P_k^\star\right|
       \left(\tau^{(j)}+1\right)\1{J_j=k}}
 &\leq
 \left(2\delta'+\frac{1}{\ell_0}\right)\E{T_k}
 +O\left(\psi^2\exp\left\{-\frac{\ell_0\delta'^2}{6\sigma_{\max}^4}\right\}\right)
 \label{eq:vlf_weighted_power_proved}\\
 &\leq\frac{2\delta'\psi}{\dG\left(\sigma_k^2\|\sigma_{k^\star}^2\right)}+o(\psi),
 \label{eq:vlf_weighted_power_limit}
\end{align}
the last step by~\eqref{eq:vlf_suboptimal_uses} and $\ell_0\geq\psi^{\frac{1}{3}}$. Letting $\delta'\downarrow0$ gives~\eqref{eq:vlf_weighted_power}, and the proof of \lemref{lem:vlf_channel_selection} is complete.

\section*{Acknowledgment}

The core ideas of this paper are the author's, including the problem formulation, the new coding schemes, and all the main arguments of the proofs. In preparing the manuscript, the author used two AI systems, Claude Opus~5.5 (Anthropic, Max effort) and GPT-6 Astra (OpenAI, Ultra effort). They helped draft, revise, and proofread the text; prepared the figures; helped develop and check the proofs of the auxiliary results; and carried out some routine calculations. In the design of the VLSF-Bandit scheme, both systems helped the author find that the ratio-of-mixtures score in~\eqref{eq:Skn_def}--\eqref{eq:Sn_def} yields a better second-order term than the score he had initially considered, the logarithm of the exponentiated information density averaged over the Jeffreys prior. The author independently verified every mathematical statement and proof in the final manuscript and checked each cited result against its original source. The author alone is responsible for the entire content of the manuscript, including all claims, proofs, and interpretations, and for any remaining errors.

\bibliographystyle{IEEEtran}
\bibliography{mac}
\end{document}